\documentclass[preprintnumbers, floatfix, letterpaper, twocolumn,aps,prd,epsfig,nofootinbib,natbib,longbibliography]{revtex4-1}
\usepackage{graphicx}
\usepackage{epstopdf}
\usepackage{latexsym}
\usepackage{amssymb}
\usepackage{amsmath}
\usepackage{color}
\usepackage{mathrsfs}
\usepackage{xparse}
\usepackage{bbding}
\usepackage{pifont}
\usepackage{comment}
\usepackage{ulem}
\usepackage{float}
\usepackage[inline]{enumitem}
\usepackage{soul}
\usepackage[caption=false]{subfig}

\let\Right\right
\let\Left\left
\makeatletter
\def\right#1{\Right#1\@ifnextchar){\!\right}{}}
\def\left#1{\Left#1\@ifnextchar({\!\left}{}}
\makeatother

\usepackage[
            pdfstartview=FitH,
            bookmarksnumbered=true,
            bookmarksopen=true,
            colorlinks,
            linkcolor=blue,
            anchorcolor=green,
            citecolor=blue
            ]{hyperref}
\usepackage{orcidlink}
\providecommand{\orcid}[1]{\orcidlink{#1}}
\begin{document}

  \renewcommand\arraystretch{2}
 \newcommand{\bq}{\begin{equation}}
 \newcommand{\eq}{\end{equation}}
 \newcommand{\bqn}{\begin{eqnarray}}
 \newcommand{\eqn}{\end{eqnarray}}
 \newcommand{\nb}{\nonumber}
 \newcommand{\lb}{\label}
 
\newcommand{\La}{\Lambda}
\newcommand{\va}{\scriptscriptstyle}
\newcommand{\be}{\nopagebreak[3]\begin{equation}}
\newcommand{\ee}{\end{equation}}

\newcommand{\ba}{\nopagebreak[3]\begin{eqnarray}}
\newcommand{\ea}{\end{eqnarray}}

\newcommand{\la}{\label}
\newcommand{\n}{\nonumber}

\newcommand{\R}{\mathbb{R}}

 \newcommand{\cb}{\color{blue}}
    \newcommand{\cc}{\color{cyan}}
        \newcommand{\cm}{\color{magenta}}
\newcommand{\rc}{\rho^{\scriptscriptstyle{\mathrm{I}}}_c}
\newcommand{\rd}{\rho^{\scriptscriptstyle{\mathrm{II}}}_c} 
\NewDocumentCommand{\evalat}{sO{\big}mm}{%
  \IfBooleanTF{#1}
   {\mleft. #3 \mright|_{#4}}
   {#3#2|_{#4}}%
}
\newcommand{\PRL}{Phys. Rev. Lett.}
\newcommand{\PL}{Phys. Lett.}
\newcommand{\PR}{Phys. Rev.}
\newcommand{\CQG}{Class. Quantum Grav.}

\title{Quasi-Normal Modes of Covariant Loop Quantum Black Holes}

\author{Chao Zhang \orcid{0000-0001-9915-8649}${}^{a, b}$}
\email{ phyzc@cup.edu.cn}

 \author{Sushant G. Ghosh\orcid{0000-0002-0835-3690}${}^{c,d}$}
 \email{sghosh2@jmi.ac.in}

\author{Anzhong Wang \orcid{0000-0002-8852-9966}${}^{e}$}
\email{ anzhong$\_$wang@baylor.edu} 

\affiliation{${}^{a}$Basic Research Center for Energy Interdisciplinary, College of Science, China University of Petroleum-Beijing, Beijing 102249, China\\
${}^{b}$Beijing Key Laboratory of Optical Detection Technology for Oil and Gas, China University of Petroleum-Beijing, Beijing 102249, China\\
${}^{c}$Centre for Theoretical Physics, Jamia Millia Islamia, New Delhi 110025, India\\
${}^{d}$Astrophysics and Cosmology Research Unit, School of Mathematics, Statistics and Computer Science,
University of KwaZulu-Natal, Private Bag 54001, Durban 4000, South Africa\\
${}^{e}$ GCAP-CASPER, Department of Physics and Astronomy, Baylor University, Waco, Texas 76798-7316, USA}

\date{\today}

\begin{abstract}

 We investigate the quasinormal modes (QNMs) and time-domain response of a massless, minimally coupled test scalar field on static, spherically symmetric covariant loop quantum black hole (cLQBH) backgrounds. Focusing on a phenomenological, Gaussian-localized holonomy profile, we obtain a localized deformation of the scalar effective potential containing both positive and negative parts. This provides a geometric realization of the bump effect, with the potential deformation and tortoise-coordinate mapping determined consistently by the same background. We characterize the deformation by its location, amplitude, and width, and study its effects using frequency-domain shooting and time-domain evolution, supplemented by matrix-pencil mode extraction and waveform reconstruction. The QNM spectrum exhibits spiral trajectories as the bump location varies, enhanced sensitivity of the overtones, and mode overtaking, whereby the branch continued from the first Schwarzschild overtone becomes longer-lived than the continued fundamental branch. Despite substantial spectral changes, the early response can remain close to its Schwarzschild counterpart, with more pronounced differences emerging later. To quantify this behavior, we introduce a local waveform-deviation time and compare it with a relative-amplitude faintness time, thereby defining an operational observation window. The deviation time depends nonmonotonically on the bump location, and suitable parameter choices allow appreciable departures to emerge before the adopted faintness threshold is reached. These results connect localized holonomy corrections with spectral sensitivity and waveform evolution, providing a basis for future gravitational-perturbation studies and assessments of gravitational-wave detectability.

\end{abstract}


\maketitle
\section{Introduction}

\renewcommand{\theequation}{1.\arabic{equation}} \setcounter{equation}{0}
\label{Introduction}

During the past century, general relativity (GR) has passed a wide range of experimental tests since its birth (see, e.g., \cite{Willbook}). These include the observation of gravitational light deflection during the solar eclipse of 1919, tests of the strong equivalence principle motivated by Nordtvedt's work in 1968 \cite{Ericbook}, and measurements of the orbital decay of the binary pulsar PSR B1913+16, discovered by Hulse and Taylor in 1974, to name a few. So far, GR remains one of the most successful theories of gravity because of its predictive accuracy over a wide range of scales. Nevertheless, the nature of dark matter, dark energy, and the quantization of gravity remain open questions. These questions motivate both further tests of GR and the development of theories that go beyond it.

The history of modified theories of gravity can be traced back to the early 1920s \cite{Ericbook}. Many such theories agree with solar system tests, where gravity is relatively weak, but can differ from GR in other regimes. Observations of binary pulsars have therefore made it possible to rule out or tightly constrain many alternatives \cite{Willbook}, while others remain viable and help us explore the nature of gravity. From the Lagrangian, or equivalently the action, point of view, one can modify GR by introducing higher derivatives \cite{Nojiri:2006ri}, adding additional fields \cite{Will:1977wq}, or relaxing assumptions such as local Lorentz invariance \cite{Jacobson:2007veq,Wang:2017brl}.

On the other hand, loop quantum gravity (LQG) provides an important approach to the quantization of gravity. By constructing a quantum theory of Riemannian geometry, it offers a framework for investigating quantum gravitational effects, including the possible resolution of classical singularities \cite{Ashtekar:2021kfp}. Ideas and methods from LQG have also been applied to symmetry-reduced cosmological models in loop quantum cosmology (LQC) \cite{Ashtekar:2011ni,Agullo:2023rqq,Li:2023dwy}. This has enabled detailed studies of quantum corrections and singularity resolution in these models \cite{Li:2022dwm,Ashtekar:2011ni,Agullo:2023rqq,Li:2023dwy}. In widely studied homogeneous LQC models, quantum geometrical effects replace the classical big bang or big crunch singularity by a regular bounce, connecting contracting and expanding branches of spacetime.

Black holes (BHs) are among the most mysterious objects in the universe and are a major prediction of GR \cite{Schwarzschild:1916uq}. Interest in BHs has grown further with the Event Horizon Telescope (EHT) observations of the shadows of M87* and Sagittarius $A^*$, the latter at the center of our own Milky Way galaxy \cite{EventHorizonTelescope:2019dse,EventHorizonTelescope:2022wkp}. Many galaxies are believed to host supermassive BHs at their centers \cite{Hawkingbook,Zhang:2021bdr}. Their strong gravitational fields and interactions with surrounding matter provide opportunities to study phenomena such as relativistic jets and tidal disruption. From the theoretical point of view, BHs and other strong-field systems also provide opportunities for testing theories of gravity beyond GR (see, e.g., \cite{Berti:2015itd,Barack:2018yly,LIGOScientific:2021sio,Zhao:2019kif,Zhang:2019iim,Zhang:2020too,Zhang:2023kzs,Carson:2020ter}). This role also extends to investigating possible quantum gravitational effects in effective BH models motivated by LQG \cite{Ashtekar:2023cod,Gambini:2022hxr}. Under suitable assumptions, the classical singularity theorems imply geodesic incompleteness, indicating limitations of the classical description \cite{MTW2017book}. Studies of effective LQG models suggest that quantum geometry can resolve the spacelike singularity inside BHs. Singularity resolution has also been extensively studied in LQC \cite{Ashtekar:2011ni,Agullo:2023rqq,Li:2023dwy}. Such studies often use effective equations to describe how quantum corrections modify the classical geometry.

The test of gravitational theories continues in the strong-field regime. In particular, the first direct detection of gravitational waves (GWs) from the coalescence of two BHs by Advanced LIGO marked the beginning of GW astronomy \cite{LIGOScientific:2016aoc}. Since then, the LIGO/Virgo/KAGRA (LVK) collaborations have reported hundreds of compact binary coalescence candidates, with catalogs and public data supporting increasingly detailed studies \cite{LIGOScientific:2026wfs,LIGOScientific:2018mvr,LIGOScientific:2019lzm,LIGOScientific:2020aai,KAGRA:2021vkt,LIGOScientific:2025slb,caltechlink}. Planned ground-based detectors and space-based missions aim to improve sensitivity and extend the frequency ranges accessible to GW observations \cite{Moore:2014lga,Gong:2021gvw}. These include the Cosmic Explorer \cite{CE}, Einstein Telescope \cite{ET}, LISA \cite{LISA}, TianQin \cite{Liu:2020eko,Shi:2019hqa,Luo:2025ewp}, Taiji \cite{Ruan:2018tsw}, and DECIGO \cite{Kawamura:2020pcg}. Together, they could probe a wider range of sources and distances. These prospects have stimulated studies of BH quasinormal modes (QNMs) \cite{Berti:2009kk,Berti:2018vdi}, extreme mass ratio inspirals (EMRIs) \cite{Zhang:2022rfr,Tu:2023xab,Zhang:2025wni}, and the effects of dark matter (DM) on BH dynamics and GW signals \cite{Zhang:2021bdr,Zhang:2022roh,Cardoso:2021wlq,Shen:2023erj,Shen:2024qbb,Maeda:2024tsg,Fier:2026yjc,Speeney:2022ryg,Zhao:2026yis,Li:2026obt,Deng:2026jgk}, to name a few.

During the ringdown stage of a coalescence, the newly formed BH approaches a stationary state. When linear perturbation theory applies, its response can be approximately described by a superposition of QNMs \cite{Regge:1957td,Teukolsky:1973ha,Schutz:1985km,Leaver:1985ax,Iyer:1986np,Chandra92book}, although nonlinear contributions can also be important, especially in the early ringdown \cite{Cheung:2022rbm,Mitman:2022qdl}. Scalar-field, vector (electromagnetic), and tensor (gravitational) perturbations and their associated QNM problems have been extensively studied in GR and in modified theories of gravity. In particular, test scalar fields provide a relatively simple setting for studying how the background geometry affects wave propagation and QNM spectra \cite{Berti:2009kk}. From the theoretical point of view, QNMs are characteristic modes of dissipative systems with generally complex frequencies $\omega$. The real part determines the angular frequency of oscillation, while the imaginary part determines the decay or growth rate. For a damped mode with time dependence $e^{-i\omega t}$, the damping time is $1/|\operatorname{Im}\omega|$. These complex frequencies carry information about the background geometry and perturbation dynamics, providing a way to test gravity in the strong-field regime. In four-dimensional GR, the standard no-hair result states that stationary, asymptotically flat electrovacuum BHs are characterized by mass, charge, and angular momentum under suitable assumptions \cite{Berti:2009kk,Berti:2018vdi}. For a specified perturbing field and mode, their QNM frequencies are therefore fixed by these parameters. Measuring multiple modes can test the consistency of these relations. The dependence of QNMs on the background geometry and perturbation sector has also been explored in specific BH models \cite{Zhang:2021bdr,Zhang:2022roh,Hou:2024odb}. An important example is the isospectrality of axial and polar gravitational perturbations of the Schwarzschild BH \cite{Chandra92book}. For more details, we refer readers to \cite{Berti:2009kk,Berti:2018vdi,Kokkotas:1999bd,Konoplya:2011qq}.

On the other hand, from the observational point of view, identifying modes beyond the dominant ringdown contribution remains challenging with current detectors \cite{Carullo:2018sfu}. Searches for overtones in GW150914 have led to extensive discussion of the effects of noise, the choice of fitting interval, and the waveform model \cite{Isi:2019aib,CalderonBustillo:2020rmh,Finch:2022ynt,Ma:2023vvr,Isi:2023nif,Carullo:2023gtf}. Evidence for a subdominant angular mode has also been reported for GW190521 \cite{Capano:2021etf}. The high signal-to-noise ratio of GW250114 has provided stronger evidence for the first overtone of the dominant quadrupolar mode and enabled more precise tests of the Kerr spectrum \cite{LIGOScientific:2025rid,LIGOScientific:2025wao,Suzuki:2026tir}. Future third-generation ground-based detectors and space-based missions are expected to improve the prospects for resolving subleading modes in sufficiently loud events \cite{Berti:2018vdi,Shi:2024ttu}. Interpreting such observations within theories beyond GR requires predictions for how changes in the background geometry and perturbation dynamics affect BH ringing. 
We study this question in an effective LQG framework using scalar perturbations as a first step. Although these perturbations do not directly describe GWs, they provide a simpler setting in which to examine how holonomy corrections affect the QNM spectrum and the time-domain response. Such studies can probe the dynamical consequences of the effective theory's holonomy modifications \cite{Bojowald:2024lhr}. This study forms part of a broader research program that we aim to extend to gravitational perturbations and their observable GW signatures.

This paper focuses on a class of covariant loop quantum black hole solutions \cite{Belfaqih:2024vfk,Belfaqih:2025eak}, obtained by applying emergent modified gravity to holonomy modifications motivated by LQG \cite{Bojowald:2023xat,Bojowald:2024naz}. To distinguish them from other loop quantum black holes (LQBHs), we refer to them as cLQBHs. In the class considered here, the solutions are characterized by two functions $\chi(r)$ and $\lambda(r)$. We restrict attention to $\chi(r)=1$ and focus on the localized profile of the dimensionless function $\lambda(r)$ in case (ii), with the previously studied decreasing profile in case (i) included for reference,
\begin{equation}
\label{eq1.1}
\begin{aligned}
\text{(i)}\quad &\lambda(r)=\frac{\sqrt{\Delta}}{r},\\
\text{(ii)}\quad &\lambda(r)=\epsilon
\left(\frac{r}{r_0}\right)^{-\frac{3M}{r_0-2M}}
\exp\!\left[-\frac{(r-r_0)^2}{2\sigma^2 M^2}\right],
\end{aligned}
\end{equation}
where $M>0$ is the mass parameter, $\Delta>0$, $\epsilon>0$, $\sigma>0$, and $r_0>2M$ (with $c=G=1$).\footnote{The parameters $\epsilon$ and $\sigma$ are dimensionless, $r_0$ has dimensions of length, and $\Delta$ has dimensions of area. Within the covariant framework, physically suitable choices of $\chi(r)$ and $\lambda(r)$ are subject to requirements including real functions with $\chi(r)\neq0$ in the region considered, finiteness and differentiability in that region, and, for asymptotically flat solutions, $\lambda(r)\to\lambda_\infty$ together with $\chi_\infty=(1+\lambda_\infty^2)^{-1/2}$ \cite{Belfaqih:2024vfk,Bojowald:2024lhr}. Both profiles in \eqref{eq1.1} are real and have $\lambda_\infty=0$, so the choice $\chi=1$ satisfies the asymptotic conditions. Case (i) is smooth throughout $r\geq2M$ and decays as $r^{-1}$. For fixed parameters satisfying the above conditions, case (ii) is positive and infinitely differentiable throughout $r\geq2M$, with finite derivatives at the horizon, and decays faster than any power of $r$. We therefore conclude that the chosen Gaussian-localized form provides a smooth and controlled representative of a localized correction, with $r_0$, $\epsilon$, and $\sigma$ controlling its location, strength, and width, respectively. Similar Gaussian profiles are widely used in studies of localized QNM deformations \cite{Cheung:2021bol,Berti:2022xfj,Boyanov:2024fgc,Zhang:2026vgd}. In our setting, however, the profile specifies the background holonomy function rather than being inserted directly into the QNM potential, so that both the effective potential and the tortoise-coordinate mapping are determined consistently by the same cLQBH geometry. With $\chi(r)=1$, it preserves asymptotic flatness and introduces no additional singularity in the BH exterior. This profile is a phenomenological choice within the emergent modified gravity framework, rather than being uniquely determined by LQG.}

By considering a massless, minimally coupled test scalar field on these fixed backgrounds \cite{Bojowald:2024lhr}, we investigate the corresponding wave dynamics and QNM spectra as a first step toward future studies of gravitational perturbations. The decreasing profile in case (i) provides a reference from previous discussions of covariant holonomy modifications \cite{Belfaqih:2024dzn}. Our focus is on case (ii), whose Gaussian factor allows us to vary the position, width, and strength of the resulting deformation of the effective potential, as we demonstrate below. The same background geometry determines both the effective potential and the relation between the areal radius $r$ and the tortoise coordinate. This enables us to examine how localized holonomy corrections affect BH ringing, with a view to future investigations of their possible observational implications \cite{Zhang:2024svj,Konoplya:2016pmh}.

It is important to note that the localized correction in case (ii) produces a bump structure in the scalar effective potential, connecting this model to the study of the bump effect on BH QNMs \cite{Cheung:2021bol}. Recent work in this direction has explored spectral sensitivity, mode excitation, and the relation between frequency-domain spectra and time-domain waveforms \cite{Berti:2022xfj,Yang:2024vor,Ianniccari:2024ysv,Oshita:2025ibu,Qian:2024iaq,Yang:2025dbn,Torres:2026uey, Jaramillo:2020tuu}. These studies have identified several characteristic features. In particular, a small bump far from the main potential peak can strongly shift QNM frequencies while leaving the early ringdown waveform nearly unchanged, and its effects on the waveform may become more apparent at later times \cite{Berti:2022xfj,Yang:2024vor}. Related studies have also examined whether localized negative or time-dependent potential deformations can trigger dynamical instability \cite{Mai:2025mim}. Some of these studies use toy models with prescribed potential bumps or simplified scattering problems. Meanwhile, other studies examine complex potential structures, including double peaks in specific gravitational theories \cite{Wang:2025mxe,Cao:2024sot}, as well as modifications produced by modeled matter environments \cite{Laeuger:2025zgb,Lan:2025brn}. These approaches provide complementary insights into the origin and consequences of localized potential structures. In this context, our chosen case (ii) provides a concrete LQBH realization of the bump effect. As noted above, the bump follows from the cLQBH geometry specified by the holonomy function $\lambda(r)$, rather than being imposed directly on the GR effective potential. This gives the bump a definite origin within an effective gravitational theory and allows systematic calculations of both the QNM spectrum and the time-domain response. It thus provides a bridge between the bump effect, BH spectroscopy, and observational tests of gravity. The delayed effects discussed above also motivate a quantitative study of when the waveform departs from its GR counterpart, using a suitable deviation criterion. The physical significance of this timescale also depends on whether the deviation occurs while the signal still has appreciable amplitude. Considering the deviation time together with the complex QNM frequencies $\omega$ and the signal decay helps assess the observational significance of the bump effect, including exploratory comparisons of frequency and damping scales with published GW measurements. These considerations motivate the study of an ``observation window'' in which deviations may become apparent before the signal becomes too weak, bringing the bump effect into a concrete framework for QNM calculations and tests of gravity.

The rest of the paper is organized as follows. Section~\ref{secII} briefly introduces the cLQBH solutions and the master equation for a test scalar field, along with the corresponding effective potential. In Sec.~\ref{secQNM}, we study QNMs in both the frequency and time domains [mainly for case (ii), with case (i) as a reference] and compare the results with their Schwarzschild counterparts. We also discuss their dependence on the model parameters and make preliminary comparisons with published LIGO data to explore possible constraints on the model. In Sec.~\ref{seckey}, we discuss the deviation time and its relation to waveform decay and consider their observational relevance, including the idea of an observation window and the role of detector properties. Finally, our conclusions and outlook are given in Sec.~\ref{secconclusion}.

Before proceeding further, we note that the scalar perturbations studied here do not directly describe the spin-2 GWs emitted during BH ringdown \cite{Berti:2009kk,Berti:2018vdi,Kokkotas:1999bd,Konoplya:2011qq}. Our main comparison is therefore between the same test scalar field on cLQBH and Schwarzschild backgrounds. Comparisons with published GW data are exploratory, while quantitative constraints from GW ringdown require gravitational perturbation equations consistent with the cLQBH dynamics and a statistical analysis that accounts for detector noise (see, e.g., \cite{Berti:2018vdi,LIGOScientific:2025wao}). For future extensions to gravitational perturbations, useful technical references include the master equations and QNM calculations for odd-parity perturbations in other effective LQBH models \cite{Bouhmadi-Lopez:2020oia,Liu:2021djf}. Work covering both odd- and even-parity perturbations has also revealed a breaking of isospectrality \cite{del-Corral:2022kbk}. Gravitational perturbation theory has also been developed at the quantum level through canonical formulations and hybrid quantization of both parity sectors on Kantowski--Sachs backgrounds relevant to BH interiors \cite{MenaMarugan:2024qnj,MenaMarugan:2025anx}. More recently, this framework has been extended to BH exteriors \cite{Lenzi:2026aay,Lenzi:2025pmk,Lenzi:2025man} and applied to gravitational QNM calculations in an effective LQBH model \cite{Garcia-Quismondo:2026tys}. More directly related to our scalar analysis, Bojowald et al.~\cite{Bojowald:2024lhr} derived scalar master equations within the cLQBH framework and compared minimally and nonminimally coupled QNMs for constant and decreasing holonomy functions using a third-order WKB approximation. Their results show that both the holonomy profile and the matter coupling can substantially affect the QNM spectrum. Their minimally coupled equations provide a basis for our analysis, with case (i) retained as the reference discussed above. Building on this work, we focus on the Gaussian-localized holonomy profile in case (ii). We study the resulting bump effect in both the frequency and time domains, linking spectral changes, waveform deviations, and their possible observational significance.


Throughout this paper, we adopt units with $c=G=1$ \cite{Bojowald:2024lhr}, where $c$ is the speed of light and $G$ is the gravitational constant. Where convenient, we further set $M=1$, completing the choice of units for length, time, and mass. All dimensional quantities must be expressed consistently in this unit system. Greek indices run from 0 to 3. Other index conventions will be specified where needed. We adopt the metric signature $(-,+,+,+)$.


\section{Spherically symmetric cLQBHs and scalar field perturbations}
\renewcommand{\theequation}{2.\arabic{equation}}
\setcounter{equation}{0}
\label{secII}

In this section, we first briefly introduce the static, spherically symmetric cLQBH solutions constructed in \cite{Belfaqih:2024vfk}, and then consider linear perturbations of a massless, minimally coupled test scalar field on these backgrounds. After rescaling the time coordinate by a constant factor, the line element can be written as 
\bqn
\lb{ds2}
d s^2 &=& \frac{1}{\chi^2(r)}
\left(-f(r) dt^2 + \frac{dr^2}{H(r)}\right)
+ r^2 d\Omega^2,
\eqn
where $r$ is the areal radius, $d\Omega^2 \equiv d\theta^2 + \sin^2\theta\,d\varphi^2$, and
\bqn
\lb{fr}
f(r) &=& 1 - \frac{2M}{r} - \frac{\Lambda}{3}r^2,\nb\\
H(r) &=& f(r)\left[1+\lambda^2(r)f(r)\right].
\eqn
Here $\chi(r)$ and $\lambda(r)$ are two real functions of $r$, with $\chi(r)\neq 0$ in the region under consideration. The constants $M$ and $\Lambda$ denote the mass parameter and cosmological constant, respectively. In Ref.~\cite{Belfaqih:2025eak}, Hawking radiation was studied for the decreasing function $\lambda(r)=\sqrt{\Delta}/r$ [cf., case (i) in \eqref{eq1.1}], with $\Lambda=0$ and $\chi(r)=1$, where $\Delta>0$ is a constant with dimensions of area related to the area scale in LQG. For these choices of $\Lambda$ and $\chi(r)$, $g_{tt}=-f(r)$ coincides with the corresponding Schwarzschild metric component \cite{Schwarzschild:1916uq}, and the metric horizon (MH) is located at $r=r_{\mathrm{MH}}\equiv 2M$ for both profiles in \eqref{eq1.1}.
The covariant framework allows considerable freedom in the choice of $\lambda(r)$, subject to regularity and the required asymptotic behavior. More specific quantum-gravity models may impose additional restrictions \cite{Belfaqih:2024vfk,Bojowald:2024lhr}. Our choice in case (ii) satisfies these geometric requirements in the exterior and produces a bump in the scalar effective potential, as discussed in the Introduction.

Bojowald et al.~\cite{Bojowald:2024lhr} studied scalar QNMs for both constant and decreasing functions $\lambda(r)$, considering minimal and nonminimal couplings. The decreasing function used in their minimally coupled analysis corresponds to case (i) in \eqref{eq1.1}, and their results provide an important benchmark for our work. However, our main focus is case (ii), in which $\lambda(r)$ has a Gaussian-like radial profile.
For the remainder of this paper, we restrict attention to
\bqn
\lb{eq2.3}
\Lambda = 0, \quad \chi(r) = 1.
\eqn

Then, following \cite{Zhang:2024svj}, we consider scalar perturbations governed by the Klein-Gordon equation 
\bqn
\lb{KGeqn}
\frac{1}{\sqrt{-g}} \partial_\mu \left( \sqrt{-g} g^{\mu \nu} \partial_\nu \Phi \right) &=& 0,
\eqn
where $\Phi=\Phi(t,r,\theta,\varphi)$ denotes the scalar field in  Boyer--Lindquist-type coordinates \cite{Urbano:2018nrs} and $g$ is the determinant of the background metric. This equation is compatible with the covariance of the theory \cite{Belfaqih:2025eak,Bojowald:2024lhr}. Separating variables as $\Phi=Y_{lm}(\theta,\varphi)\Psi(t,r)/r$, where $Y_{lm}$ are spherical harmonics, we obtain the master equation for $\Psi$,
\bqn
\lb{master1}
\frac{\partial^2}{\partial x^2}\Psi+\left(-\frac{\partial^2}{\partial t^2}-V_{\mathrm{eff}}(r)\right)\Psi &=& 0,
\eqn
where the effective potential $V_{\mathrm{eff}}$, with dimensions of $M^{-2}$, is given by \cite{Daghigh:2020fmw}
\bqn
\lb{Veff}
V_{\mathrm{eff}} &\equiv& f\frac{L}{r^2}+\frac{1}{2r}\frac{d(fH)}{dr}\nb\\
&=& V_{\mathrm{GR}} +\frac{1}{2r}\frac{d}{dr}\left[f^3(r)\lambda^2(r)\right],
\eqn
where $L\equiv l(l+1)$, $V_{\mathrm{GR}}=(r-2M)(Lr+2M)/r^4$ denotes the effective potential for a massless scalar field on the Schwarzschild background \cite{Bojowald:2024lhr}\footnote{Strictly speaking, notations such as $V_{\mathrm{eff,cLQBH}}$ and $V_{\mathrm{eff,GR}}$ would explicitly distinguish the effective potentials on different backgrounds (the same convention applies to quantities such as the tortoise coordinate $x$, to be introduced shortly). Nonetheless, for simplicity, we use $V_{\mathrm{eff}}$ to denote the effective potential for our cLQBH and $V_{\mathrm{GR}}$ to denote its Schwarzschild counterpart throughout this paper, unless clarification is needed.}, and $x$ is the tortoise coordinate defined through
\bqn
\lb{rstar}
\frac{dr}{dx} &=& \sqrt{f(r)H(r)}.
\eqn
Here a prime denotes differentiation with respect to $r$. This effective potential agrees with that given in \cite{Belfaqih:2025eak} after accounting for the different notation. Since the background is spherically symmetric, the radial equation and QNM frequencies are independent of the angular index $m$ \cite{Chandra92book}. Thus, without loss of generality, we set $m=0$. For given functions $f(r)$ and $\lambda(r)$ and a chosen value of $l$, we solve \eqref{master1} for $\Psi$.


Notice that, the master equation \eqref{master1} mainly serve the time-domain calculations for the QNMs. For the frequency-domain analysis the calculations of $\omega$, a more straightforward way is to use its frequency-domain counterpart by doing the Fourier transformation, with which we obtain \cite{Berti:2009kk, Daghigh:2020fmw}
\bqn
\lb{master2}
\frac{\partial^2}{\partial x^2}\psi+\left(\omega^2-V_{\mathrm{eff}}(r)\right)\psi &=& 0,
\eqn
where $\Psi(t, x)=e^{-i \omega t} \psi(x)$. Here, $\omega$ is the complex QNM frequency we are looking for. 

For later convenience, and following the notation used in, e.g., \cite{Cheung:2021bol,Yang:2024vor}, let us introduce the ``bump potential''
\bqn
\lb{Vbump}
V_{\mathrm{bump}}(r) &\equiv& \frac{1}{2r}\frac{d}{dr}\left[f^3(r)\lambda^2(r)\right],
\eqn
so that $V_{\mathrm{eff}}=V_{\mathrm{GR}}+V_{\mathrm{bump}}$ (notice that no additional overall amplitude factor is placed in front of $V_{\mathrm{bump}}$).
Substituting the expressions for $\lambda(r)$ in case (ii) and $f(r)$, we obtain
\begin{widetext}
\bqn
\lb{Vbump2}
V_{\mathrm{bump}} &=& -\frac{(r-r_0)f^2\lambda^2}{r^3}
\left[\frac{3M}{r_0-2M}+\frac{r(r-2M)}{M^2\sigma^2}\right] \nb\\
&=& \frac{M\left(3M^2\sigma^2+4Mr-2r^2\right)+rr_0(r-2M)}{M^2r^5\sigma^2(2M-r_0)}(r-r_0)\epsilon^2(r-2M)^2\left(\frac{r}{r_0}\right)^{-\frac{6M}{r_0-2M}}e^{-\frac{(r-r_0)^2}{M^2\sigma^2}} \nb\\
&=& \frac{\epsilon^2}{M^2}
\frac{(\tilde r-2)^2(\tilde r-\tilde r_0)
\left[3\sigma^2+\tilde r(\tilde r-2)(\tilde r_0-2)\right]}
{\tilde r^5\sigma^2(2-\tilde r_0)}
\left(\frac{\tilde r}{\tilde r_0}\right)^{-\frac{6}{\tilde r_0-2}}
e^{-\frac{(\tilde r-\tilde r_0)^2}{\sigma^2}},
\eqn
\end{widetext}
where, following \cite{Zhang:2024svj}, we have introduced the dimensionless quantities $\tilde r\equiv r/M$ and $\tilde r_0\equiv r_0/M$ to simplify the last line. In fact, using $M$ as the reference scale, the physical quantities of interest considered here can be expressed in dimensionless form, as illustrated by $r$ and $r_0$. For example, according to \eqref{Vbump2}, we can further introduce the dimensionless bump potential $\tilde V_{\mathrm{bump}}\equiv M^2V_{\mathrm{bump}}$. The numerical values of these dimensionless quantities are equivalent to those obtained by setting $M=1$ in the corresponding dimensional expressions. Taking advantage of this equivalence and the simplicity of dimensionless quantities, we shall continue to use the ``tilde'' notation to construct dimensionless expressions, or set $M=1$, whenever convenient and without causing confusion. Notice that $V_{\mathrm{bump}}$ is independent of $l$.
Also notice that, since $f^3\lambda^2$ tends to zero at both boundaries, $r\to r_{\mathrm{MH}}^{+}$ and $r\to+\infty$, while being positive at intermediate radii, it follows from \eqref{Vbump} that $V_{\mathrm{bump}}$ must contain both positive and negative parts, as shown in the figure to be given below.

\begin{figure}[h]
\includegraphics[width=0.9\linewidth]{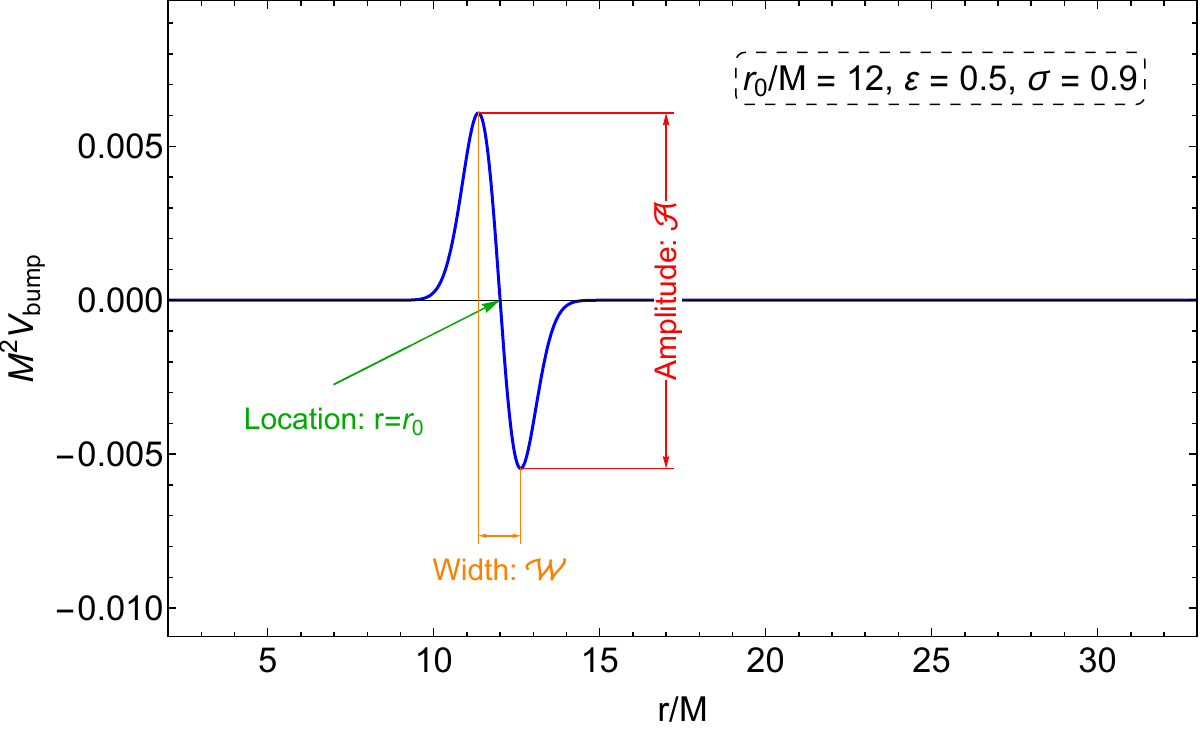}
\caption{The behavior of the bump potential $V_{\mathrm{bump}}$ [cf.~\eqref{Vbump2}] as a function of $r/M$ (plotted in its dimensionless form, $\tilde V_{\mathrm{bump}}=M^2V_{\mathrm{bump}}$) for a representative case with $r_0/M=12$, $\epsilon=0.5$ and $\sigma=0.9$. As shown in the figure, the location of the bump is described by the parameter $r_0/M$ (see the green arrow), the amplitude $\mathcal{A}$ of the bump is defined as the vertical distance between the local peak and trough (see the red marker), while the width $\mathcal{W}$ of the bump is defined as the horizontal distance between the local peak and trough (see the orange marker).}
\label{plotVbump1}
\end{figure}

We plot a representative case of $M^2V_{\mathrm{bump}}$ in Fig.~\ref{plotVbump1} by setting $r_0/M=12$, $\epsilon=0.5$, and $\sigma=0.9$ ($V_{\mathrm{bump}}$ is independent of $l$, as mentioned earlier). This figure illustrates the typical behavior of the bump potential in case (ii) [cf.~\eqref{eq1.1}]. In particular, the bump exhibits a ``sine-like'' behavior around $r=r_0$, although the height of the local peak generally differs from the depth of the local trough. In general, the bump also has a nonzero integral, $\int_{-\infty}^{+\infty}V_{\mathrm{bump}}\,dx\neq0$. More importantly, with the assistance of Fig.~\ref{plotVbump1}, we can define the location, width and amplitude of such bumps.

More specifically, the first line of \eqref{Vbump2} shows that $V_{\mathrm{bump}}(r=r_{\mathrm{MH}})=V_{\mathrm{bump}}(r=r_0)=0$. The zero crossing at $r=r_0$ naturally allows us to define this point as the center (location) of the bump, as marked by the green arrow in Fig.~\ref{plotVbump1}. In contrast, the bump's width and amplitude are not directly specified by $\sigma$ and $\epsilon$, respectively. Although $\epsilon^2$ independently scales the amplitude, changing $\sigma$ generally affects both the width and amplitude. Therefore, as illustrated in Fig.~\ref{plotVbump1} (see the red and orange markers), we introduce the dimensionless quantities $\mathcal{A}$ and $\mathcal{W}$ to describe the amplitude and width directly. Together with a given $r_0/M$, these quantities allow us to solve, usually numerically, for the corresponding $\sigma$ and $\epsilon$ using \eqref{Vbump2}. For the cases considered here, this procedure establishes a correspondence between $\{r_0/M,\epsilon,\sigma\}$ and $\{r_0/M,\mathcal{A},\mathcal{W}\}$ through the exterior local peak and trough. This enables us to investigate the individual effects of the location, amplitude and width of the bump, rather than their combined effects when varying the original parameters.

\begin{figure}[h]
\includegraphics[width=0.9\linewidth]{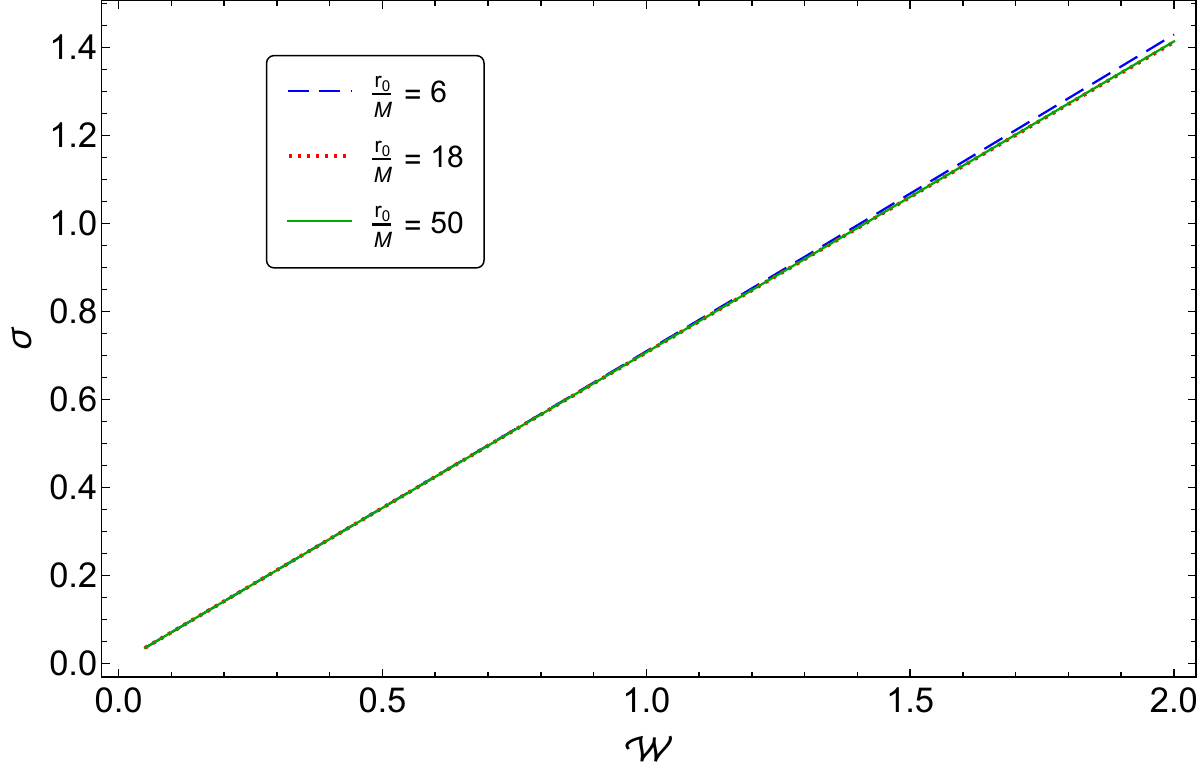}
\includegraphics[width=0.9\linewidth]{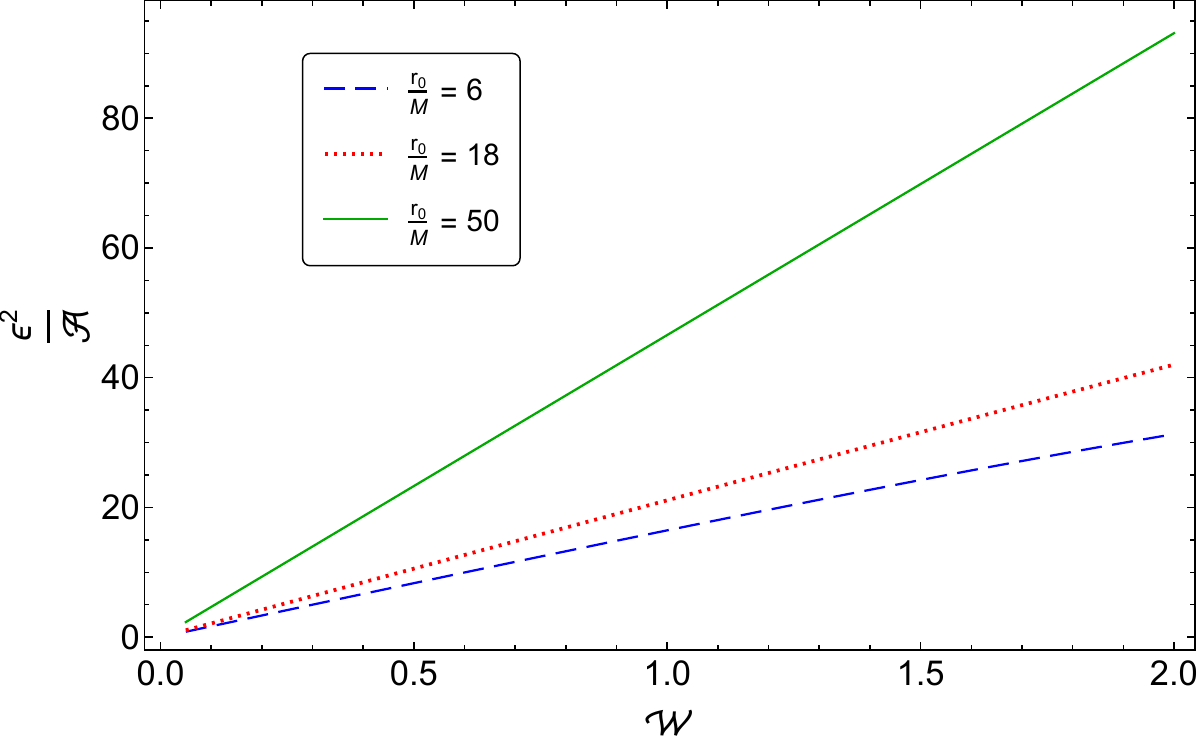}
\caption{The behavior of the dimensionless quantities $\sigma$ (top) and $\epsilon^2/\mathcal{A}$ (bottom) as functions of $\mathcal{W}$ for different choices of $r_0/M$. For a given position and width, these two panels provide the corresponding $\sigma$ and the ratio needed to determine $\epsilon$ for a prescribed amplitude $\mathcal{A}$.}
\label{plotsigma}
\end{figure}

By examining the structure of $V_{\mathrm{bump}}$ in the last line of \eqref{Vbump2}, we notice that, for a given set of $\{r_0/M,\mathcal{A},\mathcal{W}\}$, we can first determine $\sigma$ from $r_0/M$ and $\mathcal{W}$. The same pair of $\{r_0, \mathcal{W}\}$ also determines $\epsilon^2/\mathcal{A}$, whose reciprocal gives the bump amplitude for $\epsilon=1$. The amplitude information is therefore already contained in this quantity, with $\epsilon^2$ providing the remaining overall adjustment. We plot these two dimensionless quantities as functions of $\mathcal{W}$ for several choices of $r_0/M$ in the top and bottom panels of Fig.~\ref{plotsigma}, respectively. Once $\mathcal{A}$ is prescribed, obtaining $\epsilon$ from the bottom panel is a trivial algebraic step, so separate numerical plots of $\epsilon$ would add no information to this parameter mapping and are omitted here. This simplification illustrates a broader consideration in our study. The large parameter space entails substantial computational effort and can obscure the principal physical trends among numerous parameter combinations. We therefore organize the investigation around the mathematical structure and the relative importance of the parameter effects, seeking a manageable set of cases that reveals the essential behavior clearly. This consideration will also guide the subsequent analysis. Returning to Fig.~\ref{plotsigma}, we observe that both $\sigma$ and $\epsilon^2/\mathcal{A}$ depend approximately linearly on $\mathcal{W}$ for a given $r_0/M$, while $\sigma$ is insensitive to $r_0/M$.

\begin{figure*}[t]
\centering
\begin{minipage}{0.24\textwidth}
\centering
\includegraphics[width=\linewidth]{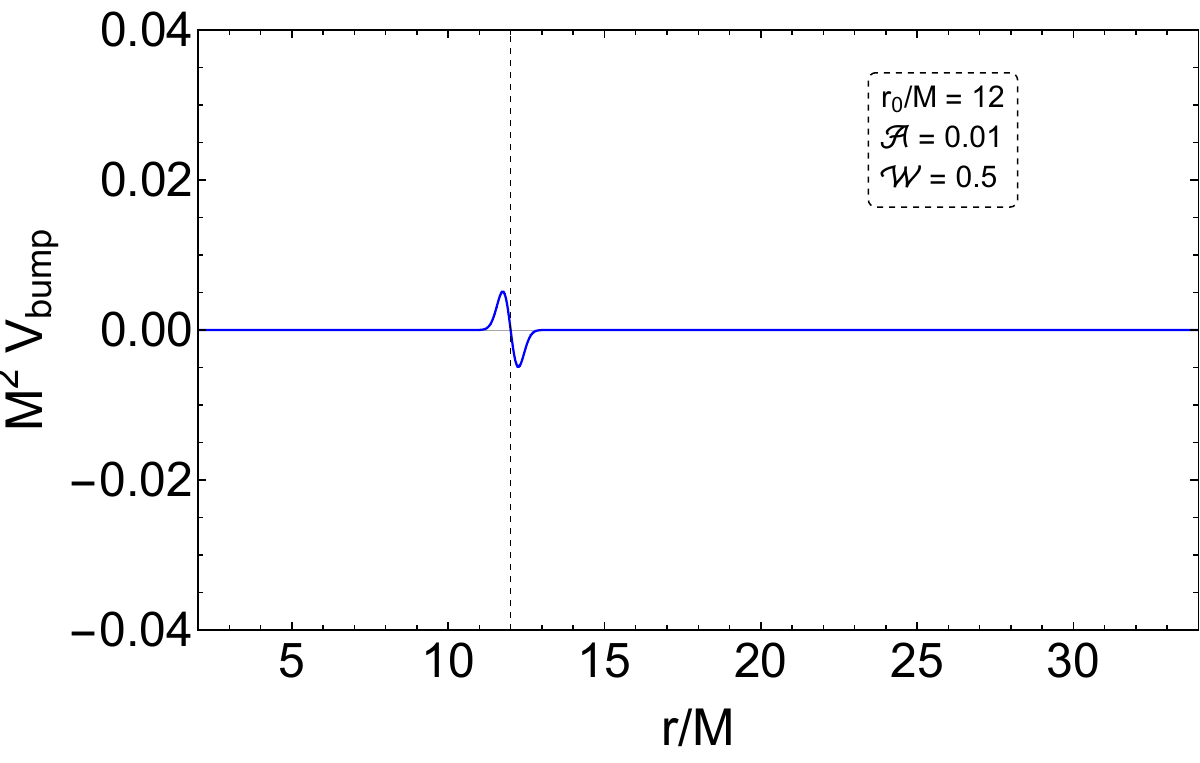}\\[-1mm]
(a1)
\end{minipage}\hfill
\begin{minipage}{0.24\textwidth}
\centering
\includegraphics[width=\linewidth]{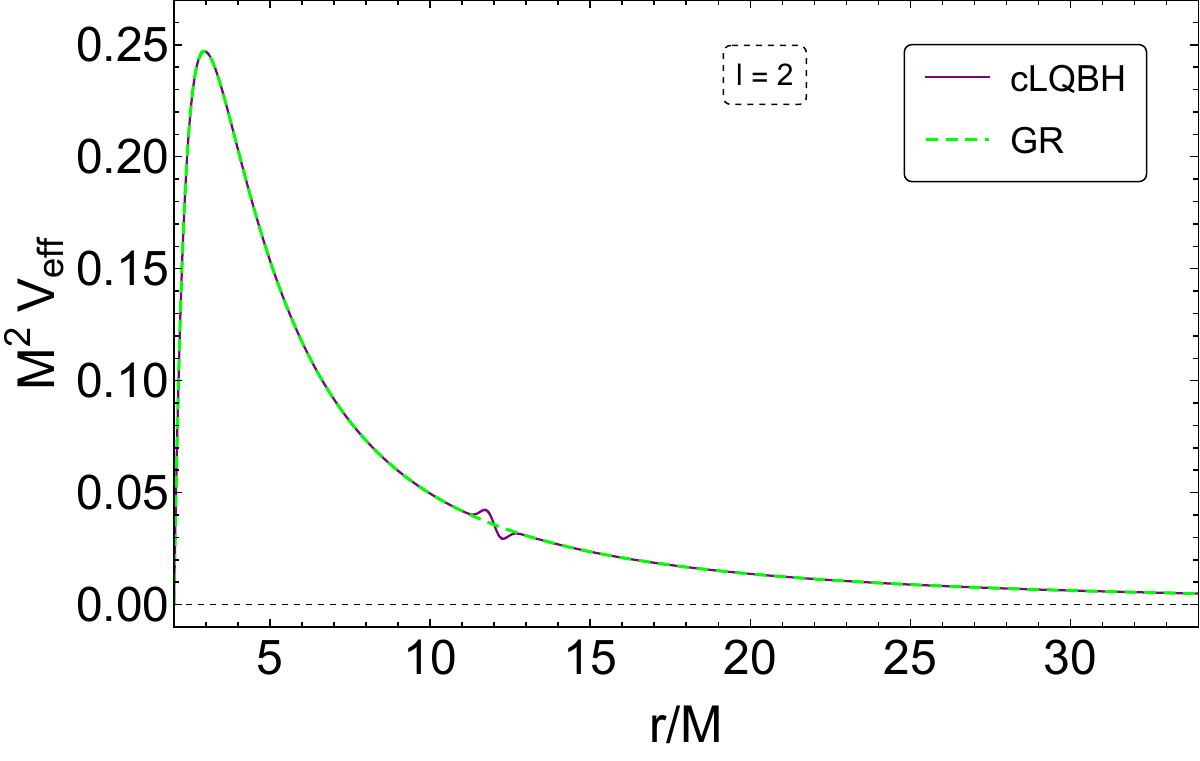}\\[-1mm]
(a2)
\end{minipage}\hfill
\begin{minipage}{0.24\textwidth}
\centering
\includegraphics[width=\linewidth]{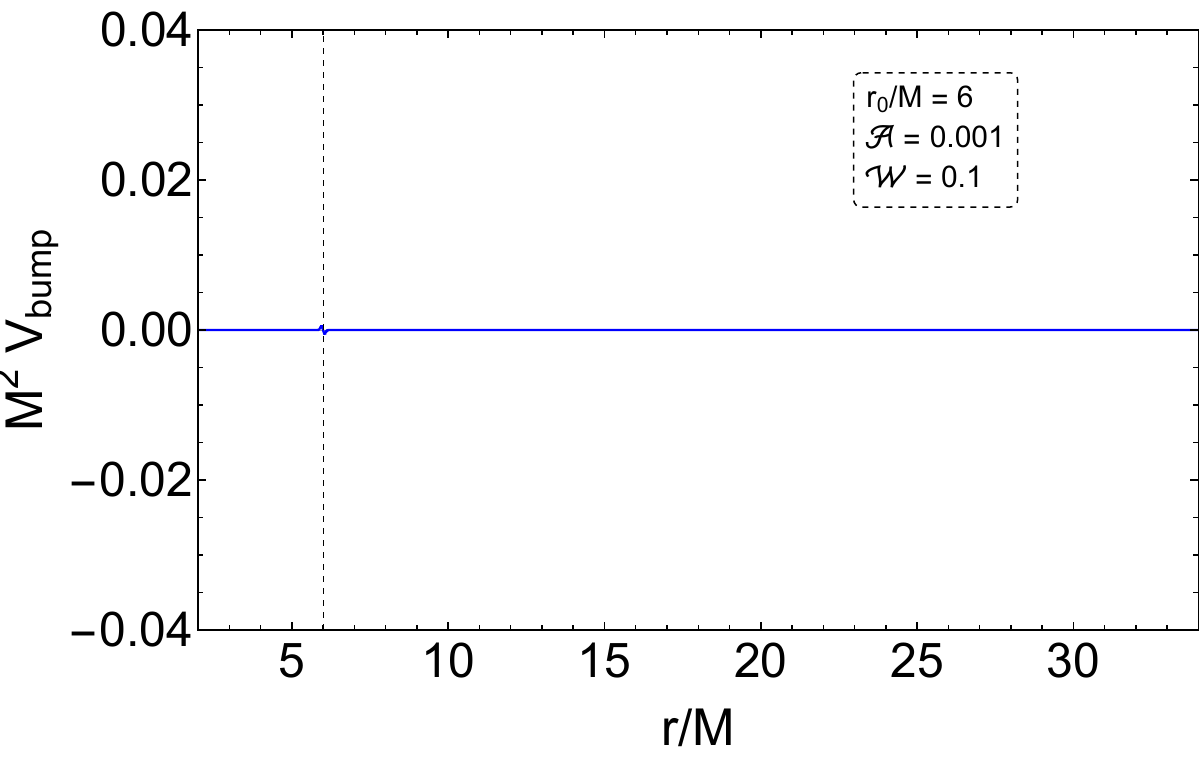}\\[-1mm]
(b1)
\end{minipage}\hfill
\begin{minipage}{0.24\textwidth}
\centering
\includegraphics[width=\linewidth]{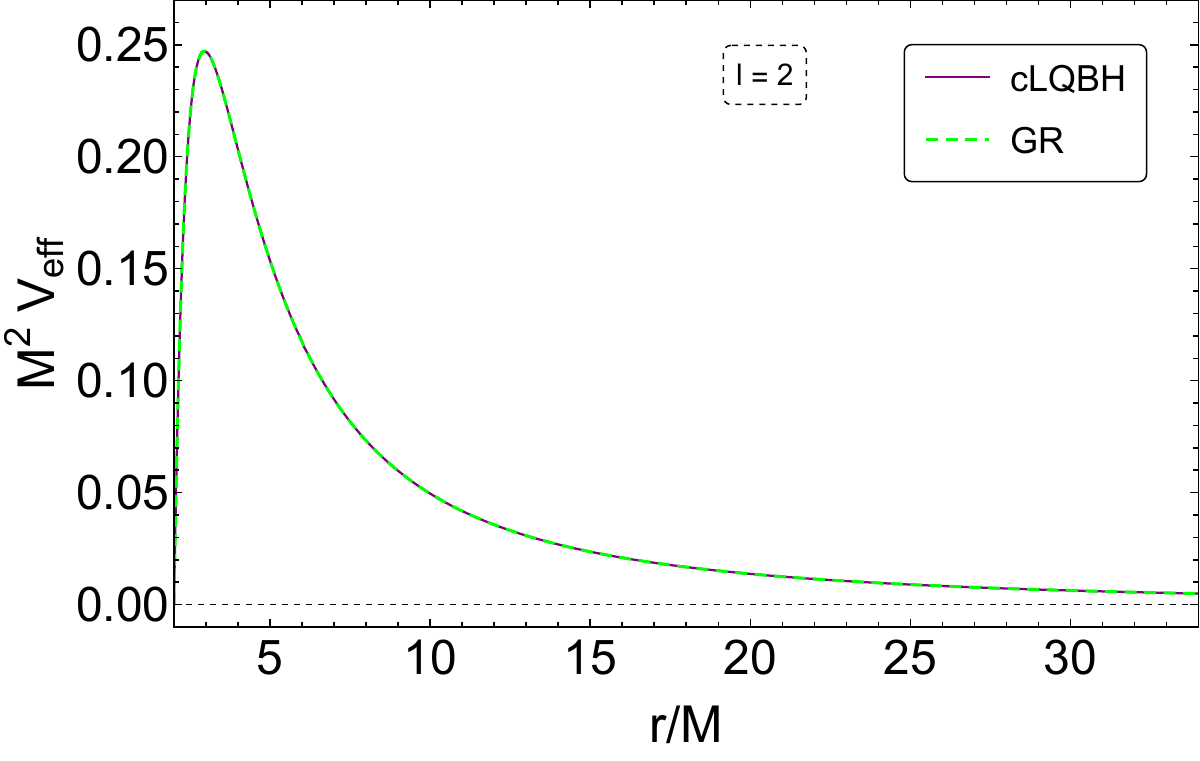}\\[-1mm]
(b2)
\end{minipage}
\par\vspace{2mm}
\begin{minipage}{0.24\textwidth}
\centering
\includegraphics[width=\linewidth]{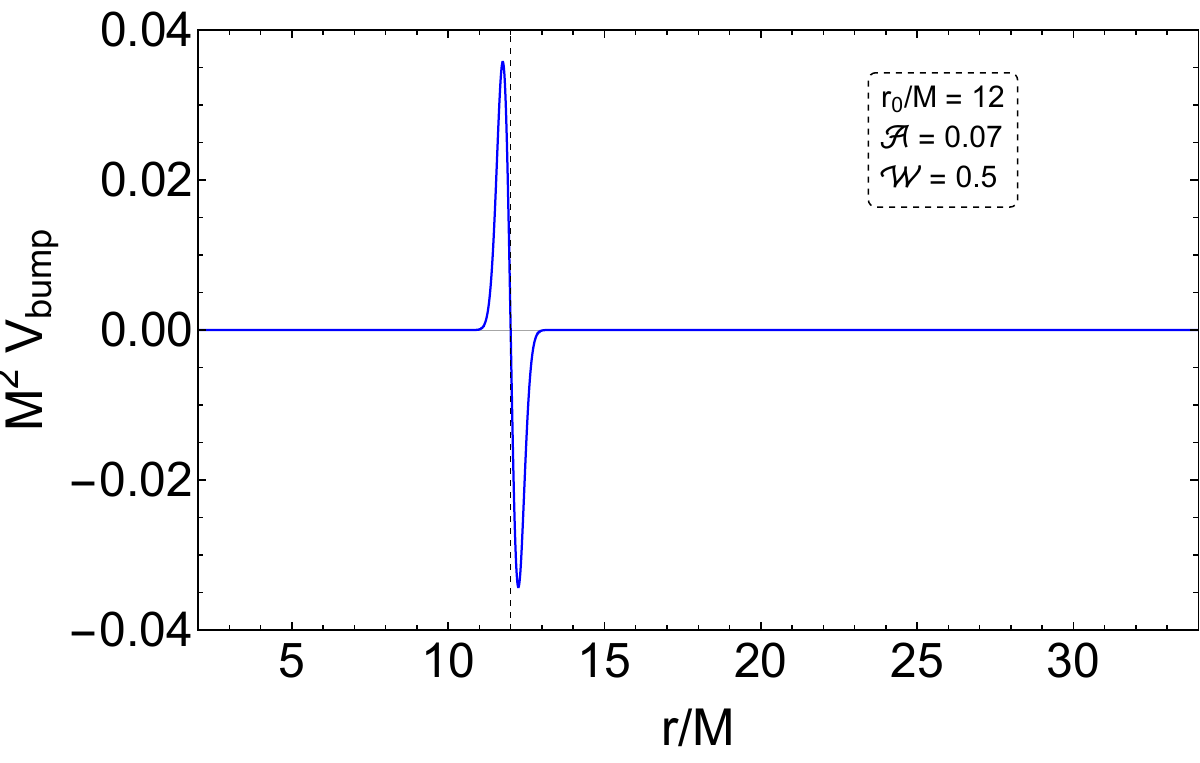}\\[-1mm]
(c1)
\end{minipage}\hfill
\begin{minipage}{0.24\textwidth}
\centering
\includegraphics[width=\linewidth]{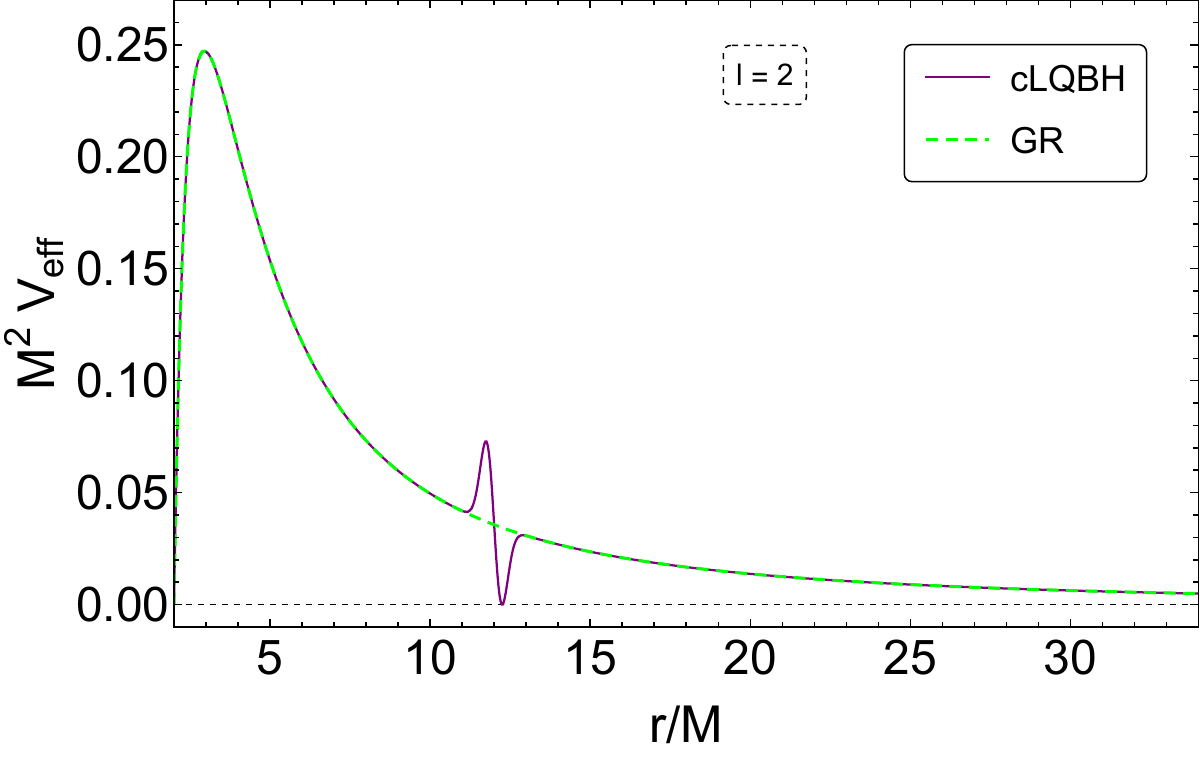}\\[-1mm]
(c2)
\end{minipage}\hfill
\begin{minipage}{0.24\textwidth}
\centering
\includegraphics[width=\linewidth]{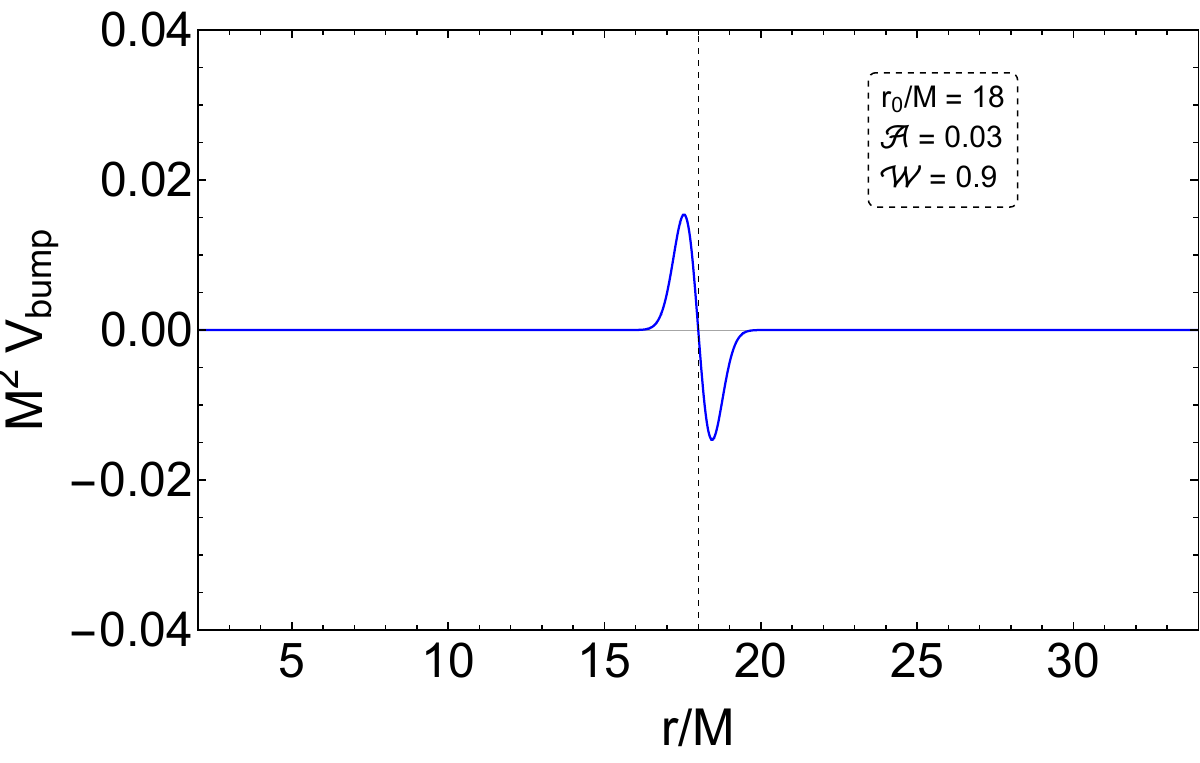}\\[-1mm]
(d1)
\end{minipage}\hfill
\begin{minipage}{0.24\textwidth}
\centering
\includegraphics[width=\linewidth]{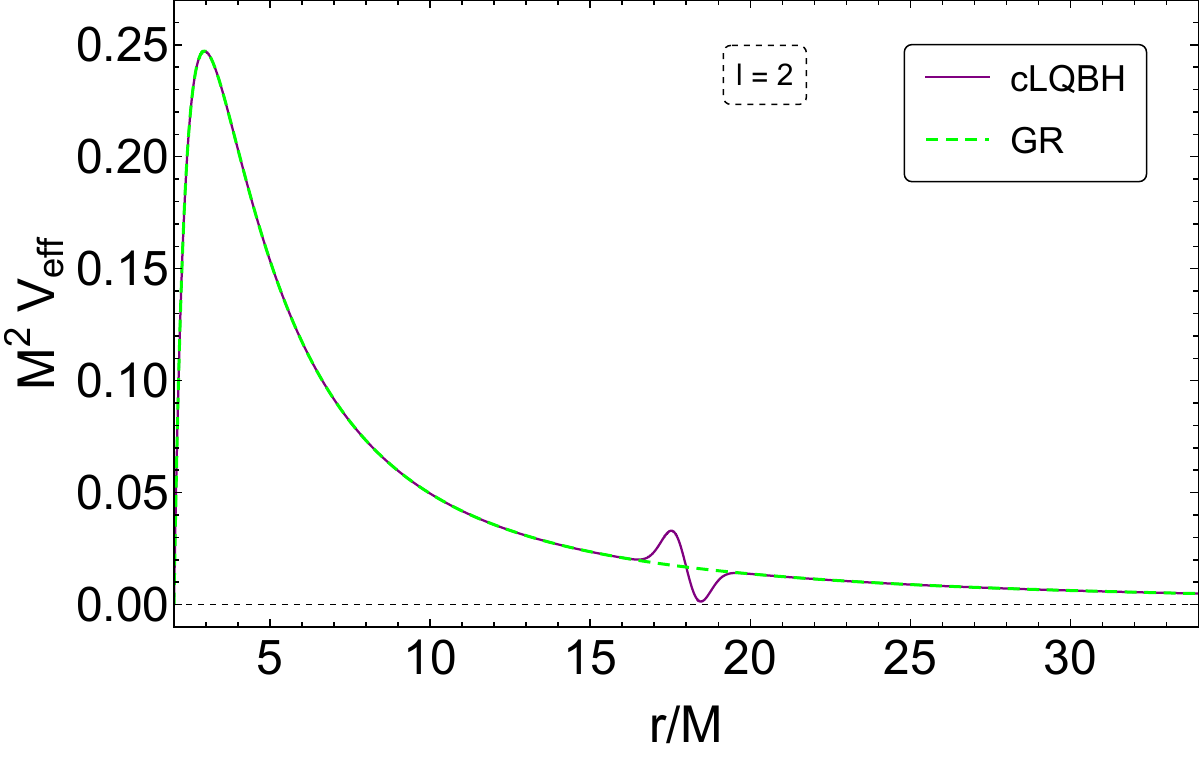}\\[-1mm]
(d2)
\end{minipage}
\par\vspace{2mm}
\begin{minipage}{0.24\textwidth}
\centering
\includegraphics[width=\linewidth]{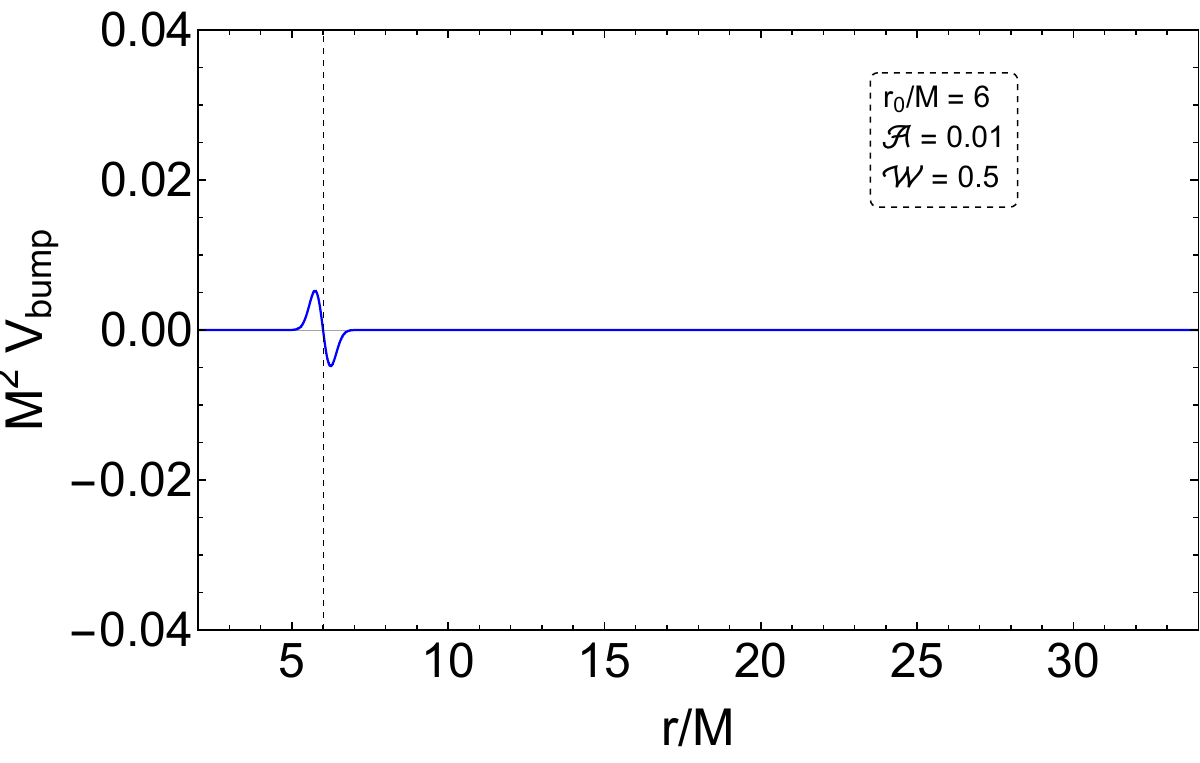}\\[-1mm]
(e1)
\end{minipage}\hfill
\begin{minipage}{0.24\textwidth}
\centering
\includegraphics[width=\linewidth]{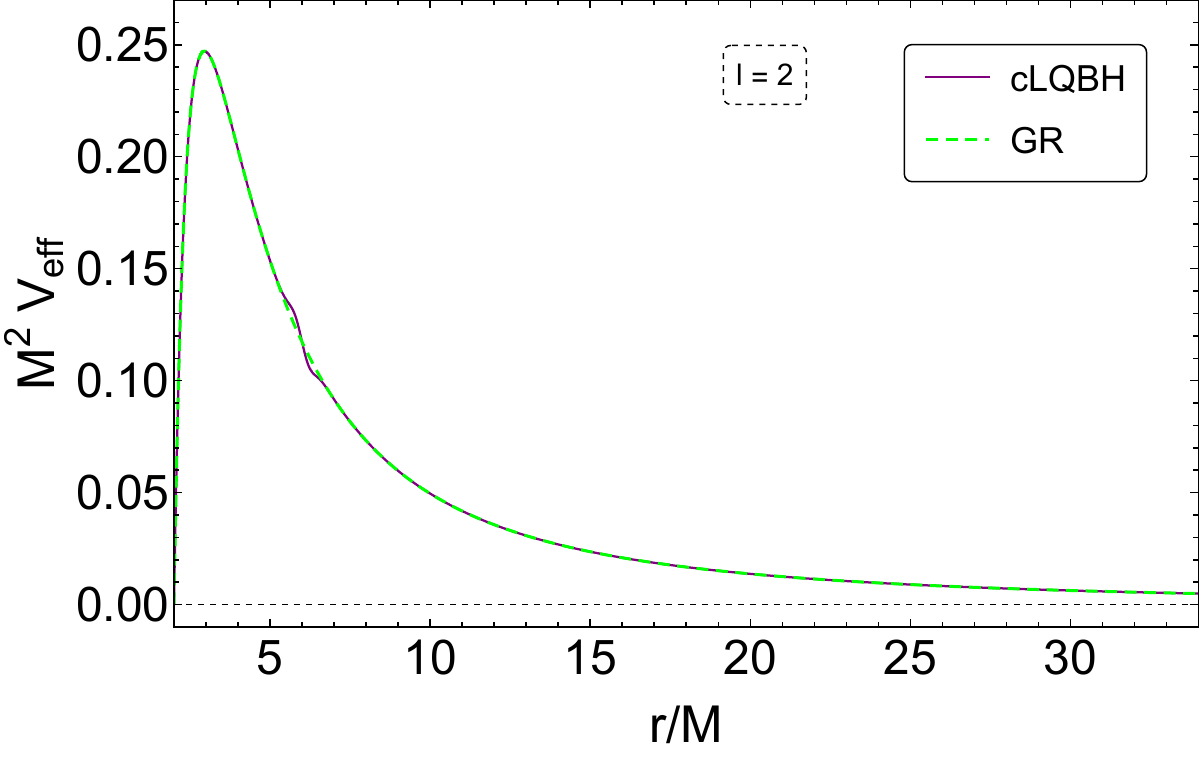}\\[-1mm]
(e2)
\end{minipage}\hfill
\begin{minipage}{0.24\textwidth}
\centering
\includegraphics[width=\linewidth]{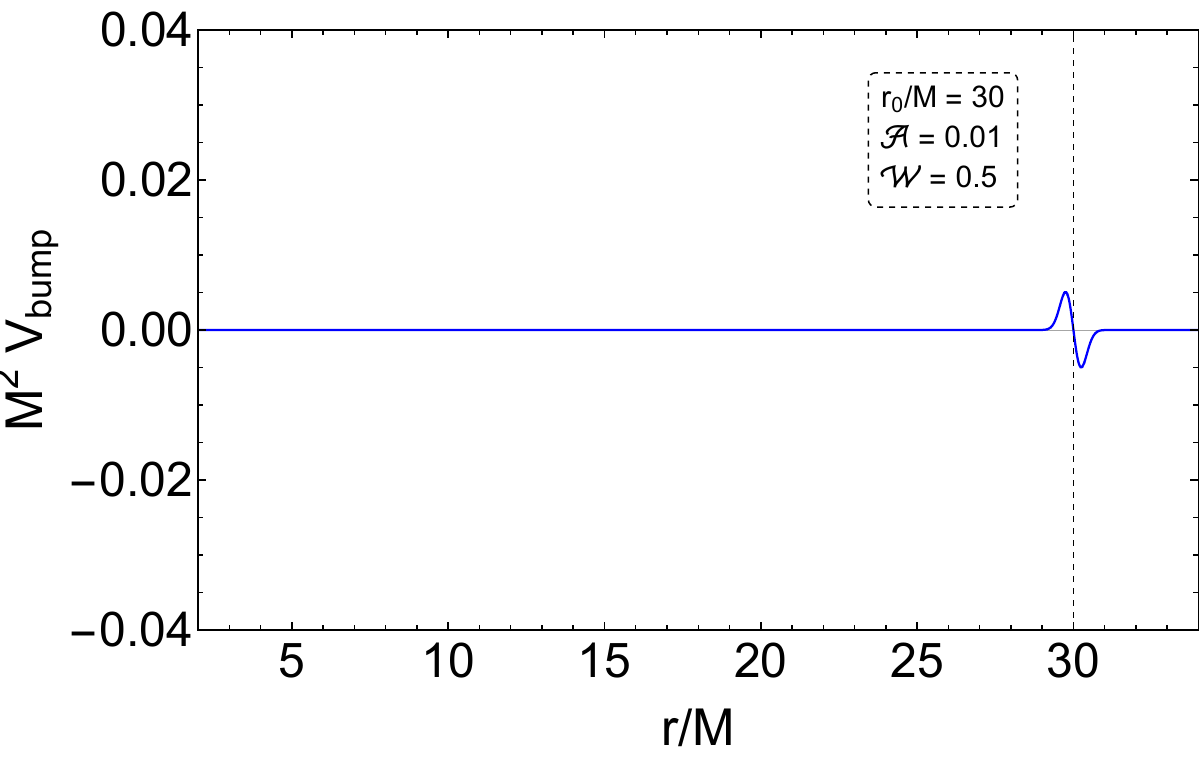}\\[-1mm]
(f1)
\end{minipage}\hfill
\begin{minipage}{0.24\textwidth}
\centering
\includegraphics[width=\linewidth]{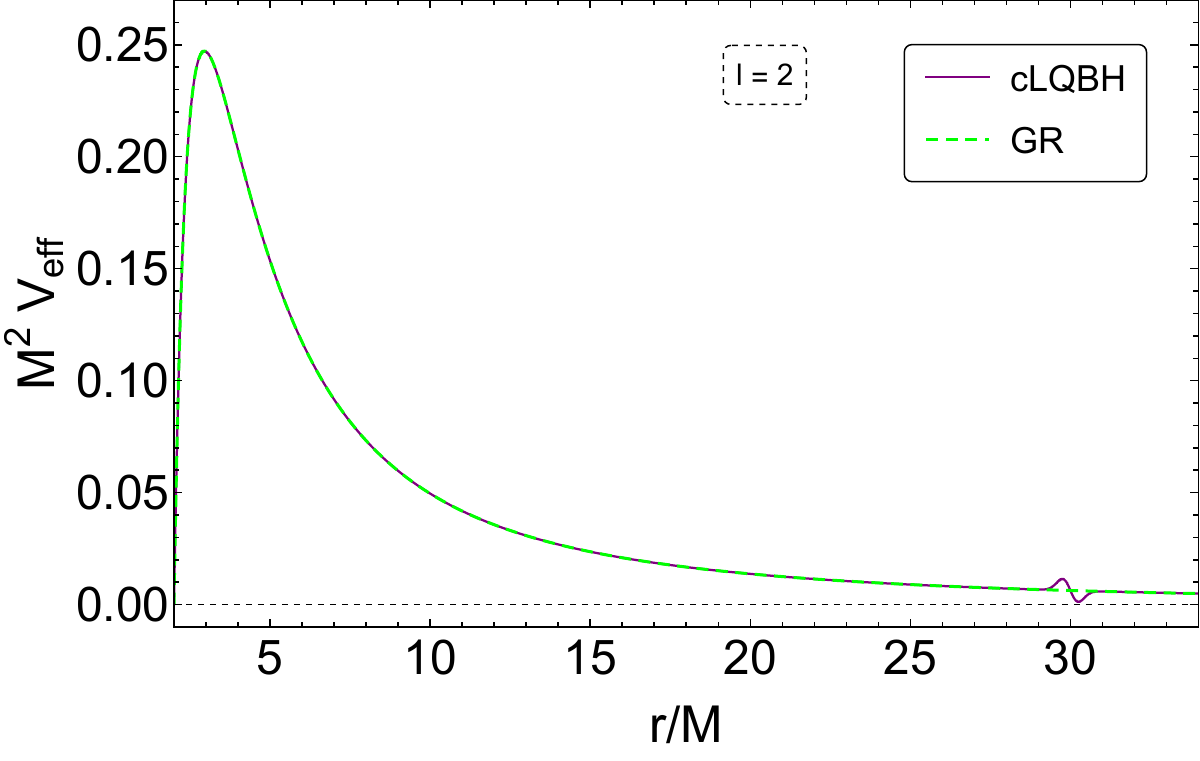}\\[-1mm]
(f2)
\end{minipage}
\par\vspace{2mm}
\begin{minipage}{0.24\textwidth}
\centering
\includegraphics[width=\linewidth]{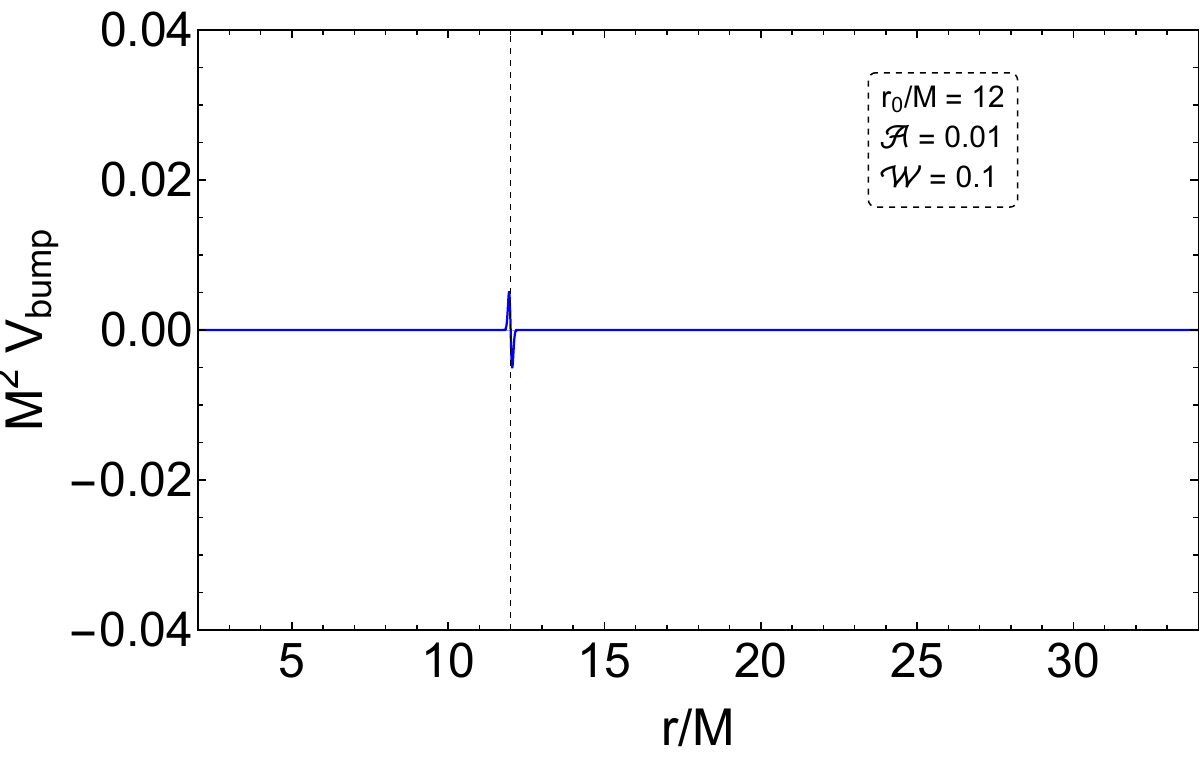}\\[-1mm]
(g1)
\end{minipage}\hfill
\begin{minipage}{0.24\textwidth}
\centering
\includegraphics[width=\linewidth]{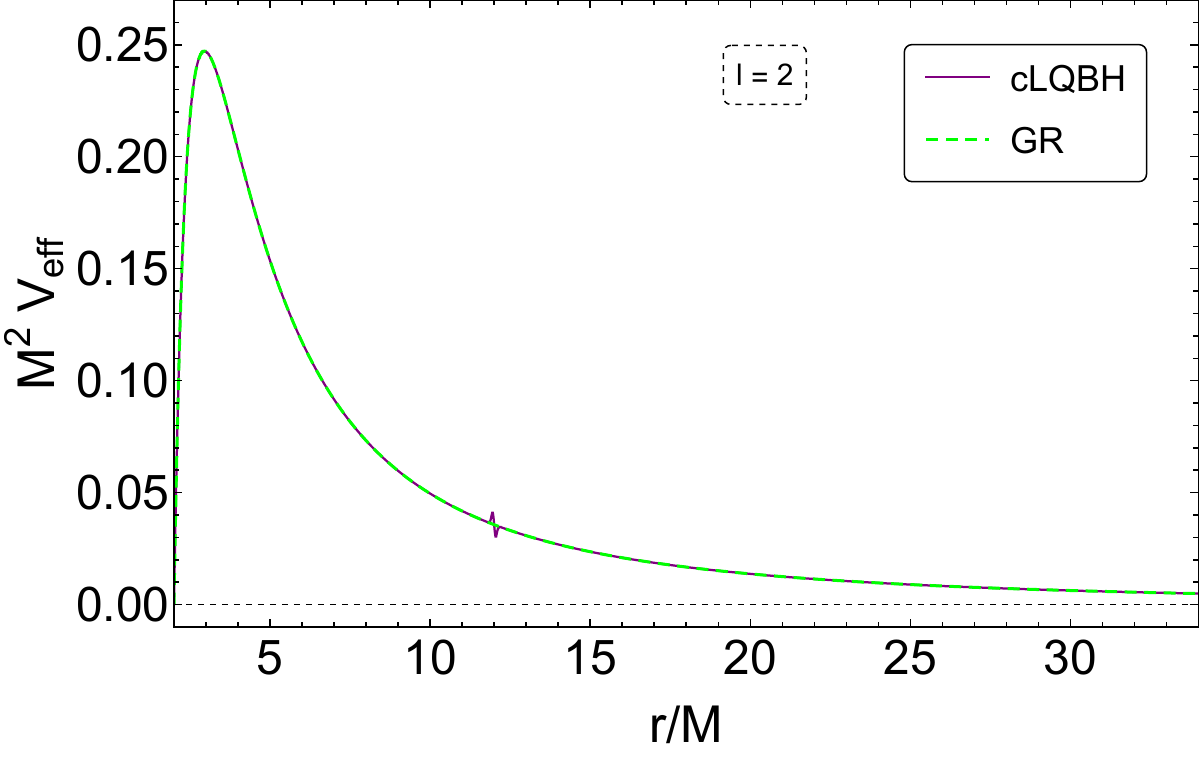}\\[-1mm]
(g2)
\end{minipage}\hfill
\begin{minipage}{0.24\textwidth}
\centering
\includegraphics[width=\linewidth]{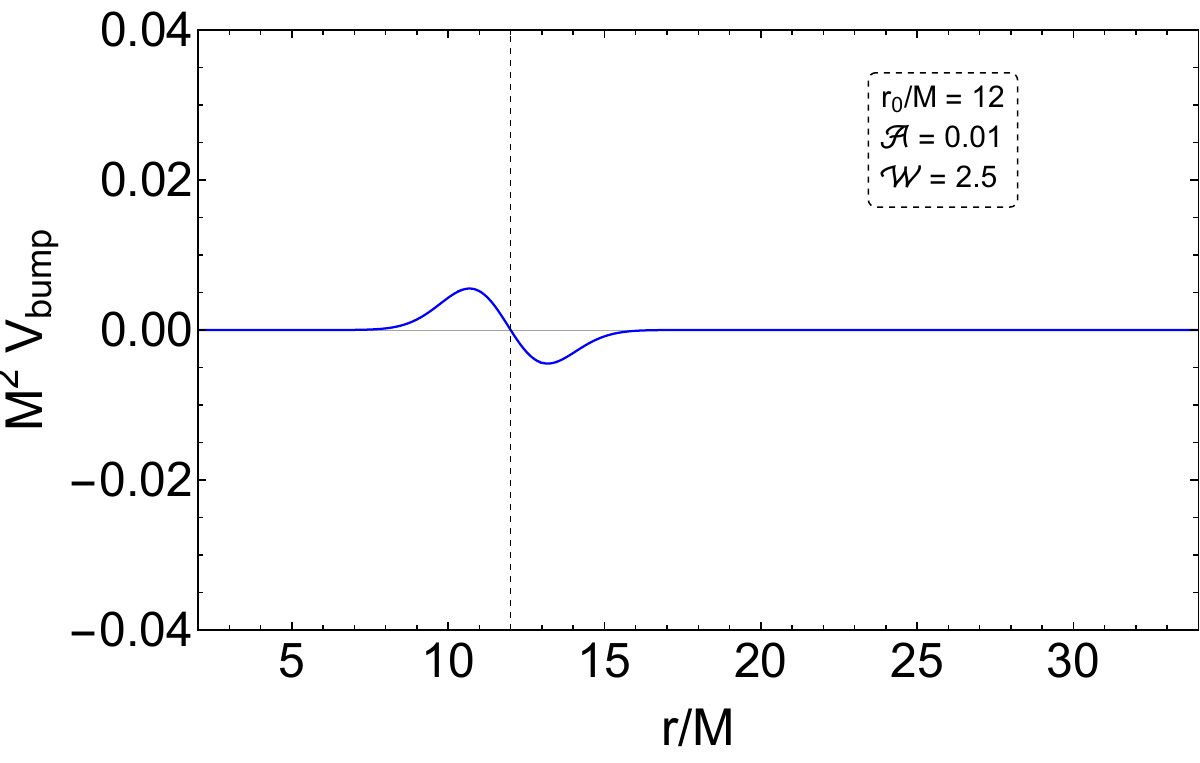}\\[-1mm]
(h1)
\end{minipage}\hfill
\begin{minipage}{0.24\textwidth}
\centering
\includegraphics[width=\linewidth]{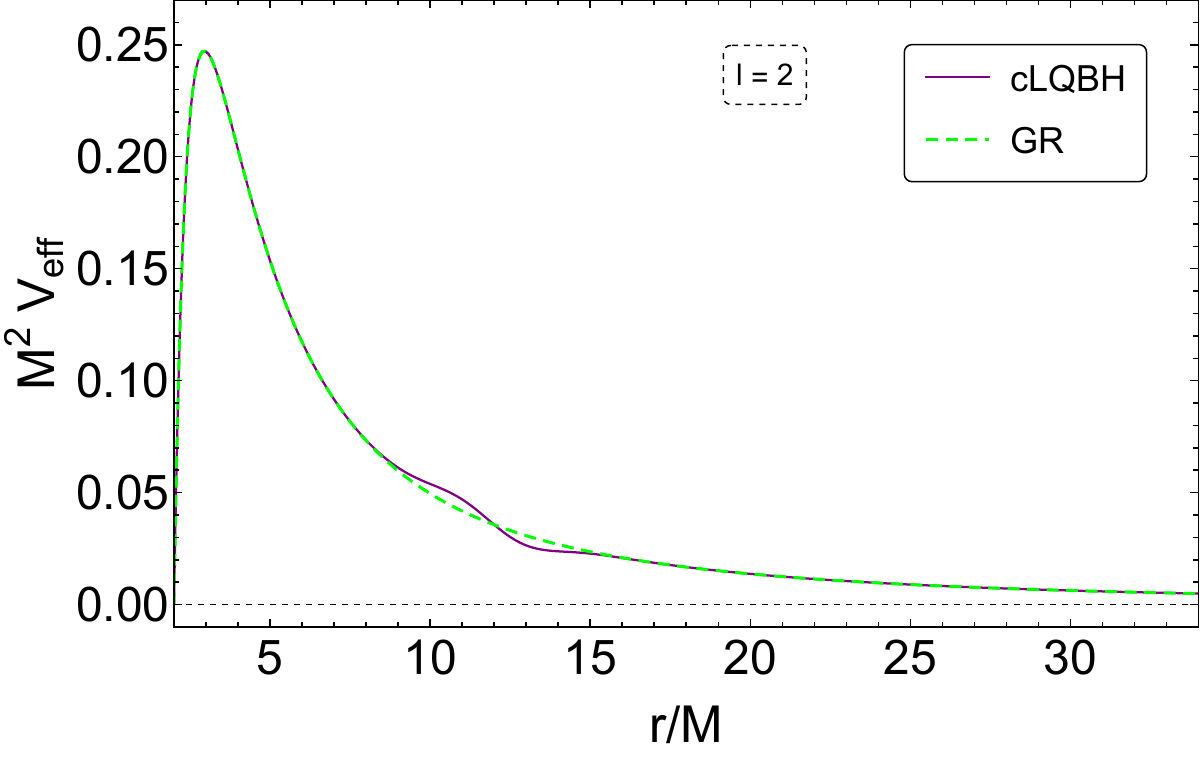}\\[-1mm]
(h2)
\end{minipage}
\caption{The dimensionless bump potentials $M^2V_{\mathrm{bump}}$ and effective potentials $M^2V_{\mathrm{eff}}$ as functions of $r/M$ for eight typical cases with $l=2$. Panels (a1)--(h1) show the bump potentials, while panels (a2)--(h2) compare the effective potentials of cLQBHs (the purple solid curves) with those of Schwarzschild BHs (the green dashed curves). The parameter choices of $\{r_0,\mathcal{A},\mathcal{W}\}$ are indicated in the corresponding bump-potential panels.}
\label{plotpotentials}
\end{figure*}

We now return to the features of $V_{\mathrm{bump}}$ and $V_{\mathrm{eff}}$. Motivated by the observational significance of the $l=2$ gravitational-wave multipole \cite{MaggioreB}, and following the conventions of, e.g., \cite{Zhang:2024svj,Cheung:2021bol,Berti:2022xfj}, our analysis will mainly focus on the representative $l=2$ case in the following. Fig.~\ref{plotpotentials} shows the behavior of the dimensionless potentials $M^2V_{\mathrm{bump}}$ and $M^2V_{\mathrm{eff}}$ for 8 carefully selected typical combinations of $\{r_0,\mathcal{A},\mathcal{W}\}$ at $l=2$ (The factor $M^2$ was multiplied to $V_{\mathrm{eff}}$ to balance its original dimension so that the overall ${\tilde V}_{\mathrm{eff}}=M^2 V_{\mathrm{eff}}$ can be dimensionless). These choices will serve as starting points for subsequent explorations of the parameter space spanned by $\{r_0,\mathcal{A},\mathcal{W}\}$. In particular, as illustrated in Fig.~\ref{plotpotentials}, this paper mainly considers cases where $V_{\mathrm{eff}}(r)>0$ throughout the exterior region $r\in(2M,+\infty)$. This is a reasonable choice, since negative regions of the effective potential can favor the development of dynamical instabilities \cite{Mai:2025mim,Doneva:2017bvd,Silva:2018qhn}, an issue beyond the scope of the present work (although a locally negative effective potential does not necessarily imply dynamical instability \cite{Ishibashi:2003ap}, and this restriction does not exclude a locally negative $V_{\mathrm{bump}}$ from our study).

\begin{figure}[h]
\includegraphics[width=0.9\linewidth]{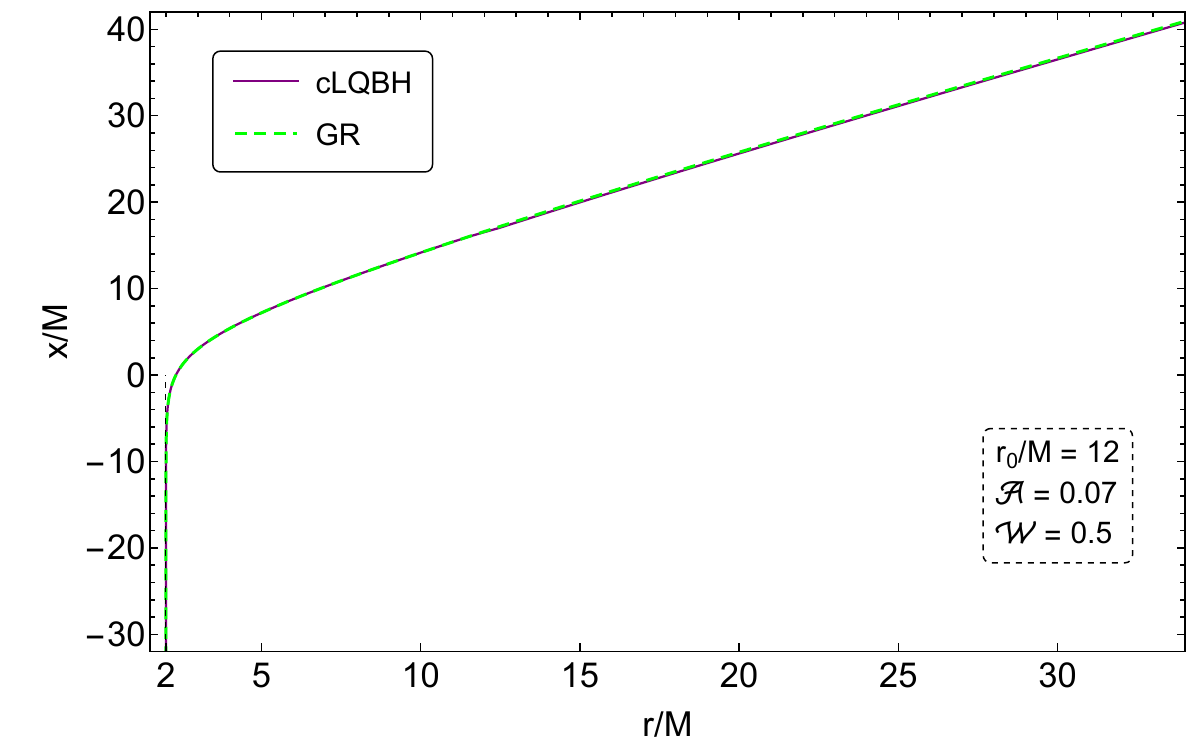}
\includegraphics[width=0.9\linewidth]{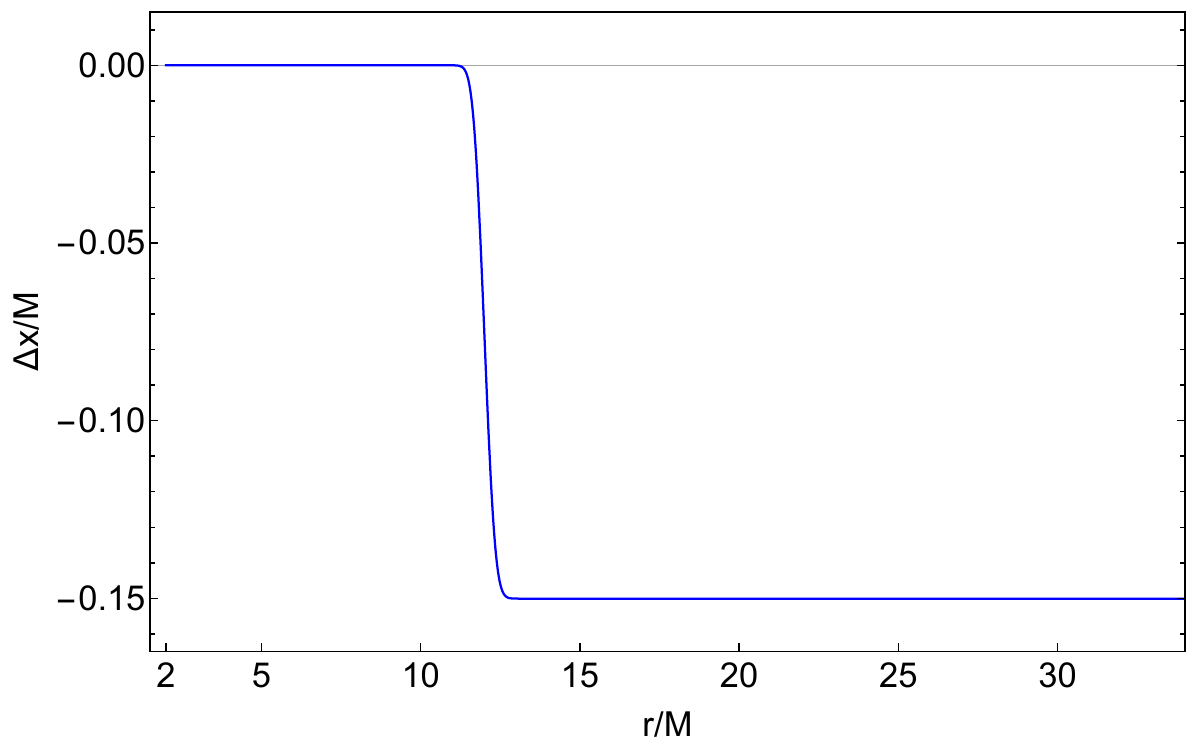}
\caption{The behavior of $x$ (top panel) and $\Delta x = x-x_{\mathrm{GR}}$ (bottom panel) as functions of $r/M$ for a typical case with $\{r_0/M,\mathcal{A},\mathcal{W}\}=\{12,0.07,0.5\}$. These quantities are shown in their dimensionless forms, $x/M$ and $\Delta x/M$, respectively. The Schwarzschild tortoise coordinate $x_{\mathrm{GR}}/M$ is also shown in the top panel for comparison. For both backgrounds, the integration constant is chosen so that $\tilde x(\tilde r=3)=3$.}
\label{plotrstar}
\end{figure}

Before proceeding to the QNM calculations and the associated analysis, let us briefly examine the behavior of the tortoise coordinate [cf.~\eqref{rstar}]. Substituting the expressions for $f(r)$ and $H(r)$, we obtain
\bqn
\lb{rstar2}
\frac{dr}{dx} &=& \sqrt{1+\frac{r-2M}{r}\epsilon^2\left(\frac{r}{r_0}\right)^{-\frac{6M}{r_0-2M}}e^{-\frac{(r-r_0)^2}{M^2\sigma^2}}} \nb\\
&& \times \left(1-\frac{2M}{r}\right) \nb\\
&=& \frac{\tilde r-2}{\tilde r}\sqrt{1+\frac{\tilde r-2}{\tilde r}\epsilon^2\left(\frac{\tilde r}{\tilde r_0}\right)^{-\frac{6}{{\tilde r}_0-2}}e^{-\frac{(\tilde r-\tilde r_0)^2}{\sigma^2}}}, \nb\\
\eqn
where the last line gives the dimensionless form, with $\tilde x\equiv x/M$ and $dr/dx=d\tilde r/d\tilde x$. For the Schwarzschild background, $dr/dx=1-2M/r$, and a convenient choice of the integration constant gives $x_{\mathrm{GR}}=r+2M\ln[(r-2M)/M]$ \cite{Chandra92book,Kokkotas:1999bd} so that $r=3M$ leads to $x_{\mathrm{GR}}=3M$ \footnote{As pointed out earlier, we use $x$ in place of $x_{\mathrm{cLQBH}}$ unless an explicit distinction is needed. The notation $x_{\mathrm{GR}}$ is introduced here to distinguish the Schwarzschild tortoise coordinate from its cLQBH counterpart. Similar treatments will be applied also to other physical quantities in the following of this paper.}. We can integrate \eqref{rstar2} numerically to obtain the relation between $r$ and $x$ for a given $\{r_0,\epsilon,\sigma\}$, or equivalently, $\{r_0,\mathcal{A},\mathcal{W}\}$. Fig.~\ref{plotrstar} shows the typical behavior of $x(r)$ for $\{r_0/M,\mathcal{A},\mathcal{W}\}=\{12,0.07,0.5\}$. To compare the cLQBH and GR coordinates, we use the freedom in the integration constant to impose $\tilde x(\tilde r=3)=3$ for both backgrounds, consistently with the expression for $x_{\mathrm{GR}}$ above. This merely fixes the origin of $x$, without affecting the physical results \cite{Zhang:2022roh}. Clearly, the $x$ maps $r/M \in (2, +\infty)$ to $x/M \in (-\infty, +\infty)$.   In the bottom panel of Fig.~\ref{plotrstar}, we further plot the dimensionless difference $\Delta x/M\equiv(x_{\mathrm{cLQBH}}-x_{\mathrm{GR}})/M$ as a function of $r/M$. Together, the two panels show that the bump has a small influence on $x$ in this example, while $\Delta x$ remains approximately constant away from the bump region on either side of $r=r_0$. The location $r=r_0$ marks the center of the transition between these two nearly constant values. Furthermore, Fig.~\ref{plotrstar} illustrates the mapping between $r$ and $x$, which will be used in the subsequent QNM calculations.


\section{Quasi-Normal Modes of cLQBHs}
\renewcommand{\theequation}{3.\arabic{equation}} \setcounter{equation}{0}
\label{secQNM}

At this point, we are at the stage of calculating the corresponding
QNMs from the master equations given above. Specifically, we can execute the
time-domain calculations by starting from \eqref{master1}, and the
frequency-domain ones by starting from its counterpart, \eqref{master2}. In
principle, for any given $l$ and $\{r_0, \mathcal{A}, \mathcal{W}\}$, by
integrating \eqref{master1} we can obtain the temporal evolution of
$\Psi(t, x_{\mathrm{obs}})$ (where $x_{\mathrm{obs}}$ is the location of the
observer to be assigned below) with suitable physical boundary conditions to
be described at appropriate places. Similarly, for the frequency-domain
analysis, one can further extract a set of $\omega$ (the full notation is
$\omega_{ln}$, with $l, n=0, 1, 2, 3, ...$ and no $m$ index since the
background is non-rotating, though we in general omit these subscripts and
simply write $\omega$ \cite{Berti:2009kk, Bojowald:2024lhr}), either from
$\Psi(t, x_{\mathrm{obs}})$ or by directly solving \eqref{master2}.

In the literature, various methods are available for handling such QNM problems. Examples include the WKB approximation \cite{Schutz:1985km,Iyer:1986np,Konoplya:2003ii,Matyjasek:2026yiu} (with the Pad\'e improvement \cite{Matyjasek:2017psv,Konoplya:2019hlu}), the continued fraction method (CFM) \cite{Leaver:1985ax,Konoplya:2011qq,Leaver:1990zz}, the matrix method \cite{Zhang:2022roh,Lin:2019mmf,Lin:2022ynv}, the pseudospectral method \cite{Dias:2014eua,Jansen:2017oag}, the shooting method \cite{Chandrasekhar:1975zza}, and the finite difference method (FDM) \cite{Zhang:2024svj,Yang:2022btw,Huang:2021qwe,Pazos-Avalos:2004uyd}, together with the Runge–Kutta method \cite{Yang:2024vor,Chen:2026fad}. In what follows, we adopt the ones suitable for our calculations in practice. It is worth noting that, while testing these methods, we performed some preliminary calculations and compared them with the case (i) results of \cite{Bojowald:2024lhr}. These comparisons served to test our numerical implementation and to improve our understanding of the scalar-field dynamics in cLQBHs. Together with checks against selected results from the studies of the bump effect~\cite{Yang:2024vor}, they guided the development of the numerical codes and the computational workflow employed below.

\subsection{Frequency-domain analysis}
\label{subsecA}

Let us first pay attention to the frequency-domain calculations and relevant
analysis for the cLQBH mentioned above. As usual, we shall treat the $l=2$
case as the representative one. However, the selection of the suitable
physical parameters $\{r_0, \mathcal{A}, \mathcal{W}\}$ can be a more
complicated task. First of all, as mentioned earlier, the current work is
focusing on the regime where $V_{\mathrm{eff}}(r)>0$. This leads to some
implicit constraints on $\{r_0, \mathcal{A}, \mathcal{W}\}$. By referring to
the cases considered in Figs. \ref{plotsigma} and \ref{plotpotentials}, we
shall mainly focus on some subspaces of the parameters that satisfy
${\tilde r}_0\in(5, 50)$, $\mathcal{A}\in(0,0.1)$ and $\mathcal{W}\in(0,5)$.

We first notice that we have $V_{\mathrm{eff}}(x) \to 0$ at the two
boundaries of $x$ [cf., \eqref{Veff}], viz., $x\to-\infty$ (${\tilde
r}\to2^{+}$) and ${x}\to+\infty$ (${\tilde r}\to+\infty$). According to
\eqref{master2}, that brings us $\psi \propto e^{\pm i \omega x}$ at the two
boundaries. The physical boundary conditions, i.e., the pure-ingoing-wave
condition at the metric horizon and the pure-outgoing-wave condition at the
spatial infinity, can be translated as the following Robin boundary conditions
\cite{Bojowald:2024lhr,Yang:2024vor,Kinoshita:2023iad}
\begin{equation}
\psi(x)\sim
\begin{cases}
e^{-i\omega x}, & x\to-\infty,\\
e^{+i\omega x}, & x\to+\infty.
\end{cases}
\label{BC}
\end{equation}
Therefore, we can pick out the correct branch of $\psi(x)$ as the boundary
conditions according to \eqref{BC} in the following calculations.

The shooting method \cite{Chandrasekhar:1975zza, Zhang:2022fbz,Pani:2013pma,Doneva:2010ke} mentioned above is well suited to our mathematical problem. Choosing this method can make the calculations for $\omega$ straightforward, clean and manageable. Following \cite{Chandrasekhar:1975zza}, we introduce $\omega=\omega_R+i\omega_I$ and $\phi=\phi_R+i\phi_I$, where $\psi(x) = e^{i \int \phi(x) dx }$. Substituting these quantities into \eqref{master2} leads to the Riccati equations \cite{Konoplya:2011qq}
\bqn
\lb{shooting}
\frac{d\phi_R}{dx} &=& 2\phi_R\phi_I-2\omega_R\omega_I,\nb\\
\frac{d\phi_I}{dx} &=& \omega_R^2-\omega_I^2-\phi_R^2+\phi_I^2-V_{\mathrm{eff}}(r(x)).
\eqn
Here, the subscripts $R$ and $I$ denote the real and imaginary parts, respectively.
The coupled complex ordinary differential equations (ODEs) given by \eqref{shooting} are the equations we are going to integrate in our \textit{Mathematica} code (and its \textit{Python} counterpart) in practice. To fulfill the integrations, let us construct the numerical boundary values for the near-horizon as well as the far-zone regimes by using \eqref{BC} and the Frobenius method \cite{Butkov1968book}. 

In practice, to reduce our computational labor, as mentioned earlier, it is
very convenient to work with the dimensionless form of the ODEs or,
equivalently, by setting $M=1$. That is, we shall work on top of the
following equation
\bqn
\lb{master3}
\frac{\partial^2\psi}{\partial\tilde{x}^2}
+\left[{(M \omega)^2}-{\tilde V}_{\mathrm{eff}} (\tilde{r})\right]\psi &=& 0.
\eqn
To be specific, we expand $\psi$ near the ${\tilde r} \to +\infty$ regime by
\bqn
\lb{expand1}
\psi({\tilde r}) &=& e^{+i \omega x} \sum_{k=0}^{+\infty}  {\alpha_k} {\xi^k},
\eqn
where $\xi \equiv M/r$ and the dimensionless factors $\alpha_k$ will be
determined later. Notice that here we have $\omega x=(M \omega)(x/M)
={\tilde \omega}{\tilde x}$. Similarly, we expand $\psi$ near the
${\tilde r} \to 2^{+}$ regime by
\bqn
\lb{expand2}
\psi({\tilde r}) &=& e^{-i \omega x} \sum_{k=0}^{+\infty}  {\beta_k} {\zeta^k},
\eqn
where $\zeta \equiv (r-r_{\mathrm{MH}})/M$ and the dimensionless factors
$\beta_k$ will be determined later.
It is worth mentioning here that one can never expand the polynomials in
\eqref{expand1} or \eqref{expand2} to the infinite powers in reality.
Instead, we shall expand them to a certain finite order by measuring the
tolerance of the numerical error for the calculations of $\omega$. At the
same time, due to the relation between $\phi$ and $\psi$, we have the
freedom to normalize \eqref{expand1} and \eqref{expand2} to meet our
purposes by properly choosing integration constants. Thus, without loss of
generality, let us set $\alpha_0=\beta_0=1$.

After that, we insert the polynomials in \eqref{expand1} and \eqref{expand2}
into the ODE \eqref{master3}. By solving that order by order, we can arrive
at some recursion relations of the coefficients, which read
\bqn
\lb{alphak}
{2ik\tilde{\omega}} \alpha_k &=&
{\left[k(k-1)-L\right]\alpha_{k-1} -2(k-1)^2\alpha_{k-2}}, \nb\\
\eqn
and
\bqn
\lb{betak}
&& \sum_{j=0}^{k-1}
\left[j(j-1)\bar{A}_{k-j+2}+j\bar{B}_{k-j+1}+\bar{C}_{k-j}\right]\beta_j \nb\\
&=& -\frac{k(k-4i\tilde{\omega})}{4} \beta_k,
\eqn
where
\bqn
\lb{ABCbar3}
\bar{A}_j &=&\left. \frac{1}{j!} \frac{d^j p^2(\tilde r)}{d\tilde r^j} \right|_{\tilde r=2},\nb\\
\bar{B}_j &=&\left. \left[\frac{1}{2j!} \frac{d^{j+1}p^2(\tilde r)}{d\tilde r^{j+1}}
-\frac{2i\tilde{\omega}}{j!} 
\frac{d^j p(\tilde r)}{d\tilde r^j} \right] \right|_{\tilde r=2},\nb\\
\bar{C}_j &=&-\frac{1}{j!} \left. \frac{d^j {\tilde V}_{\mathrm{eff}}}{d {\tilde r}^j} \right|_{{\tilde r}=2},
\eqn
with $k \in \mathbb{N}^{+}$ and $p(r) = p(\tilde r) \equiv d\tilde r/d\tilde x = dr/dx$ [cf., \eqref{rstar}]. Notice that the factors $\bar{A}_j$, $\bar{B}_j$ and $\bar{C}_j$ here are nothing but the coefficients for the Taylor expansions of the corresponding functions.


With the above in hand, for any given $l$ and $\{r_0, \mathcal{A}, \mathcal{W}\}$, we can generate the numerical value of $\psi(x=x_-)$ [so that $\phi(x_-)$] by using \eqref{expand2} and integrate the ODEs \eqref{shooting} from a chosen $x_-$ [with $r_-=r(x=x_-)$ close to the radius of the metric horizon]  to the matching point located at, say, $x=x_m$, at which we obtain $\phi_- \equiv \phi(x=x_m^-)$.  Similarly, we can generate the numerical value of $\psi(x=x_+)$ [so that $\phi(x_+)$] by using \eqref{expand1} and integrate the ODEs \eqref{shooting} from a chosen $x_+$ [with $r_+=r(x=x_+)$ to be sufficiently large] to the same matching point located at $x=x_m$ (a typical choice of $x_m$ is to put its corresponding $r_m$ near the radial position of the main peak of $V_{\mathrm{eff}}$), at which we obtain $\phi_+ \equiv \phi(x=x_m^+)$. 
\footnote{As usual, here we continue to mix the usages of $r$ and $\tilde r$, $x$ and $\tilde x$, etc. As implied by the content from above, the introduction of those ``tilde quantities'' can bring us conveniences in executing the calculations in practice and for narrations, and the same philosophy applies to the trick of setting $M=1$. In the mathematical expressions, the tilde and non-tilde quantities are strictly distinguished. In the narrations, however, we shall loosely refer to both by their original dimensional names, unless such a mixing causes confusions.}
For each iteration, we pick up a reasonable test value of $\omega$ for the numerical integration and check the condition $|\phi_+-\phi_-|\lesssim \epsilon_0$ (which is equivalent to using the Wronskian condition \cite{Butkov1968book,Volkel:2022ewm}), where $\epsilon_0$ is a small number to be selected to meet the accuracy requirement of our calculations. 
We iteratively adjust the trial value of $\omega$ until the condition $|\phi_+-\phi_-|\lesssim \epsilon_0$ is met (in practice, a typical choice is $\epsilon_0=2\times10^{-9}$ in units with $M=1$).
At that stage, the latest test $\omega$ will be the QNM frequency we are looking for.

At the beginning part of this subsection we have mentioned roughly the parameter space under consideration for this current work (although we are not going to explore every corner of the space mentioned in there). That is to cover two anchor points of the full parameter space, i.e., the one in the vicinity of the GR limit and the one far beyond that (while keeping the deviation as a finite number, as an extremely huge discrepancy is meaningless for our purposes). Here are some considerations regarding the chosen parameter space.

As mentioned earlier, the case (i) [cf., \eqref{eq1.1}] considered in \cite{Bojowald:2024lhr} represents a crucial benchmark for our current study. To do a more direct comparison and by taking the same manner [cf., \eqref{Veff}], we may first construct the counterpart of our $V_{\mathrm{bump}}$ for case (i) by introducing a $\delta V_{\mathrm{i}} \equiv V_{\mathrm{i}}-V_{\mathrm{GR}}$, where $V_{\mathrm{i}}$ denotes the effective potential for the case (i). For the physical parameter $\Delta$, the authors of \cite{Bojowald:2024lhr} considered the case where $\Delta/(4 M^2) \in [0, 1]$. That brings the amplitude of $M^2 \delta V_{\mathrm{i}}$ to be able to reach the order of ${\cal O}(10^{-3})$. However, they actually did not set up an upper limit for this factor. In fact, by considering the ultralight BHs \cite{Carr:2015nqa,Domenech:2023mqk,Sasaki:2025zao,Modesto:2009ve}, the amplitude of $M^2 \delta V_{\mathrm{i}}$ can in principle reach the order of approximately ${\cal O}(10^{-1})$. This gives us the reason to mainly focus on a sufficiently large $\mathcal{A}$ while keeping it no more than the order of ${\cal O}(10^{-1})$, which can bring us to the balanced point between finding an observable deviation from that of GR (at least for some theoretical purposes) and avoiding taking into account unnecessary extreme cases.

On the other hand, since the bump effect on QNM has been investigated intensively during the past several years \cite{Cheung:2021bol,Berti:2022xfj,Yang:2024vor,Yang:2025dbn,Konoplya:2024wds}, we shall refer to works for some clues in determining a reasonable range for $r_0$ that is worth exploring. These studies imply that a typical choice of $r_0/M$ can be of the order ${\cal O}(10)-{\cal O}(100)$. However, we shall not attach equal significance to these choices of $r_0$, since this current work is not a comprehensive study of the bump effect nor purely an extension in this direction. Under the requirement of $V_{\mathrm{eff}}>0$, the allowed range for $\mathcal{A}$ will be too small for a large $r_0$, which will dilute the contribution of the corresponding $\mathcal{A}$ to certain degree. Therefore, for our purpose, as seen from Fig.\ref{plotpotentials}, we will mainly focus on the values of $r_0/M$ smaller than 30. In fact, our choices can make some special contributions to the area of the bump effect, as most of the existing works have not paid enough attention to the small $r_0$ regime. Our chosen range for $r_0$ can link the near-zone cases to some of the typical intermediate $r_0$'s existing in the literature. Similarly, we can find the counterpart of our $\mathcal{W}$ in the literature. A typical size for it ranges approximately from 1 to 3. Nonetheless, to extend our study to the ``sharp peak'' regime, we will allow $\mathcal{W}$ to be smaller than 1 \cite{Torres:2026uey,Daghigh:2026lhk}.

\begin{figure*}[t]
\centering
\begin{minipage}{0.34\textwidth}
\centering
\includegraphics[width=\linewidth]{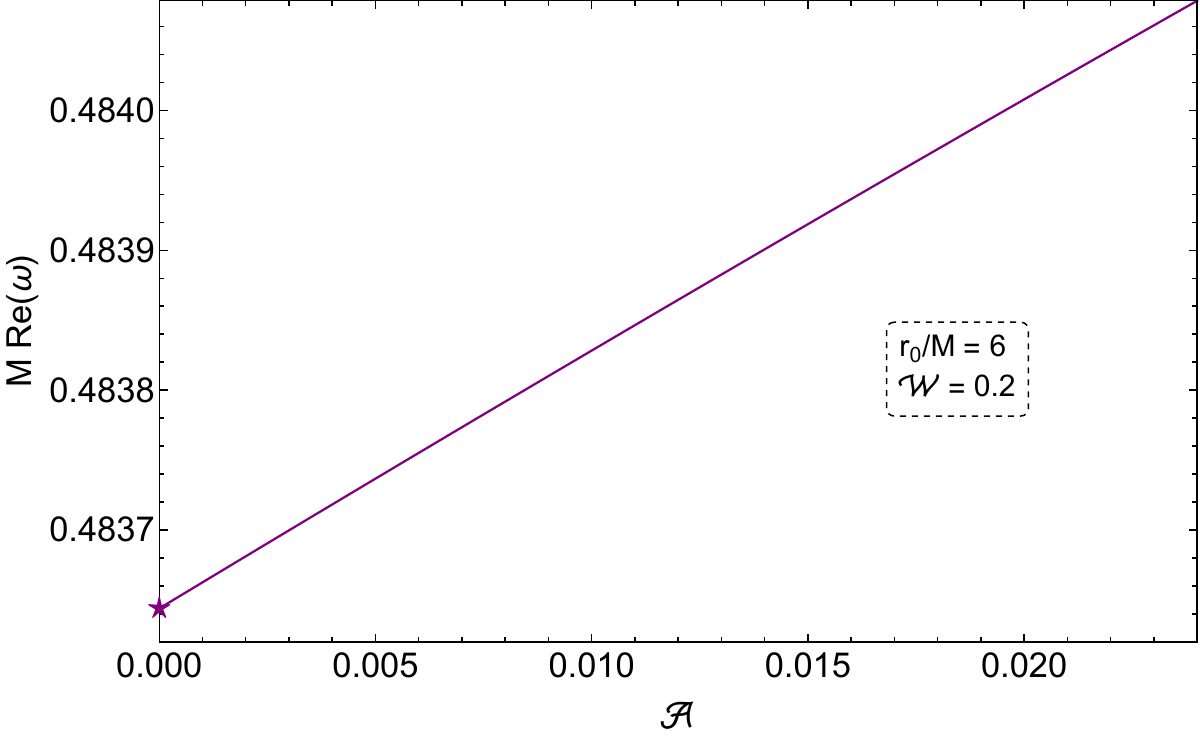}
\end{minipage}\hspace{0.02\textwidth}
\begin{minipage}{0.34\textwidth}
\centering
\includegraphics[width=\linewidth]{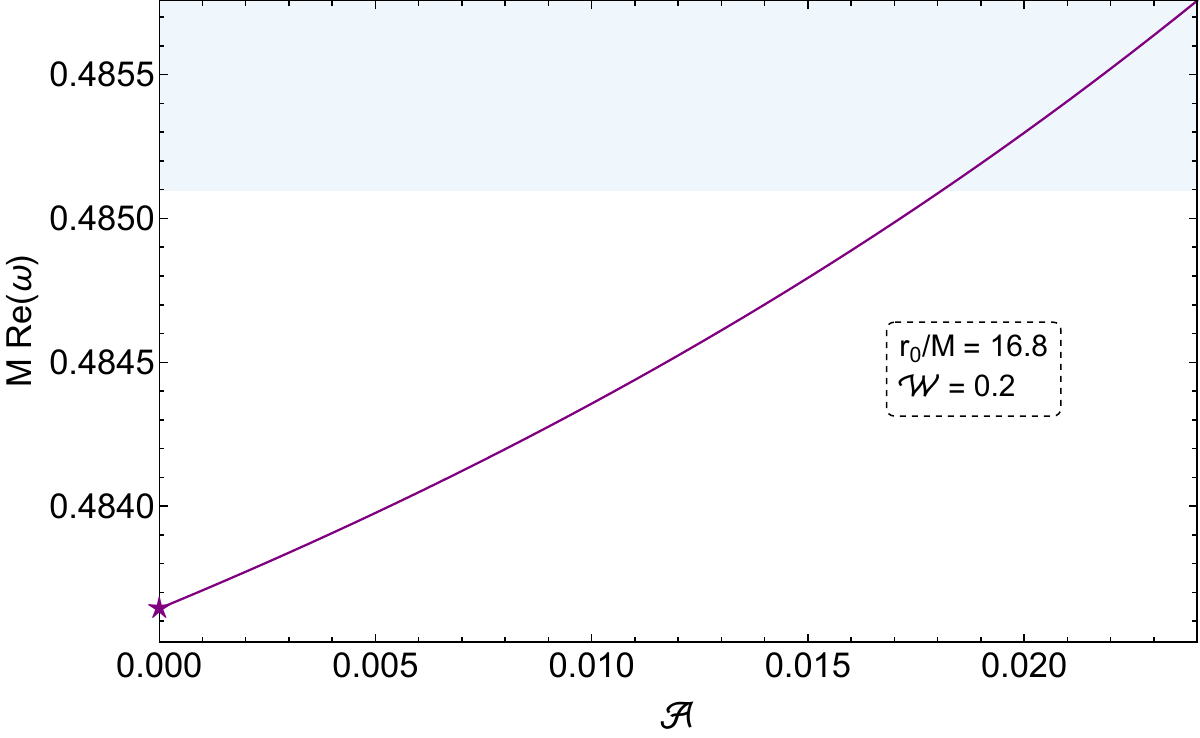}
\end{minipage}
\par\vspace{2mm}
\begin{minipage}{0.34\textwidth}
\centering
\includegraphics[width=\linewidth]{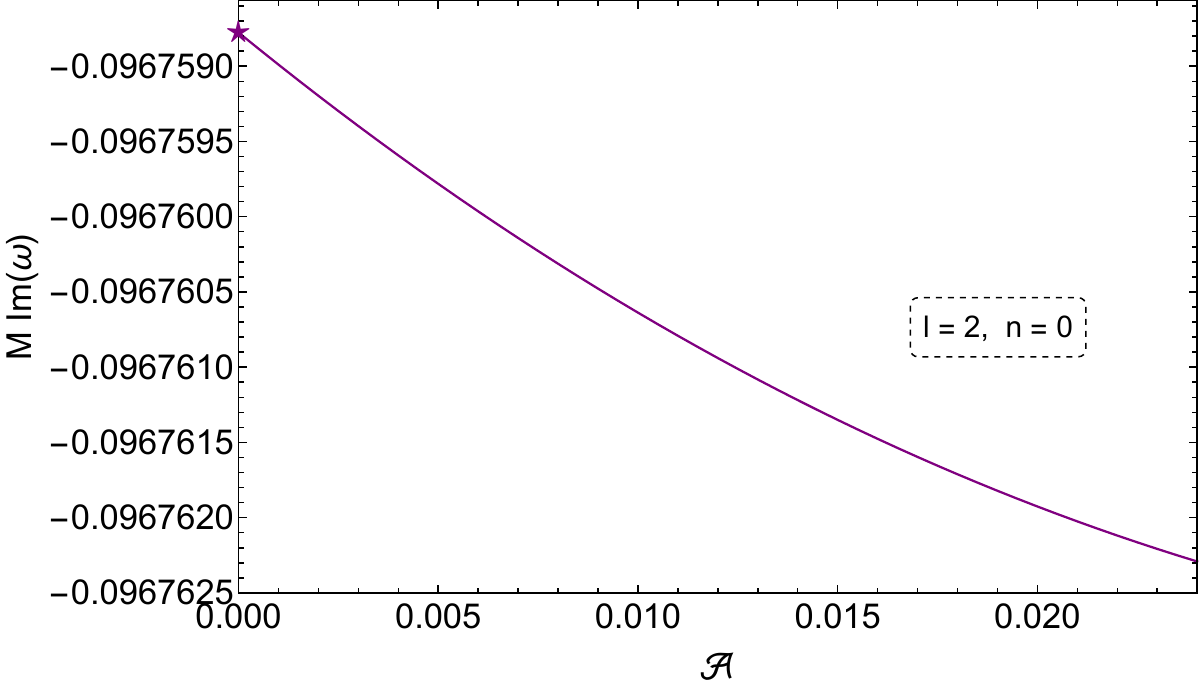}
\end{minipage}\hspace{0.02\textwidth}
\begin{minipage}{0.34\textwidth}
\centering
\includegraphics[width=\linewidth]{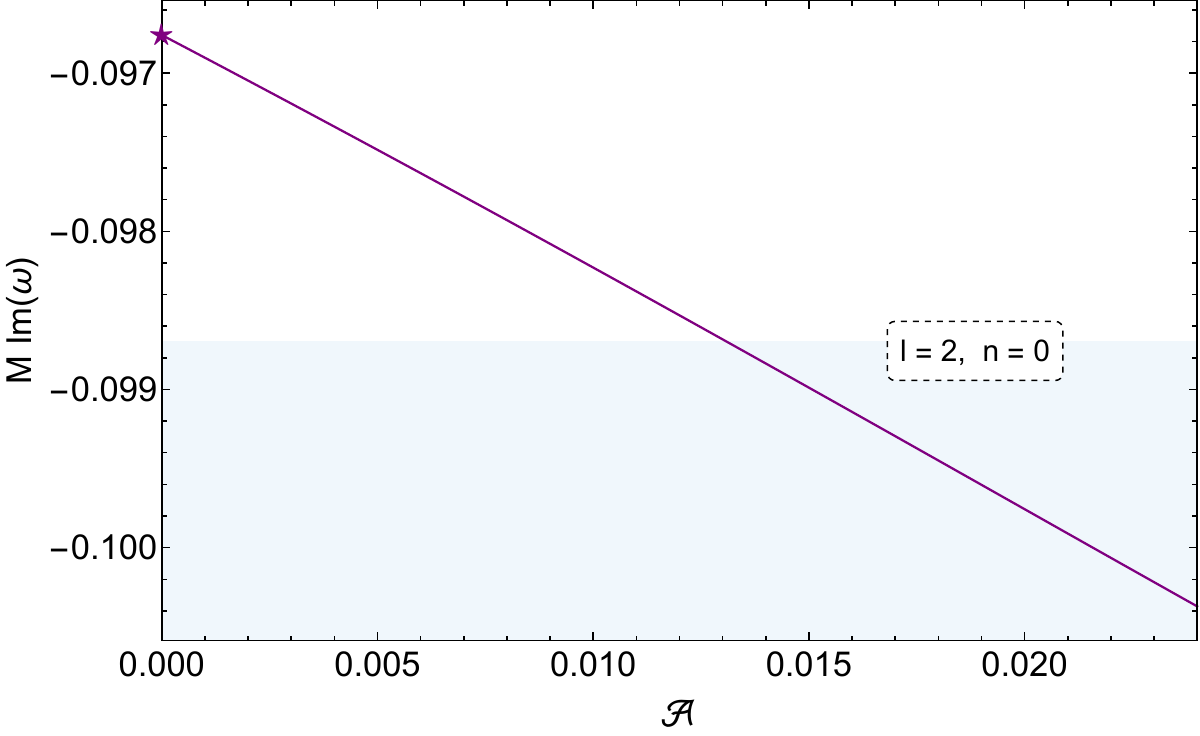}
\end{minipage}
\par\vspace{3mm}
\begin{minipage}{0.34\textwidth}
\centering
\includegraphics[width=\linewidth]{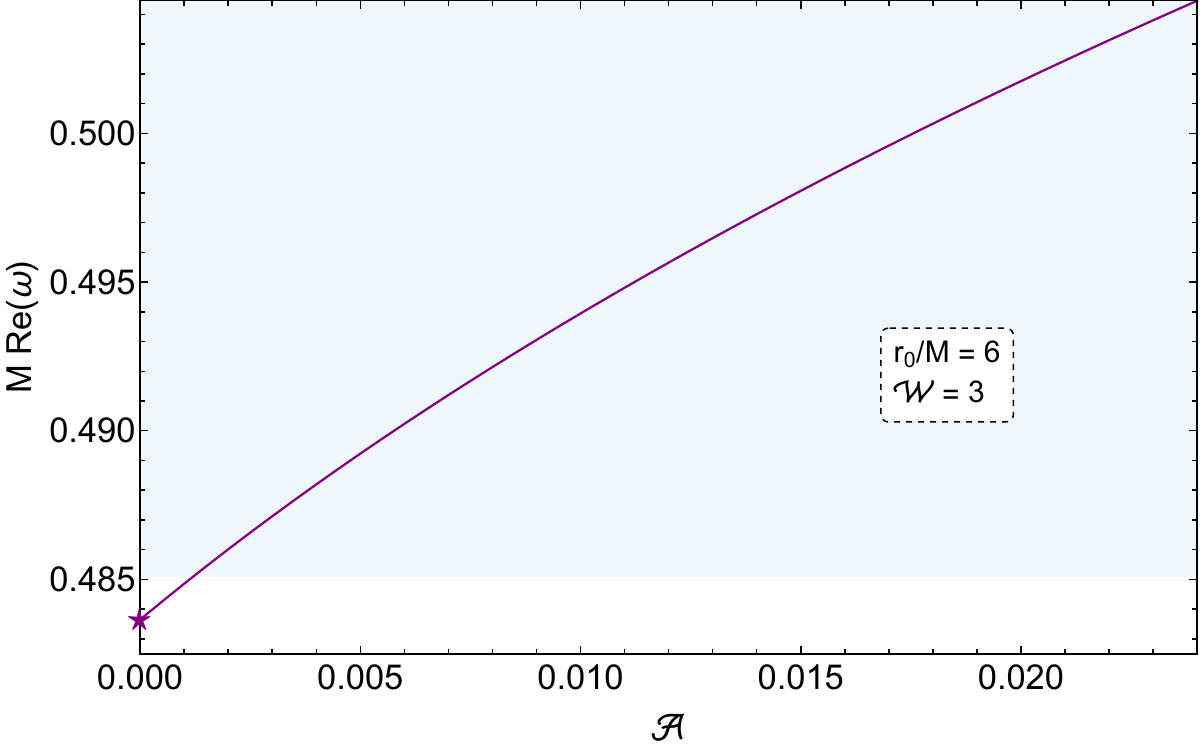}
\end{minipage}\hspace{0.02\textwidth}
\begin{minipage}{0.34\textwidth}
\centering
\includegraphics[width=\linewidth]{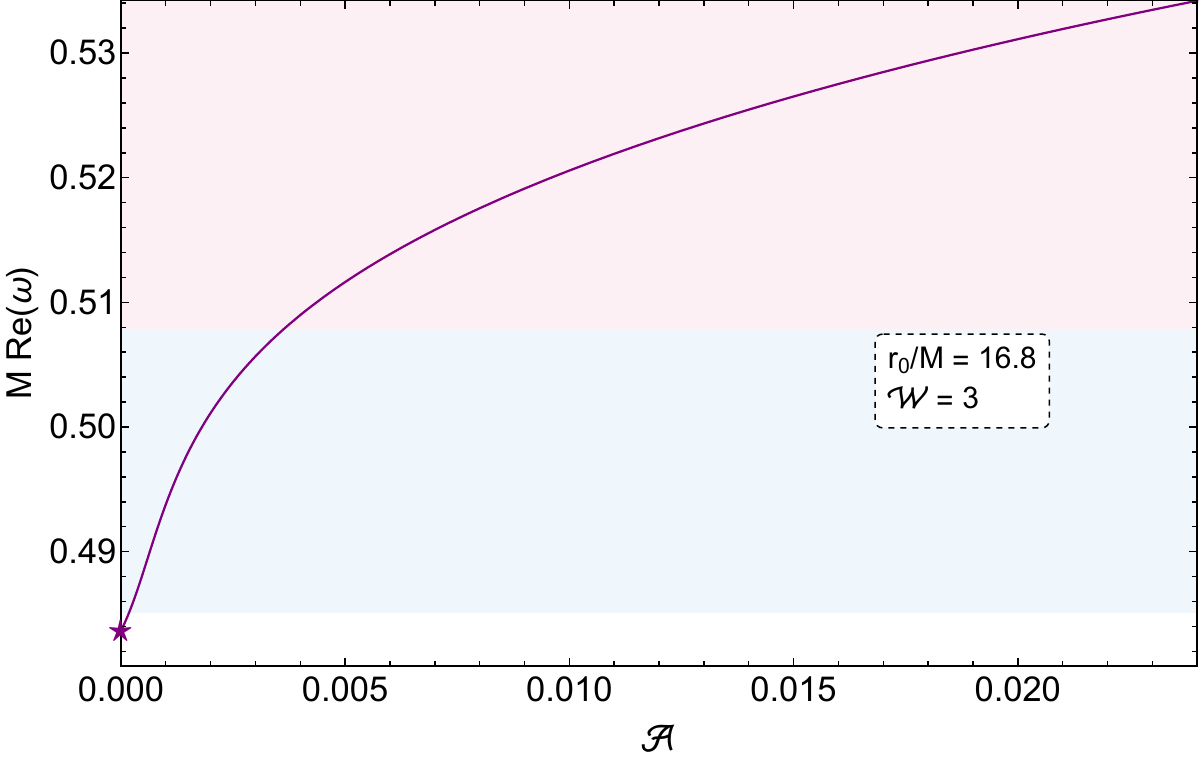}
\end{minipage}
\par\vspace{2mm}
\begin{minipage}{0.34\textwidth}
\centering
\includegraphics[width=\linewidth]{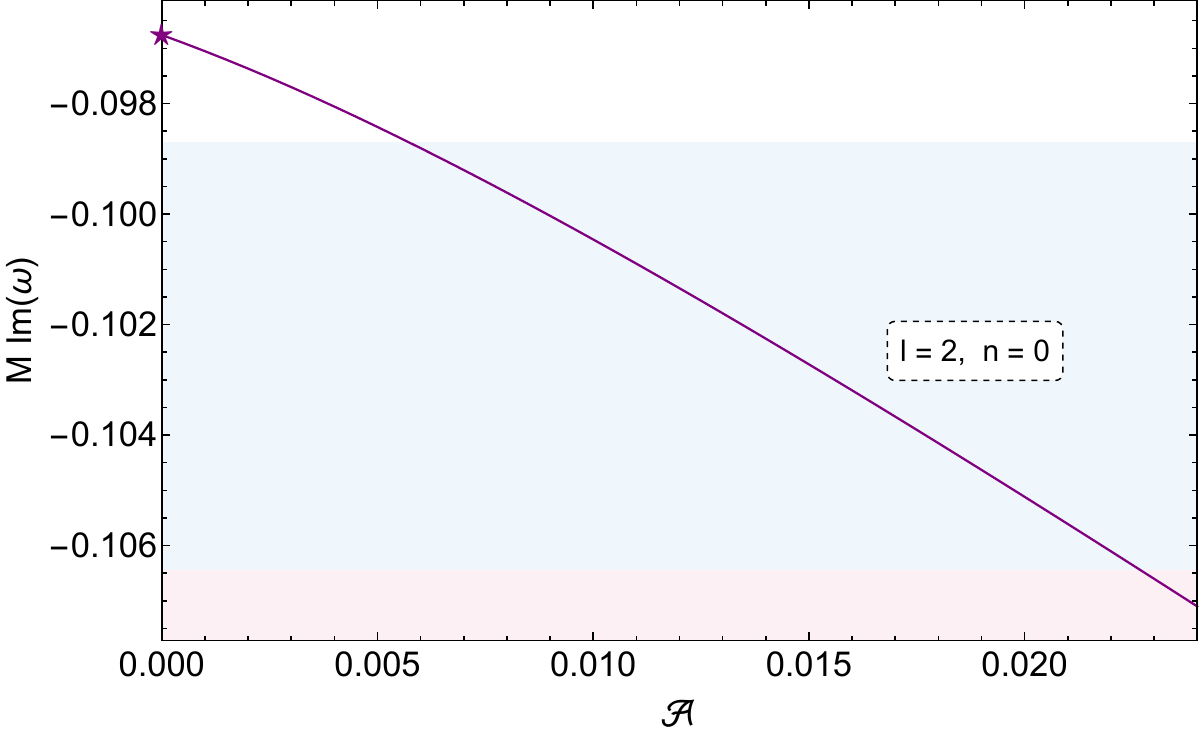}
\end{minipage}\hspace{0.02\textwidth}
\begin{minipage}{0.34\textwidth}
\centering
\includegraphics[width=\linewidth]{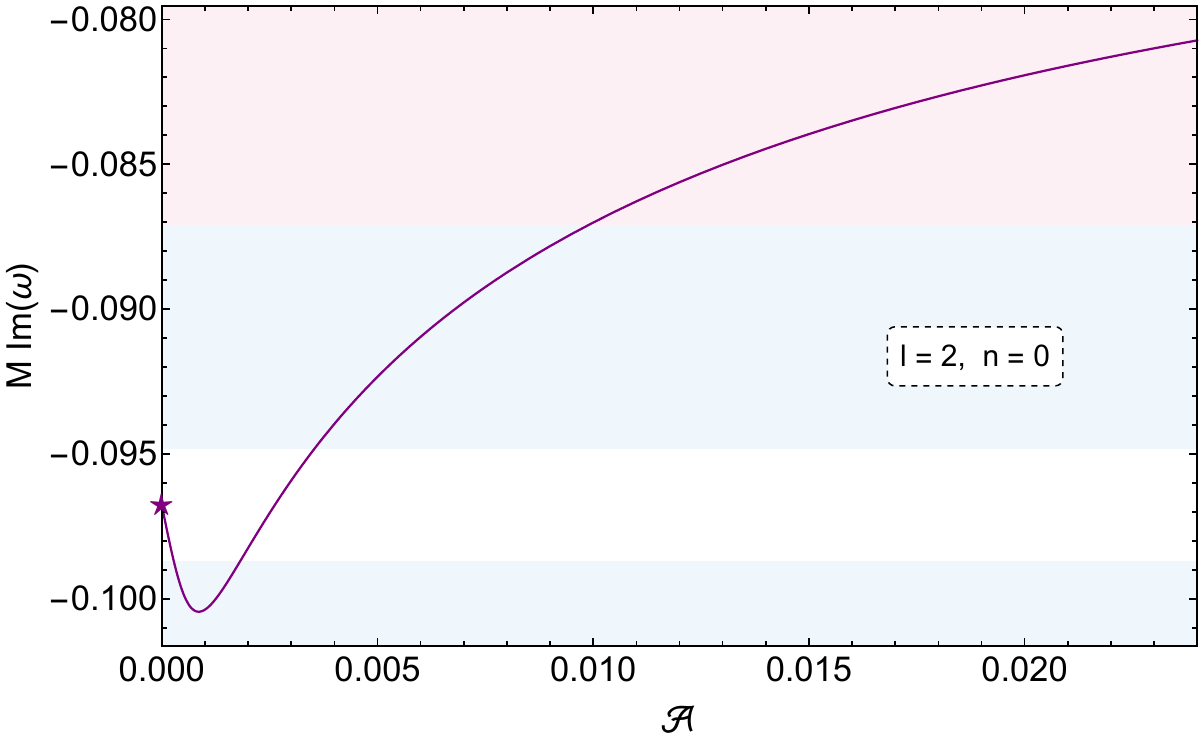}
\end{minipage}
\caption{The real and imaginary parts of $M\omega_{20}$ for our cLQBH [case (ii)] as functions of $\mathcal{A}$ for four representative parameter choices. Here, the subscript ``20'' explicitly denotes $l=2$ and $n=0$. The first and third rows show the real parts, while the second and fourth rows show the corresponding imaginary parts. For each pair of graphs, the fixed values of $r_0/M$ and $\mathcal{W}$ are indicated in the real-part panel. In each panel, the purple star marks the corresponding value in the GR limit. The light blue and light pink regions indicate deviations from GR that lie outside the approximate ranges adopted from TianQin forecasts and LIGO observations, respectively, and the pink shading takes precedence where they overlap. The shaded regions provide approximate observational reference ranges and do not imply observational exclusions.}
\label{plotomegal2n0}
\end{figure*}

\begin{figure}[h]
\includegraphics[width=0.9\linewidth]{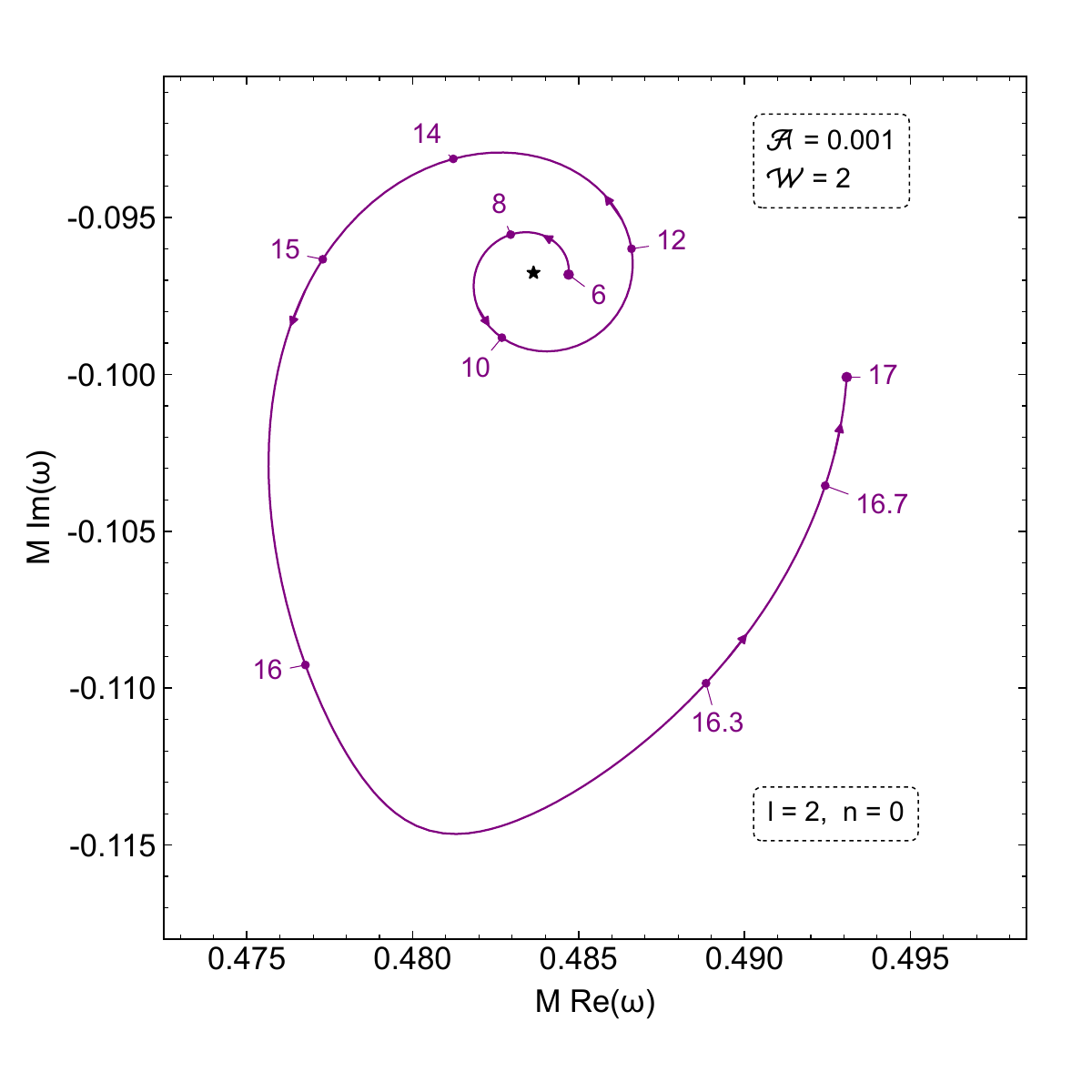}
\caption{The trajectory of $M \omega_{20}$ in the complex plane with a varying $r_0/M \in [6, 17]$. Here we set $\mathcal{A}=0.001$ and $\mathcal{W}=2$. Note that we use the black pentagram to mark the location of $M \omega_{20}$ in GR, for which we have $M \omega_{20} \approx 0.483644-0.096759i$ \cite{Matyjasek:2017psv}. The purple arrows on the curve indicate the increasing direction of $r_0$.} 
\label{plotspiral}
\end{figure}

\begin{figure}[h]
\includegraphics[width=0.9\linewidth]{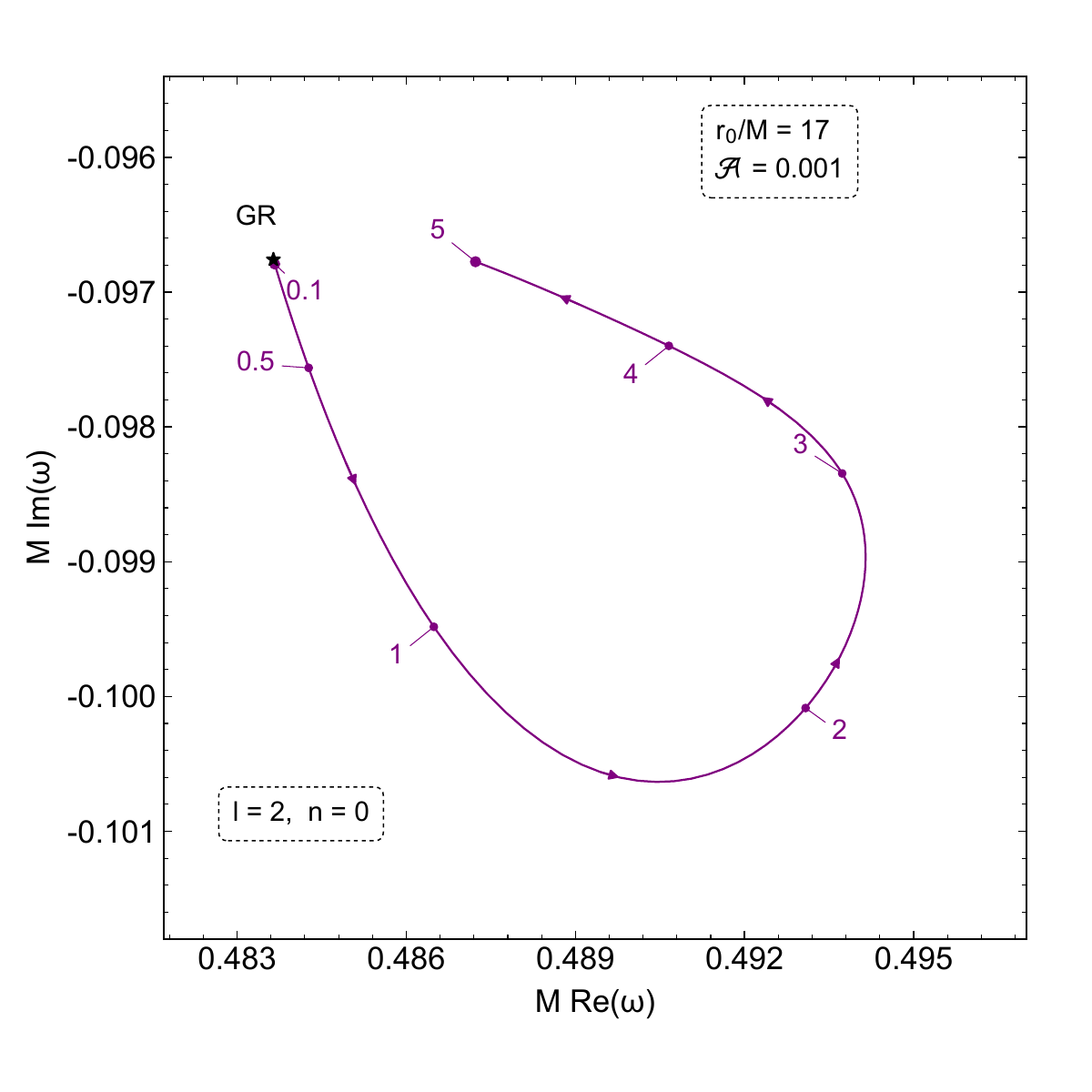}
\includegraphics[width=0.9\linewidth]{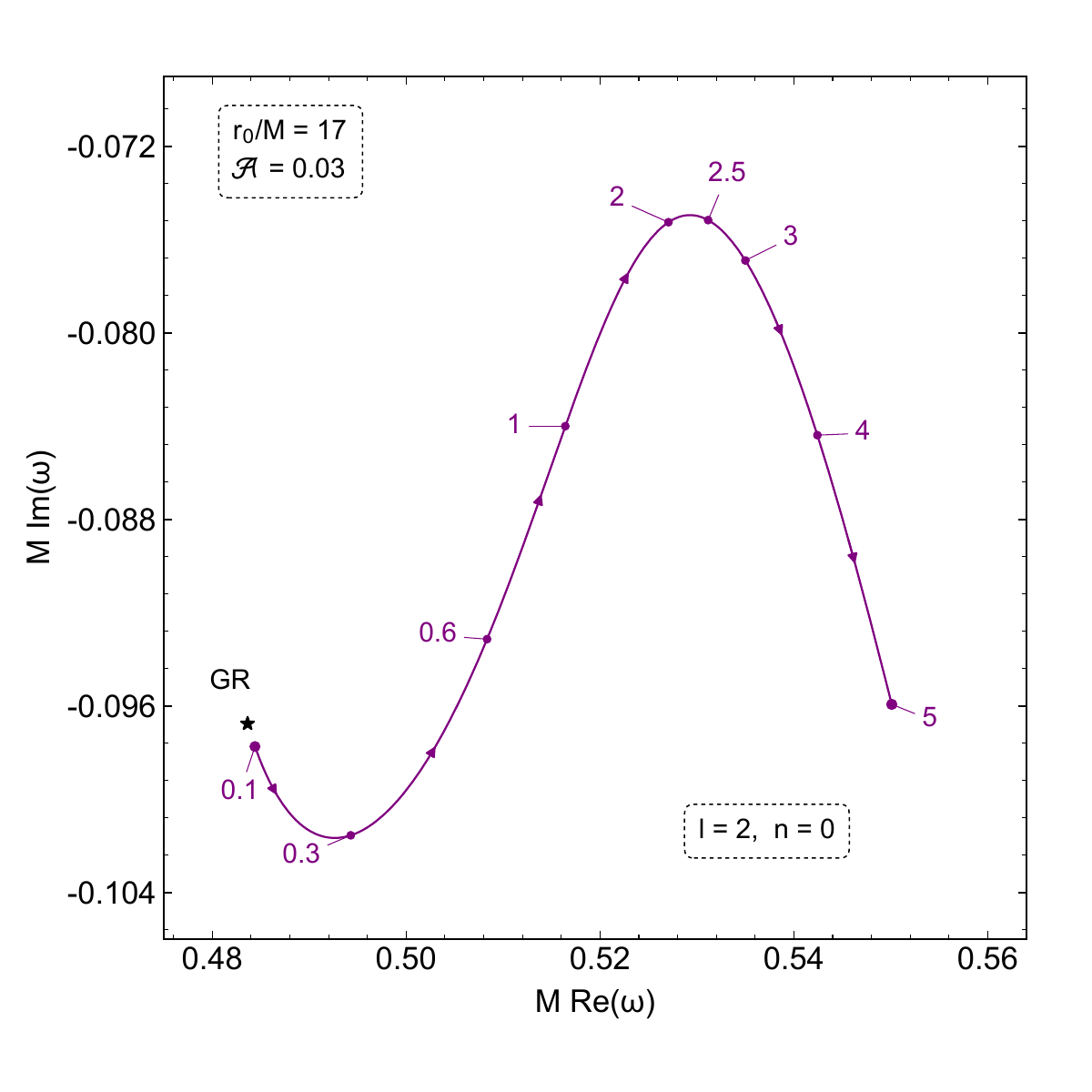}
\caption{The trajectory of $M \omega_{20}$ in the complex plane with a varying $\mathcal{W} \in [0.1, 5]$. Here we set $\mathcal{A}=0.001$ for the upper panel and $\mathcal{A}=0.03$ for the lower panel. In addition, we fix $r_0/M=17$ for both panels. Note that we use the black pentagram to mark the location of $M \omega_{20}$ in GR, for which we have $M \omega_{20} \approx 0.483644-0.096759i$ \cite{Matyjasek:2017psv}. The purple arrows on the curve indicate the increasing direction of $\mathcal{W}$.} 
\label{plottrajectoryofvaryingW}
\end{figure}

Knowing all of these, we plot out the real and imaginary parts of the dimensionless quantity $M \omega$ (by setting $l=2$ and $n=0$) for four representative parameter choices, i.e., $\{r_0/M, \mathcal{W}\}=\{6, 0.2\}, \{16.8, 0.2\}, \{6, 3\}, \{16.8, 3\}$, as a function of the amplitude $\mathcal{A}$ in Fig.\ref{plotomegal2n0}. In this way, we cover four typical scenarios, namely, a sharp/wide bump located in the near/intermediate zone. By observing Fig.\ref{plotomegal2n0} for the four independent cases, we notice that the amount of deviation from GR for $\omega$ in general increases monotonically with $\mathcal{A}$, which agrees with our intuition. Similarly, a wider as well as farther bump tends to drag $M \omega$ away from its GR counterpart. Indeed, we also notice that, in the relatively far zone (e.g., $r_0/M \gg 2$), the behavior of $M \omega$ can be more complicated. That is partially due to the fact that a relatively large $r_0$ is strongly connected to some fancy behaviors of $\omega$, as revealed by the existing literature \cite{Cheung:2021bol,Yang:2024vor,Qian:2024iaq,Ianniccari:2024ysv} and Fig.\ref{plotspiral}. Our results from Fig.\ref{plotspiral} show that this sort of spiral pattern exists not only for positive bumps (see, e.g., \cite{Leung:1999iq}) but also for mixed ones. In contrast, the trajectory of $\omega$ will tend to be more regular with a changing $\mathcal{W}$, as seen from Fig.\ref{plottrajectoryofvaryingW}.
As a connection to some of the previous works (including ours) \cite{Zhang:2022fbz,Mark:2017dnq}, we would also like to point out another issue arising from Fig.\ref{plotspiral}. Replacing the purely ingoing condition with a partially reflective boundary near the horizon, characterized by a generally complex reflection coefficient, provides another way to modify QNM trajectories \cite{Wu:2025wbp}. Motivated also by effective descriptions of localized potential perturbations in terms of reflectivity \cite{Oshita:2025ibu}, it would be interesting to investigate whether suitable variations of the reflection amplitude and phase can reproduce spiral trajectories resembling Fig.\ref{plotspiral}, and whether a correspondence with the bump parameters can be established, (especially when some sort of degeneracy can occur). We leave this possible connection between potential deformations and modified boundary conditions for future investigation.

It is worth mentioning here that Fig.\ref{plotomegal2n0} compares our results with approximate reference ranges motivated by LIGO observations and TianQin forecasts. The LVK Collaboration recently reported constraints on departures from the Kerr QNM spectrum using the event GW250114 \cite{LIGOScientific:2025wao}. Although our calculations concern scalar perturbations of spherically symmetric BHs, previous studies suggest that, at least in specific models, the fractional deviations from GR for scalar and gravitational perturbations can be of similar orders of magnitude \cite{del-Corral:2022kbk,Pani:2026yzi}, an expectation also adopted in our earlier work \cite{Zhang:2024svj}. Comparisons between nearly static and moderately rotating BHs \cite{Chung:2024vaf} suggest that such an expectation may remain reasonable in specific models. In particular, \cite{Pani:2026yzi} directly compares scalar-QNM deviations with GW250114-inspired reference bands. These studies motivate our approximate comparison and help identify promising parameters for future gravitational and rotating extensions. For this purpose, we define
\bqn
\lb{deltaomega}
\delta\omega_R &\equiv& \frac{\operatorname{Re}(\omega_{\mathrm{cLQBH}})-\operatorname{Re}(\omega_{\mathrm{GR}})}{|\operatorname{Re}(\omega_{\mathrm{GR}})|}\times100\%,\nb\\
\delta\omega_I &\equiv& \frac{\operatorname{Im}(\omega_{\mathrm{cLQBH}})-\operatorname{Im}(\omega_{\mathrm{GR}})}{|\operatorname{Im}(\omega_{\mathrm{GR}})|}\times100\%,
\eqn
where each frequency is compared with its corresponding GR value at the same mass and mode indices, and the subscript ``cLQBH'' is  added expediently to distinguish it from its GR counterpart. A positive $\delta\omega_I$ indicates slower damping than in GR. Thus, we compare fractional deviations without equating the absolute scalar and gravitational frequencies.

Using the GW250114 constraints for the dominant $(l,m,n)=(2,2,0)$ mode \cite{LIGOScientific:2025wao} and translating them into the fractional deviations defined above, we adopt the rounded reference intervals $\delta\omega_R\in[0,5\%]$ and $\delta\omega_I\in[-10\%,10\%]$. Applying the same approach to the TianQin forecasts \cite{Shi:2019hqa}, we obtain the approximate ranges $\delta\omega_R\in[-0.3\%,0.3\%]$ and $\delta\omega_I\in[-2\%,2\%]$. In Fig.\ref{plotomegal2n0}, the light blue and light pink regions indicate deviations outside the adopted TianQin and LIGO ranges, respectively, and the pink shading takes precedence where they overlap. These shaded regions highlight parameter choices with potentially observable deviations, but do not establish observational exclusions. A more accurate assessment requires gravitational perturbations of rotating cLQBHs and a dedicated analysis of the GW data, which we leave to future work.\footnote{This is a sensible strategy for our current purposes. As shown in the literature, gravitational QNM calculations for LQBHs often involve additional assumptions about the perturbation dynamics, such as describing quantum corrections through effective matter or neglecting their contributions \cite{Bouhmadi-Lopez:2020oia,del-Corral:2022kbk}. Applying these prescriptions to our cLQBH requires further justification, whereas the minimally coupled scalar equation is directly determined by the specified background geometry \cite{Bojowald:2024lhr}. Scalar perturbations therefore provide a well-defined first step for identifying the relevant physical features and developing numerical tools for more complete GW ringdown calculations in future work.}

\begin{table}[t]
\centering
\setlength{\tabcolsep}{8pt}
\setlength{\fboxsep}{0pt}
\setlength{\fboxrule}{\arrayrulewidth}
\caption{The cLQBH's fundamental scalar QNM frequencies $\omega$ ($l=2$ and $n=0$) and their fractional deviations from GR, defined in \eqref{deltaomega}, for $r_0/M=16.8$, $\mathcal{W}=3$ and various $\mathcal{A}$. The frequencies are rounded to six decimal places.}
\label{tableomegaIV}
\fbox{%
\begin{tabular}{c|c|c|c}
\noalign{\vskip\doublerulesep}
\hline
$\mathcal{A}$ & $M\omega_{20}$ & $\delta\omega_R$ & $\delta\omega_I$ \\
\hline
$0.001$ & $0.493776-0.100377i$ & $+2.09\%$ & $-3.74\%$ \\
$0.002$ & $0.501119-0.098249i$ & $+3.61\%$ & $-1.54\%$ \\
$0.004$ & $0.508980-0.093964i$ & $+5.24\%$ & $+2.89\%$ \\
$0.006$ & $0.513861-0.090959i$ & $+6.25\%$ & $+5.99\%$ \\
$0.008$ & $0.517547-0.088741i$ & $+7.01\%$ & $+8.29\%$ \\
$0.010$ & $0.520572-0.087016i$ & $+7.64\%$ & $+10.1\%$ \\
$0.012$ & $0.523168-0.085624i$ & $+8.17\%$ & $+11.5\%$ \\
$0.016$ & $0.527515-0.083496i$ & $+9.07\%$ & $+13.7\%$ \\
$0.020$ & $0.531110-0.081933i$ & $+9.81\%$ & $+15.3\%$ \\
$0.024$ & $0.534193-0.080730i$ & $+10.5\%$ & $+16.6\%$ \\
\hline
\noalign{\vskip\doublerulesep}
\end{tabular}%
}
\end{table}

\begin{table*}[t]
\centering
\footnotesize
\setlength{\tabcolsep}{2.5pt}
\setlength{\fboxsep}{0pt}
\setlength{\fboxrule}{\arrayrulewidth}
\renewcommand{\arraystretch}{1.2}
\caption{The scalar QNM frequencies $\omega$ for $l=2$ and $n=0,1,2$, together with their fractional deviations from the corresponding GR values, defined in \eqref{deltaomega}, for $r_0/M=12$, $\mathcal{W}=2$ and various $\mathcal{A}$. The mode labels follow continuous branches from GR as $\mathcal{A}$ increases. The GR reference values can be denoted as $M\omega_{20}^{\mathrm{GR}}\approx0.483644-0.096759i$, $M\omega_{21}^{\mathrm{GR}}\approx0.463851-0.295604i$ and $M\omega_{22}^{\mathrm{GR}}\approx0.430544-0.508558i$ \cite{Matyjasek:2017psv}. Frequencies are rounded to six decimal places.}
\label{tableomegal2n012}
\resizebox{\textwidth}{!}{%
\fbox{%
\begin{tabular}{c|c|c|c|c|c|c|c|c|c}
\noalign{\vskip\doublerulesep}
\hline
$\mathcal{A}$ & $M\omega_{20}$ & $\delta\omega_R$ & $\delta\omega_I$ & $M\omega_{21}$ & $\delta\omega_R$ & $\delta\omega_I$ & $M\omega_{22}$ & $\delta\omega_R$ & $\delta\omega_I$ \\
\hline
$0.001$ & $0.486596-0.095992i$ & $+0.6\%$ & $+0.8\%$ & $0.402113-0.179523i$ & $-13\%$ & $+39\%$ & $0.573272-0.229971i$ & $+33\%$ & $+55\%$ \\
$0.007$ & $0.497036-0.090412i$ & $+2.8\%$ & $+6.6\%$ & $0.395466-0.121371i$ & $-15\%$ & $+59\%$ & $0.603011-0.192020i$ & $+40\%$ & $+62\%$ \\
$0.013$ & $0.502612-0.086606i$ & $+3.9\%$ & $+10\%$ & $0.394626-0.104457i$ & $-15\%$ & $+65\%$ & $0.614212-0.182056i$ & $+43\%$ & $+64\%$ \\
$0.019$ & $0.506556-0.084011i$ & $+4.7\%$ & $+13\%$ & $0.394905-0.094796i$ & $-15\%$ & $+68\%$ & $0.621742-0.176830i$ & $+44\%$ & $+65\%$ \\
$0.025$ & $0.509673-0.082123i$ & $+5.4\%$ & $+15\%$ & $0.395548-0.088193i$ & $-15\%$ & $+70\%$ & $0.627542-0.173550i$ & $+46\%$ & $+66\%$ \\
$0.031$ & $0.512276-0.080685i$ & $+5.9\%$ & $+17\%$ & $0.396323-0.083260i$ & $-15\%$ & $+72\%$ & $0.632301-0.171299i$ & $+47\%$ & $+66\%$ \\
$0.037$ & $0.514525-0.079552i$ & $+6.4\%$ & $+18\%$ & $0.397138-0.079370i$ & $-14\%$ & $+73\%$ & $0.636352-0.169671i$ & $+48\%$ & $+67\%$ \\
$0.043$ & $0.516509-0.078637i$ & $+6.8\%$ & $+19\%$ & $0.397955-0.076190i$ & $-14\%$ & $+74\%$ & $0.639887-0.168452i$ & $+49\%$ & $+67\%$ \\
$0.049$ & $0.518288-0.077885i$ & $+7.2\%$ & $+20\%$ & $0.398753-0.073520i$ & $-14\%$ & $+75\%$ & $0.643025-0.167517i$ & $+49\%$ & $+67\%$ \\
$0.055$ & $0.519902-0.077257i$ & $+7.5\%$ & $+20\%$ & $0.399526-0.071233i$ & $-14\%$ & $+76\%$ & $0.645847-0.166788i$ & $+50\%$ & $+67\%$ \\
\hline
\noalign{\vskip\doublerulesep}
\end{tabular}%
}%
}
\end{table*}

\begin{figure}[h]
\includegraphics[width=0.9\linewidth]{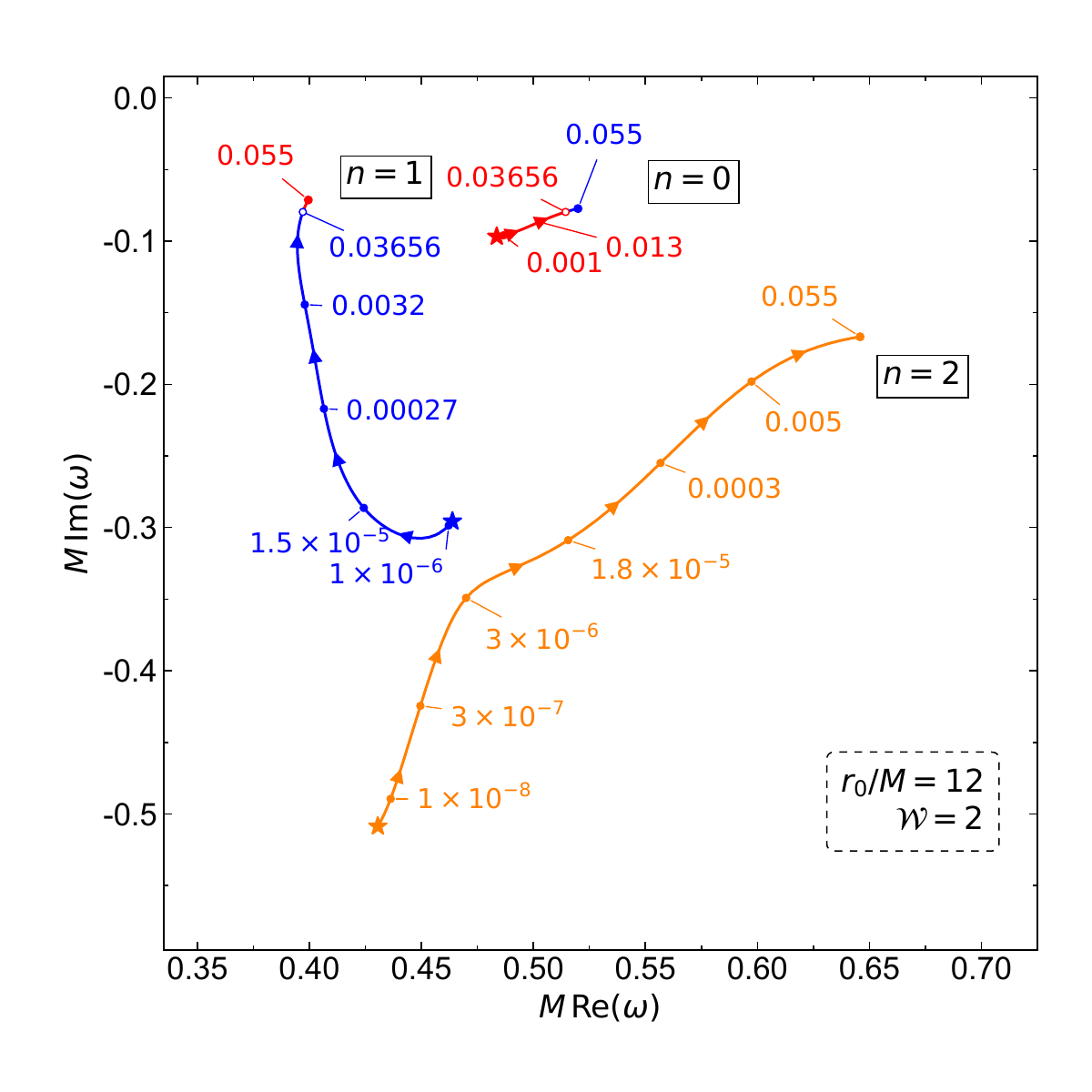}
\caption{The trajectories of $M\omega$ for our cLQBH with $l=2$ and $n=0,1,2$ as $\mathcal{A}$ varies over $[0,0.055]$, with $\mathcal{W}=2$ and $r_0/M=12$. On each curve, the colors indicate the lifetime ranking among the three modes at the same $\mathcal{A}$: red denotes the longest-lived mode, blue the second longest-lived mode, and orange the shortest-lived mode. A smaller $|\operatorname{Im}(\omega)|$ corresponds to a lower damping rate and a longer lifetime. The pentagrams mark the corresponding GR QNM frequencies at the beginning of each curve (in GR, we have $M\omega_{20}\approx0.483644-0.096759i$, $M\omega_{21}\approx0.463851-0.295604i$, and $M\omega_{22}\approx0.430544-0.508558i$ \cite{Matyjasek:2017psv}). The figure shows the existence of overtaking of the $n=0$ mode by the $n=1$ mode, with the latter becoming the new longest-lived mode as $\mathcal{A}$ increases. This overtaking occurs at $\mathcal{A}\approx0.03656$.}
\label{plottrajectoryl2n012}
\end{figure}

\begin{figure}[h]
\includegraphics[width=0.9\linewidth]{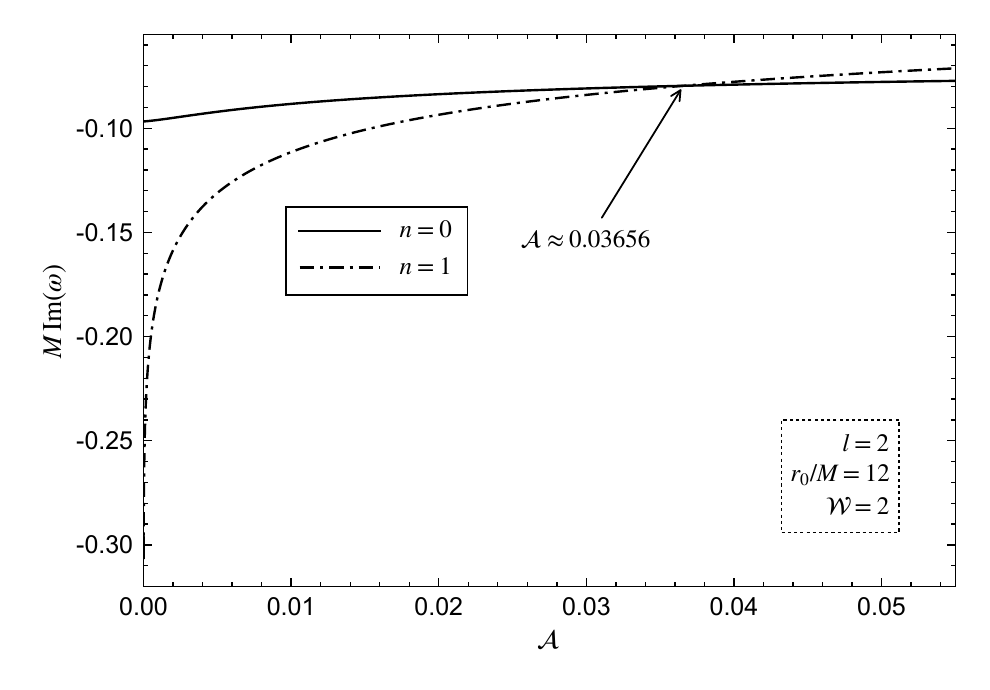}
\caption{The imaginary parts $M\operatorname{Im}(\omega)$ for our cLQBH with $l=2$, $n=0$ (solid curve), and $n=1$ (dot-dashed curve), as $\mathcal{A}$ varies over $[0,0.055]$, with $\mathcal{W}=2$ and $r_0/M=12$. Overtaking occurs at $\mathcal{A}\approx0.03656$.}
\label{plottrajectoryl2n012B}
\end{figure}

Since the $\omega$'s in the right-hand panels of Fig.\ref{plotomegal2n0} exhibit various types of behavior, covering both the small-deviation and the large-deviation regimes, we further list the exact values of the corresponding $\omega$ up to six decimal places in Table \ref{tableomegaIV}. From there we confirm that the deviation from GR can indeed reach a remarkable level when the physical parameters are sufficiently large. 
On the other hand, by mainly referring to Fig.\ref{plotomegal2n0}, we want to carefully select another combination of $\{r_0, \mathcal{W}\}$ and extend our calculation of $\omega$ to cover the $n=1, 2$ modes. Our purpose is to see the different tendencies among the different modes with a varying $\mathcal{A}$. At the same time, for this part, we will allow the deviation from GR to be sufficiently large so that the trajectory of $\omega$ in the complex plane can be clear enough (the behavior of $\omega$ for the small-deviation cases can be found in, e.g., some panels of Fig.\ref{plotomegal2n0}). Therefore, we choose $\{r_0/M, \mathcal{W}\}=\{12, 2\}$ for the calculations of $\omega$ and put the results in Table \ref{tableomegal2n012}, where we have $\mathcal{A} \in [0.001, 0.055]$ and we exhibit the dimensionless quantity $M\omega$ up to six decimal places.

In addition, a companion figure for this table is shown in Fig.\ref{plottrajectoryl2n012}, exhibiting the trajectories of the three modes. Here, we show the behavior of $\omega$ for $\mathcal{A} \in [0, 0.055]$, where $\mathcal{A}=0$ corresponds to the GR limit (in GR, we have $M\omega_{20} \approx 0.483644-0.096759i$, $M\omega_{21} \approx 0.463851-0.295604i$, and $M\omega_{22} \approx 0.430544-0.508558i$ \cite{Matyjasek:2017psv}). Combining the observations from Table \ref{tableomegal2n012} and Fig.\ref{plottrajectoryl2n012} yields a wealth of useful physical information. First of all, we notice that the changes in the $n=1$ and $n=2$ modes are more sensitive to $\mathcal{A}$ compared to the $n=0$ one, which has already been noticed in the literature \cite{Yang:2024vor,Gong:2023ghh}. They can quickly and dramatically deviate from their GR counterparts in the complex plane. Secondly, it is quite clear that, for $\mathcal{A}$ in a certain regime, the $n=1$ mode can overtake the $n=0$ mode and become the one with the longest lifetime. Evidence for this is that, in Fig.\ref{plottrajectoryl2n012}, the red curve, which marks the mode (among the three) with the longest lifetime and the smallest damping rate, switches from the curve for $n=0$ to the $n=1$ one when $\mathcal{A}$ becomes large enough (an additional confirmation of this is shown in Fig.\ref{plottrajectoryl2n012B}). This kind of overtaking effect was noticed by some authors previously \cite{Cheung:2021bol,Destounis:2023ruj,Boyanov:2024fgc,Destounis:2023nmb}. 
At least for the modes considered here, the results show that the change in $\operatorname{Im}(\omega)$ is more sensitive than that in $\operatorname{Re}(\omega)$ with increasing $\mathcal{A}$. Such a tendency is consistent with the corresponding results for case (i) \cite{Bojowald:2024lhr}. 
At the same time, we would like to emphasize that no exceptional point is observed for the parameter choices considered here, as the tracked QNM frequencies do not coalesce in the complex plane. Nevertheless, the non-Hermitian nature of the QNM problem suggests that exceptional points may (sometimes easily) be found by appropriately tuning the parameters $\{r_0,\mathcal{A},\mathcal{W}\}$ \cite{Yang:2025dbn,PanossoMacedo:2025xnf}. A systematic search for such points lies beyond the scope of this work and is left for future investigation.
\subsection{Time-domain analysis}
\label{subsecB}

We now extend our investigation of cLQBH QNMs to the time domain by evolving the master equation~\eqref{master1} using the method of lines~\cite{Moin2010book,Chen:2026fad}. We combine fourth-order spatial finite difference method (FDM) with classical fourth-order Runge-Kutta (RK4) time integration~\cite{Ikeda:2021uvc}, and apply Sommerfeld-type boundary treatment at the boundary grid points~\cite{Moin2010book}. We refer to this scheme as the FR4S method\footnote{Preliminary tests with a second-order finite-difference scheme reproduce the main waveform features with only small differences from the FR4S results for selected parameter choices, suggesting that a simpler second-order scheme may suffice when the required accuracy is modest. Nevertheless, we adopt the more elaborate FR4S scheme, combining fourth-order spatial differencing with RK4 time integration, to reduce discretization errors and improve the accuracy of the quantitative waveform analysis below~\cite{Pazos-Avalos:2004uyd,Ikeda:2021uvc}.}.
Introducing $\Pi(t,x)\equiv\partial_t\Psi(t,x)$, we first rewrite the master equation as two coupled partial differential equations (PDEs),
\bqn
\partial_t\Psi&=&\Pi,\nb\\
\partial_t\Pi&=&\partial_x^2\Psi-V_{\mathrm{eff}}(r(x))\Psi.
\label{partialPi}
\eqn

To discretize these equations, we choose a spatial interval $[x_{\min},x_{\max}]$ and an evolution time interval $[0,t_{\max}]$, with a finite inner boundary $x_{\min}<0$ in the near-MH region and an outer boundary $x_{\max}$ in the far exterior beyond the bump. For each calculation, $t_{\max}$ is chosen sufficiently large to follow the late-time behavior of the wave function $\Psi$. We use a uniform rectangular lattice in the $(t,x)$ plane, as illustrated in \cite{Linge2017book},
\bqn
x_j&=&x_{\min}+j\,\delta x,\qquad j=0,\ldots,N_x,\nb\\
t_k&=&k\,\delta t,\qquad k=0,\ldots,N_t,
\label{tdgrid}
\eqn
where $\delta x=(x_{\max}-x_{\min})/N_x$ and $\delta t=t_{\max}/N_t$ are the spatial and time step sizes, respectively. Thus, $N_x$ and $N_t$ count the numbers of spatial and time intervals, respectively. Their values are chosen to cover the required domain and duration at the desired resolution. The indices $j$ and $k$ label the spatial grid points and time levels, respectively, which avoids confusion with the QNM overtone index $n$. At fixed $x_j$, we write $\Psi_j(t)=\Psi(t,x_j)$ and $\Pi_j(t)=\Pi(t,x_j)$ for the time-dependent fields, and $\Psi_j^k=\Psi_j(t_k)$ and $\Pi_j^k=\Pi_j(t_k)$ for their values at $t_k$. Along each grid line at fixed $x=x_j$, the potential has the time-independent value $V_j=V_{\mathrm{eff}}(r(x_j))$, where $r(x)$ is obtained numerically from Eq.~\eqref{rstar}, with a specific example shown in Fig.~\ref{plotrstar}.

With the lattice specified, we initialize both fields at all spatial grid points on the $k=0$ time level using a Gaussian pulse with vanishing initial time derivative~\cite{Chen:2026fad,Yang:2024vor,Papadopoulos:1998dk},
\bqn
\Psi_j^0&=&\exp\!\left[-\frac{(x_j-x_c)^2}{M^2}\right],\nb\\
\Pi_j^0&=&0.
\label{PsiPiinitial}
\eqn
Here $x_c$ denotes the pulse center. We adopt $x_c/M=5$ as our default choice in this work, which places the pulse center to the left of the bump in the tortoise coordinate, $x_c<x(r_0)$, for the parameter choices considered. To facilitate a direct comparison between the cLQBH and Schwarzschild evolutions of $\Psi$, we use the same initial data and observation prescription in both cases, thereby avoiding discrepancies arising from differences in these settings rather than from the underlying geometries \cite{Silva:2026jih}.

We next use fourth-order spatial FDM to discretize the spatial derivative in Eq.~\eqref{partialPi}. At the interior points, $2\leq j\leq N_x-2$, the centered stencil gives~\cite{Pazos-Avalos:2004uyd,LeVeque2007book}
\bqn
\frac{d\Psi_j}{dt}&=&\Pi_j,\nb\\
\frac{d\Pi_j}{dt}&=&\frac{-\Psi_{j+2}+16\Psi_{j+1}-30\Psi_j+16\Psi_{j-1}-\Psi_{j-2}}{12\,\delta x^2}\nb\\
&& -V_j\Psi_j.
\label{tdstencil}
\eqn
The spatial truncation error is $\mathcal{O}(\delta x^4)$. This discretization produces a system of ODEs in time. Both here and in the boundary equations below, the time derivatives are obtained by evaluating the right-hand sides of \eqref{tdstencil} through arithmetic operations on the field values at the current time level or RK4 stage.

The centered stencil is unavailable at the four outermost grid points, $x=x_0,x_1,x_{N_x-1},x_{N_x}$. To close their evolution equations, we impose the Sommerfeld relations $\partial_t\Psi=\partial_x\Psi$ at the left boundary and $\partial_t\Psi=-\partial_x\Psi$ at the right one~\cite{Moin2010book}. These relations approximate purely outgoing propagation at the finite computational boundaries, thereby reducing spurious reflections. Differentiating them with respect to $t$ yields the same relations for $\Pi$. We then apply fourth-order one-sided FDM to the spatial derivatives in these Sommerfeld relations, using neighboring points toward the domain interior~\cite{LeVeque2007book}. Writing $\mathbf U_j(t)=(\Psi_j(t),\Pi_j(t))^{\mathrm T}$, the resulting boundary evolution equations are
\bqn
\frac{d\mathbf U_j}{dt}&=&\frac{-25\mathbf U_j+48\mathbf U_{j+1}-36\mathbf U_{j+2}}{12\,\delta x}\nb\\
&&\quad+\frac{16\mathbf U_{j+3}-3\mathbf U_{j+4}}{12\,\delta x},\quad j=0,1,\nb\\[3pt]
\frac{d\mathbf U_j}{dt}&=&-\frac{25\mathbf U_j-48\mathbf U_{j-1}+36\mathbf U_{j-2}}{12\,\delta x}\nb\\
&&\quad+\frac{16\mathbf U_{j-3}-3\mathbf U_{j-4}}{12\,\delta x},\quad j=N_x-1,N_x, \nb\\
\label{tdboundary}
\eqn

Together, Eqs.~\eqref{tdstencil} and~\eqref{tdboundary} determine the time derivatives of both fields at every spatial grid point. At a given time level $k$, we substitute the known numerical values $\Psi_j^k$ and $\Pi_j^k$ for $\Psi_j$ and $\Pi_j$ into the corresponding right-hand sides and evaluate the time derivatives by arithmetic operations. We collect the pairs $\mathbf U_j^k=(\Psi_j^k,\Pi_j^k)^{\mathrm T}$ for $j=0,\ldots,N_x$ vertically into a column vector $\mathbf U^k$, which therefore contains both fields over the entire spatial grid at this time level. We denote the numerical operation that evaluates these right-hand sides by $\mathbf F$. For any supplied field-value vector $\mathbf U$, the output $\mathbf F(\mathbf U)$ contains the corresponding time derivatives in the same order, calculated using fourth-order FDM at the interior points and Sommerfeld relations combined with fourth-order FDM at the four outermost points. This operation is applied to the current grid values and to the intermediate values constructed during RK4. Starting from the known state $\mathbf U^k$, classical RK4 then advances both fields at all spatial grid points together to the next time level through~\cite{Moin2010book,LeVeque2007book},
\bqn
\mathbf K_1&=&\mathbf F(\mathbf U^k),\nb\\
\mathbf K_2&=&\mathbf F\!\left(\mathbf U^k+\frac{\delta t}{2}\mathbf K_1\right),\nb\\
\mathbf K_3&=&\mathbf F\!\left(\mathbf U^k+\frac{\delta t}{2}\mathbf K_2\right),\nb\\
\mathbf K_4&=&\mathbf F\!\left(\mathbf U^k+\delta t\,\mathbf K_3\right),\nb\\
\mathbf U^{k+1}&=&\mathbf U^k+\frac{\delta t}{6}\bigl(\mathbf K_1+2\mathbf K_2+2\mathbf K_3+\mathbf K_4\bigr).~~~~
\label{tdrk}
\eqn
Here $\mathbf K_1$ gives the current-state rates, while $\mathbf K_2$ and $\mathbf K_3$ use two estimates at the same half-step time and $\mathbf K_4$ uses a full-step estimate, all constructed from $\mathbf U^k$. Each evaluation of $\mathbf F$ uses one unchanged stage state in both the interior and boundary formulas. The final RK4 combination supplies both fields on the complete next time level, including the boundary points. Starting from $\mathbf U^0$, we repeat this procedure for $k=0,\ldots,N_t-1$, recalculating all four stage-rate vectors at each step. In the end, the time-domain response will be illustrated by plots of $\ln|\Psi(t,x_{\mathrm{obs}})|$, where $x_{\mathrm{obs}}<x_{\max}$ stands for the location of the observer. 
For many of the calculations below, $x_{\mathrm{obs}}/M=150$~\cite{Yang:2024vor} provides a convenient reference location, and as long as the observer stays sufficiently far outside the bump, our major conclusions are insensitive to the precise choice of $x_{\mathrm{obs}}$.

Finally, for the nonnegative potentials considered here, the time step in the FR4S method is chosen to satisfy the following Courant-type condition for numerical stability of the interior discretization~\cite{Zhang:2023kzs,Moin2010book,LeVeque2007book},
\bqn
\frac{\delta t}{\delta x}&<&\sqrt{\frac{8}{16/3+\delta x^2V_{\max}}},\nb\\
V_{\max}&=&\max_j V_j.
\label{tdcourant}
\eqn
A typical starting choice is $\delta x/M=0.1$ and $\delta t/M=0.05$, giving $\delta t/\delta x=1/2$. This choice satisfies Eq.~\eqref{tdcourant} with a safety margin in the benchmark cases and the extended parameter samples, and the condition is rechecked whenever the parameters or grid spacing change. Beyond this step-size restriction, the spatial and temporal resolutions and the boundary locations are calibrated through convergence tests and independent time-domain comparisons during the calculations. In particular, convergence can be checked by halving both $\delta x$ and $\delta t$ and comparing the resulting waveforms. These checks assess the stability and accuracy of the complete FR4S evolution, including the Sommerfeld boundary treatment, over the waveform interval of interest and also help determine the required evolution time $t_{\max}$.

We now proceed to the time-domain analysis using the FR4S method. Specifically, we shall plot $\ln|\Psi(t,x_{\mathrm{obs}})|$ [cf.~\eqref{master1}] and seek to extract physically useful information from these waveforms. Representative choices of $\{r_0,\mathcal{A},\mathcal{W}\}$ will be considered to illustrate the physical features and trends, focusing mainly on $l=2$. As a first example, the parameters in the last row of Table~\ref{tableomegal2n012} are chosen to allow a direct comparison with the corresponding frequency-domain results. This provides an initial link between the time-domain and frequency-domain analyses, whose complementary information can help reveal the underlying physics and yield a more complete account of the cLQBH response. For these calculations, we adopt units with $c=G=M=1$, which fixes the units of length, mass, and time in terms of the BH mass. As discussed in the previous subsection, this practical simplification leaves the dimensionless numerical results unchanged while making both the presentation and the numerical implementation more straightforward.

\begin{figure}[h]
\includegraphics[width=0.95\linewidth]{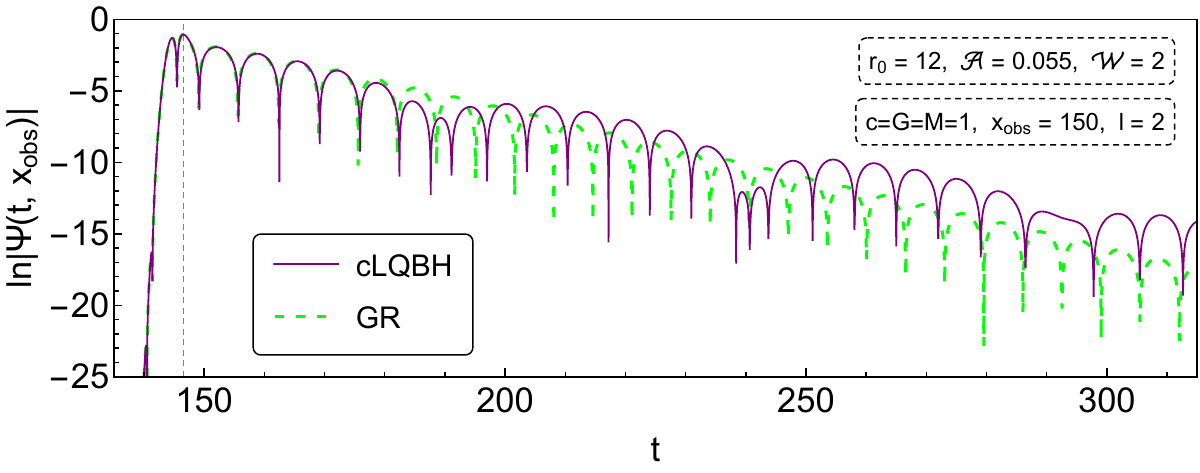}
\caption{The temporal evolution of the wave function $\Psi$ for the cLQBH [case (ii)] at the location $x=x_{\mathrm{obs}}$, for which we set $l=2$ and $\{r_0, \mathcal{A}, \mathcal{W}\}=\{12, 0.055, 2\}$. The GR counterpart is included for comparison. For convenience of presentation and numerical setup, we adopt units with $c=G=M=1$, which involves no loss of generality. The vertical dashed gray line marks $t=t_{\mathrm{peak}}$, at which $|\Psi_{\mathrm{GR}}(t, x_{\mathrm{obs}})|$ reaches its maximum \cite{Berti:2007fi}.
}
\label{plotFR4S1}
\end{figure}

The temporal evolution of the wave function $\Psi$ for the cLQBH [cf.~\eqref{master1}] is exhibited in Fig.~\ref{plotFR4S1} through $\ln|\Psi(t,x_{\mathrm{obs}})|$. Its GR counterpart is included in the same figure for comparison. As mentioned above, we adopt units with $c=G=M=1$ when solving \eqref{master1} and producing this plot (so that quantities like $t$ appear to be dimensionless). This choice involves no loss of generality and simplifies both the presentation and the numerical setup in the computational codes. For Fig.~\ref{plotFR4S1}, we set $l=2$ and $\{r_0,\mathcal{A},\mathcal{W}\}=\{12,0.055,2\}$, corresponding to the last row of Table~\ref{tableomegal2n012}. The observer is located at $x=x_{\mathrm{obs}}=150$~\cite{Yang:2024vor}.

By examining Fig.~\ref{plotFR4S1}, we see that the two time-domain waveforms $\Psi(t,x_{\mathrm{obs}})$~\cite{Berti:2022xfj}, one for the cLQBH and the other for GR, almost coincide at early times, consistent with previous studies of localized potential perturbations~\cite{Yang:2024vor}. Therefore, as shown in the figure, $t=t_{\mathrm{peak}}$ roughly marks the common maximum of the two curves. In contrast, the late-time cLQBH waveform $\Psi_{\mathrm{cLQBH}}$ exhibits a more intricate structure, with oscillation and decay patterns that differ from those of its GR counterpart $\Psi_{\mathrm{GR}}$. Here, these subscripts are introduced temporarily to distinguish the two backgrounds. More importantly, these differences are consistent with the shifts in the complex QNM frequencies $\omega$ found in the frequency-domain analysis, whose real and imaginary parts determine the oscillation periods and damping rates of the individual modes, respectively.

By adopting the so-called matrix pencil method (MPM)~\cite{HuaSarkar1990,Almunif2020,Berti:2007dg}, we can further extract dominant QNM frequencies from the signed waveform data underlying Fig.~\ref{plotFR4S1}.
\footnote{The main extraction settings are the fitting window $[t_a,t_b]$, the sampling time step $\delta t_s$, the pencil parameter $\mathfrak l$~\cite{Berti:2007dg}, and the retained rank $\mathfrak p$ in the singular value decomposition (SVD), which sets the number of exponential terms before screening. For this example, $[t_a,t_b]=[250,310]$, $\delta t_s=4\delta t=0.2$ [cf. Eq.~\eqref{tdgrid}], and $\mathfrak p=10$. The sampling time step and retained rank were selected \textit{a posteriori} by requiring the extracted frequencies to remain stable under changes of these settings. The pencil parameter is $\mathfrak l=\lfloor(N_s+1)/3\rfloor=100$~\cite{Berti:2007dg} for the current case, where $N_s=(t_b-t_a)/\delta t_s$ counts the sampling steps, $N_s+1$ is the number of samples, and $\lfloor\cdot\rfloor$ denotes the integer floor. The late-time fitting window can be adjusted case by case, and its suitability is assessed through the fitting residuals and the stability of the extracted frequencies~\cite{Dorband:2006gg}.}
For ease of reference, we denote the candidate complex frequencies by $\hat\omega_{lk}$. The index $k$ distinguishes their labeling from the frequency-domain branch index $n$, so no correspondence between them is assumed before comparison. After obtaining the candidate frequencies, we vary the fitting settings to screen out spurious roots and retain stable candidates, using a consistency check analogous to those employed in the matrix method~\cite{Lin:2019mmf} and the CFM~\cite{Daghigh:2022uws}. Following this screening, the two dominant modes for Fig.~\ref{plotFR4S1} are found to be $\hat\omega_{20}\simeq0.399526-0.071233i$ and $\hat\omega_{21}\simeq0.519903-0.077257i$. A much weaker component consistent with the frequency-domain $n=2$ mode is also recovered, but its frequency is not quoted because it is extracted with lower precision.
Here the retained modes with positive real parts are listed in order of increasing $|\operatorname{Im}\hat\omega_{lk}|$, so a smaller $k$ corresponds to a longer lifetime, with $k=0$ labeling the longest-lived mode~\cite{Berti:2009kk}. Comparison with Table~\ref{tableomegal2n012} identifies $\hat\omega_{20}$ with $\omega_{21}$ and $\hat\omega_{21}$ with $\omega_{20}$, and the differences in both the real and imaginary parts are smaller than $10^{-5}$. Whenever small differences arise, the frequency-domain values serve as the reference because they have undergone systematic convergence tests and independent cross-checks. Such differences can result from the finite fitting window and non-QNM contributions in waveform-based extraction, particularly for weak components~\cite{Berti:2007dg,Zhu:2023mzv}. The prominence of the longer-lived modes is not surprising, since the late-time fitting window favors the less damped excited QNMs, whose contributions decay more slowly during ringdown while the more strongly damped components fade. In particular, the continued $n=1$ branch outlives the $n=0$ branch, providing further time-domain evidence of the overtaking already identified in the frequency-domain analysis~\cite{Cheung:2021bol}.


As a complementary check of the frequencies identified among the candidates $\hat\omega_{lk}$ against the waveform underlying Fig.~\ref{plotFR4S1}, we reconstruct the waveform using the modes retained after numerical screening \cite{Borah:2026sja}. The initial MPM model contains $\mathfrak p$ exponential components. After screening out spurious candidates by varying the fitting settings, each retained positive-real-frequency mode $\hat\omega_{lk}$ is considered together with its companion $-\hat\omega_{lk}^{*}$. The positive-real-frequency members of these pairs are numbered $k=0,\ldots,\mathfrak q$ in order of increasing $|\operatorname{Im}\hat\omega_{lk}|$, in accordance with the convention above. Thus $\mathfrak q$ is the largest index in the retained set, which contains $\mathfrak q+1$ mode pairs, so that $2(\mathfrak q+1)\leq\mathfrak p$. For the oscillatory modes considered here, the general form of the reconstructed real-valued waveform is~\cite{Leaver:1986gd}
\bqn
\Psi_{\mathrm{rec}}(t,x_{\mathrm{obs}})
&=&\sum_{k=0}^{\mathfrak q}\left[C_{lk}e^{-i\hat\omega_{lk}(t-t_a)}
+C_{lk}^{*}e^{+i\hat\omega_{lk}^{*}(t-t_a)}\right]\nb\\
&=&2\operatorname{Re}\left[
\sum_{k=0}^{\mathfrak q}C_{lk}e^{-i\hat\omega_{lk}(t-t_a)}
\right],
\label{Psirec}
\eqn
where the superscript $*$ denotes complex conjugation. Each complex coefficient $C_{lk}$ determines the amplitude and phase of the corresponding mode at $t=t_a$, the beginning of the fitting window~\cite{Berti:2007dg}. The retained modes need not have comparable extraction accuracy or amplitudes. In practice, the reconstruction may therefore use only those modes determined with sufficient accuracy and carrying appreciable contributions, without necessarily including every term in Eq.~\eqref{Psirec}. In the late-time fitting window considered here, the less damped modes also have larger relative amplitudes and are more accurately extracted, so these considerations favor the same dominant modes.

To determine these coefficients, we fit the reconstructed waveform in Eq.~\eqref{Psirec} to the original FR4S data at the sampling times $t=t_a+j\delta t_s$. For this purpose, the sampled values and the unknown coefficients are collected into the column vectors
\bqn
\mathbf y&\equiv&(y_0,\ldots,y_{N_s})^{\mathrm T},\qquad
y_j\equiv\Psi(t_a+j\delta t_s,x_{\mathrm{obs}}),\nb\\
\mathbf C&\equiv&(C_{l0},C_{l0}^{*},\ldots,C_{l\mathfrak q},C_{l\mathfrak q}^{*})^{\mathrm T},
\label{mpmvectors}
\eqn
where $\mathrm T$ denotes transposition. At each sampling time, Eq.~\eqref{Psirec} expresses the reconstructed value as a linear combination of the coefficients in $\mathbf C$. The exponential factors multiplying these coefficients therefore form a numerical matrix $\mathbb{E}$, whose elements are
\bqn
(\mathbb{E})_{j+1,2k+1}&=&e^{-i\hat\omega_{lk}j\delta t_s},\nb\\
(\mathbb{E})_{j+1,2k+2}&=&e^{+i\hat\omega_{lk}^{*}j\delta t_s},
\label{mpmbasis}
\eqn
where $j=0,\ldots,N_s$ and $k=0,\ldots,\mathfrak q$. Consequently, $\mathbb{E}\mathbf C$ gives the reconstructed values at the sampling times, and the fitting relation becomes
\bqn
\mathbb{E}\mathbf C&\simeq&\mathbf y.
\label{mpmfit}
\eqn
The matrix $\mathbb{E}$ and the data vector $\mathbf y$ consist entirely of known numbers, obtained from the retained MPM frequencies and the original FR4S samples. With the frequencies $\hat\omega_{lk}$ held fixed, we determine the coefficients $C_{lk}$ by minimizing
\bqn
\sum_{j=0}^{N_s}
[(\mathbb{E}\mathbf{C})_{j+1}-y_j]^2.
\label{mpmsquares}
\eqn
Here $(\mathbb{E}\mathbf{C})_{j+1}$ is the reconstructed waveform value at $t=t_a+j\delta t_s$, and $y_j$ is the corresponding FR4S value defined in Eq.~\eqref{mpmvectors}. This is a linear least-squares fit~\cite{Berti:2007dg,Zhu:2023mzv}. Only the coefficients are fitted, and each $C_{lk}^{*}$ is constrained to be the complex conjugate of $C_{lk}$.
For plotting, as mentioned above, we may further retain only the terms with sufficiently accurate frequencies and appreciable amplitudes in Eq.~\eqref{Psirec}, keeping their fitted coefficients unchanged. This truncation gives an approximate reconstruction, without implying that the omitted modes are absent. Its accuracy is assessed by comparing the resulting waveform directly with the FR4S data.
We note that this coefficient fit, together with the other linear algebra operations in MPM, can be carried out using the built-in functions of, e.g., \textit{Mathematica}.

\begin{figure}[h]
\includegraphics[width=0.95\linewidth]{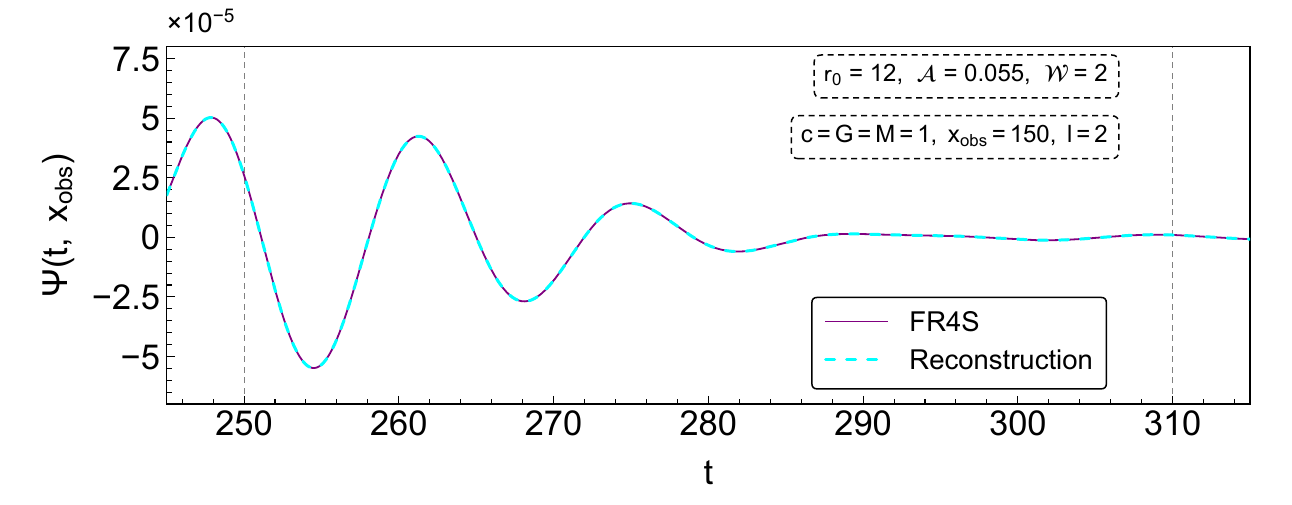}
\caption{A comparison between $\Psi(t, x_{\mathrm{obs}})$ (the solid purple curve, calculated from the FR4S method) and $\Psi_{\mathrm{rec}}(t, x_{\mathrm{obs}})$ [the dashed cyan curve, cf.~\eqref{Psirec}] over the time interval around the fitting window $[t_a, t_b]$, for the case shown in Fig.~\ref{plotFR4S1}. The vertical dashed line on the left marks $t_a$, while the one on the right marks $t_b$.
}
\label{plotMPM1}
\end{figure}

Substituting the resulting numerical coefficients into Eq.~\eqref{Psirec}, with the sum restricted to the included modes when appropriate, gives an explicit reconstructed waveform as a function of $t$. This waveform can be compared directly with the original FR4S waveform, without an additional normalization. Accordingly, for the case shown in Fig.~\ref{plotFR4S1}, we plot in Fig.~\ref{plotMPM1} the FR4S waveform $\Psi(t,x_{\mathrm{obs}})$ together with its reconstruction from Eq.~\eqref{Psirec}, where $t$ is expressed in dimensionless form through the units with $c=G=M=1$ adopted earlier. The two curves agree very well, even though only two modes, $\hat\omega_{20}$ and $\hat\omega_{21}$, are used for the reconstruction. In fact, a closer examination of the contributions of the different modes, measured by $|C_{2k}|/\max\{|C_{20}|,|C_{21}|,\ldots,|C_{2\mathfrak q}|\}$, shows that these two modes make comparable contributions to $\Psi_{\mathrm{rec}}$, while each remaining fitted component contributes less than $0.03\%$ of the largest one. This explains why these two modes are sufficient to provide an adequate truncated reconstruction over the interval shown.
On the other hand, the comparable contributions of these two modes suggest that their interference explains the more intricate oscillation and decay patterns in Fig.~\ref{plotFR4S1}, consistent with similar effects observed in other multi-mode ringdown waveforms~\cite{Dorband:2006gg}.

\begin{figure*}[t]
\centering
\begin{minipage}{0.48\textwidth}
\centering
\includegraphics[width=\linewidth]{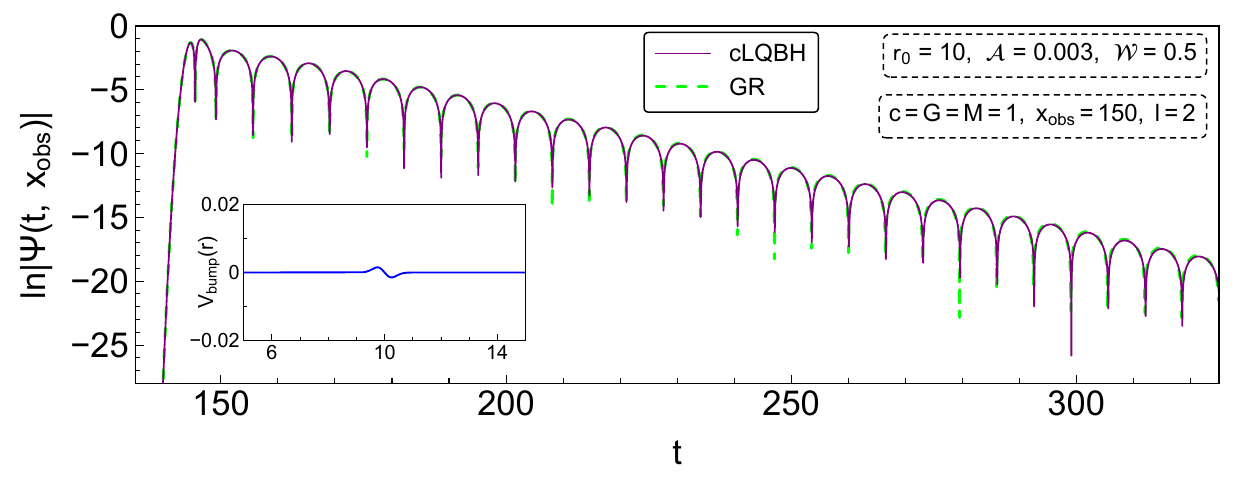}
\end{minipage}\hspace{0.02\textwidth}
\begin{minipage}{0.48\textwidth}
\centering
\includegraphics[width=\linewidth]{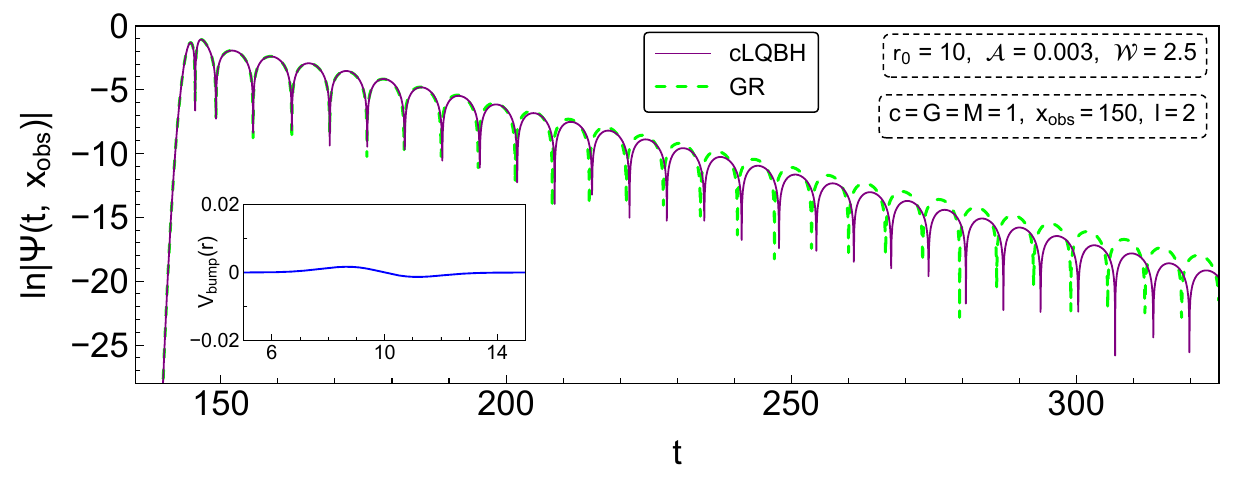}
\end{minipage}
\par\vspace{2mm}
\begin{minipage}{0.48\textwidth}
\centering
\includegraphics[width=\linewidth]{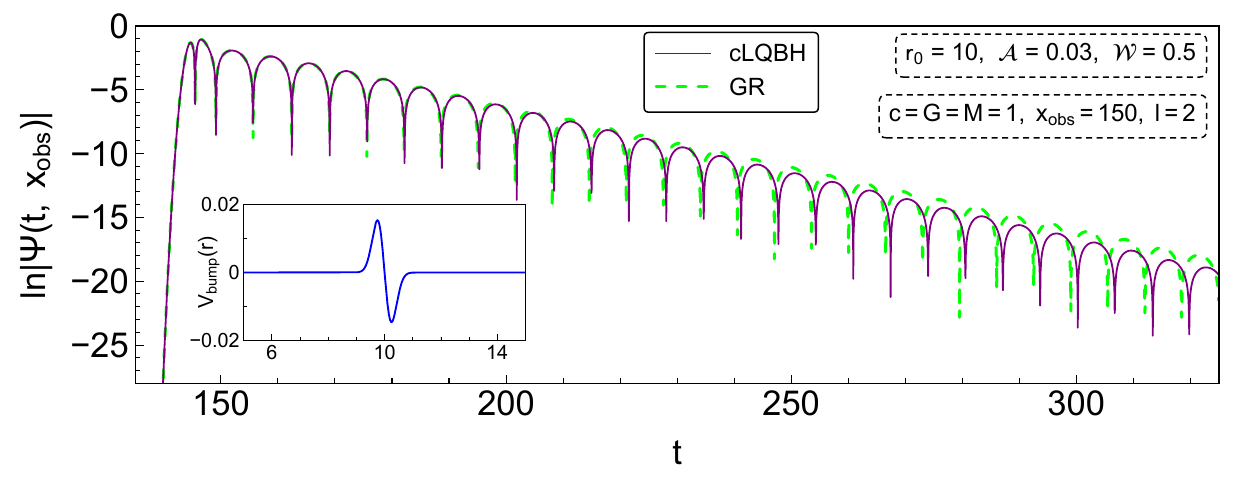}
\end{minipage}\hspace{0.02\textwidth}
\begin{minipage}{0.48\textwidth}
\centering
\includegraphics[width=\linewidth]{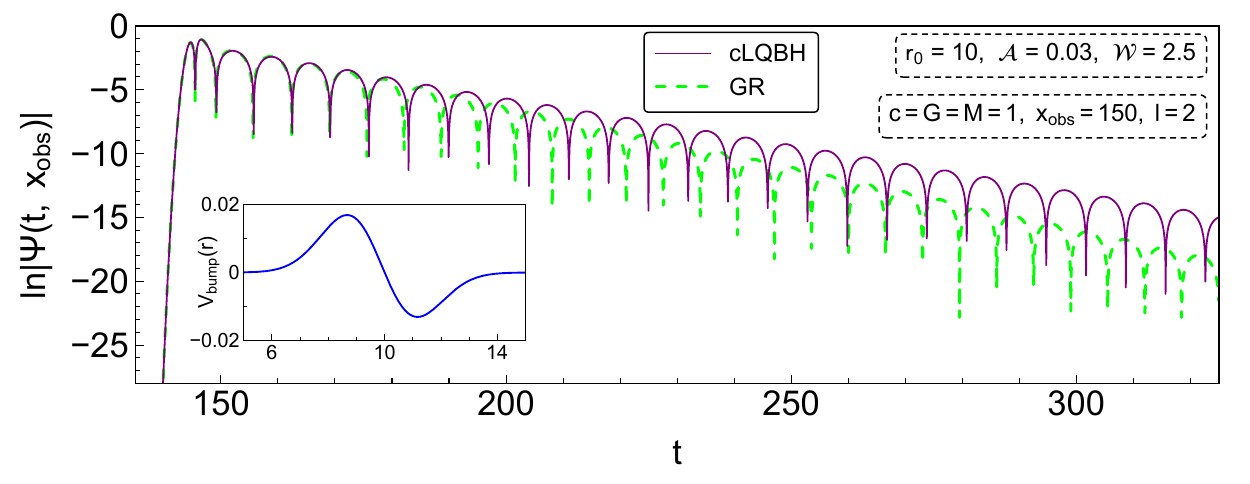}
\end{minipage}
\caption{Time-domain waveforms $\ln|\Psi(t,x_{\mathrm{obs}})|$ obtained using the FR4S method for four representative bump configurations, with $l=2$, $r_0=10$, and $x_{\mathrm{obs}}=150$ in units of $c=G=M=1$. The upper and lower rows correspond to $\mathcal{A}=0.003$ and $0.03$, respectively, while the left and right columns correspond to $\mathcal{W}=0.5$ and $2.5$. The purple solid and green dashed curves show the cLQBH and Schwarzschild results, respectively, obtained with the same initial data and observation prescription. Each inset shows the corresponding $V_{\mathrm{bump}}(r)$ as a blue solid curve, with $r$ along the horizontal axis.
}
\label{plotFR4S2}
\end{figure*}

\begin{figure*}[t]
\centering
\begin{minipage}{0.48\textwidth}
\centering
\includegraphics[width=\linewidth]{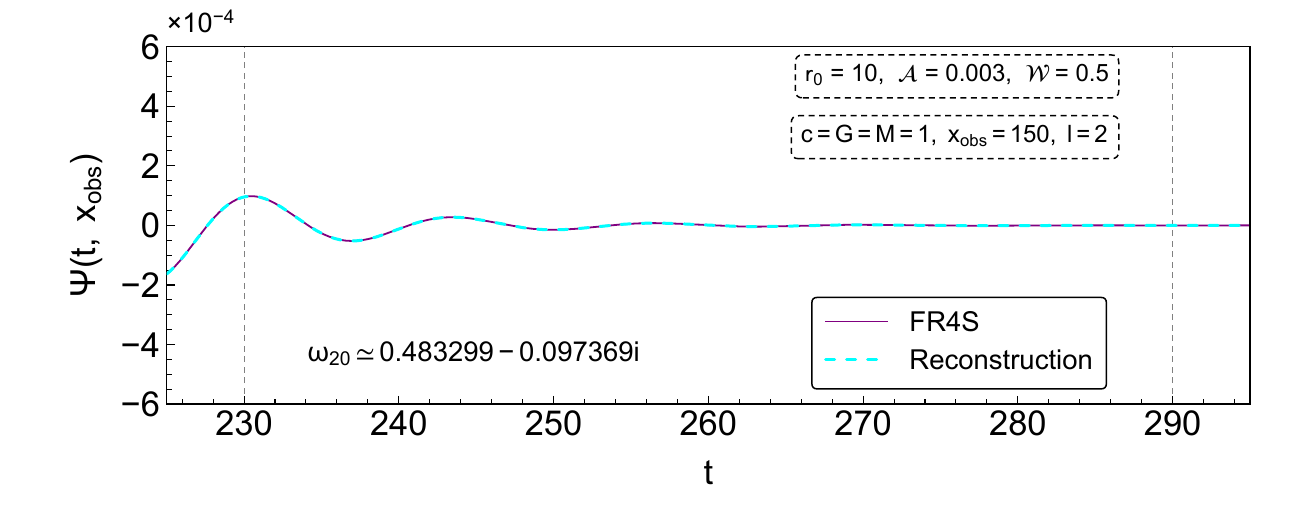}
\end{minipage}\hspace{0.02\textwidth}
\begin{minipage}{0.48\textwidth}
\centering
\includegraphics[width=\linewidth]{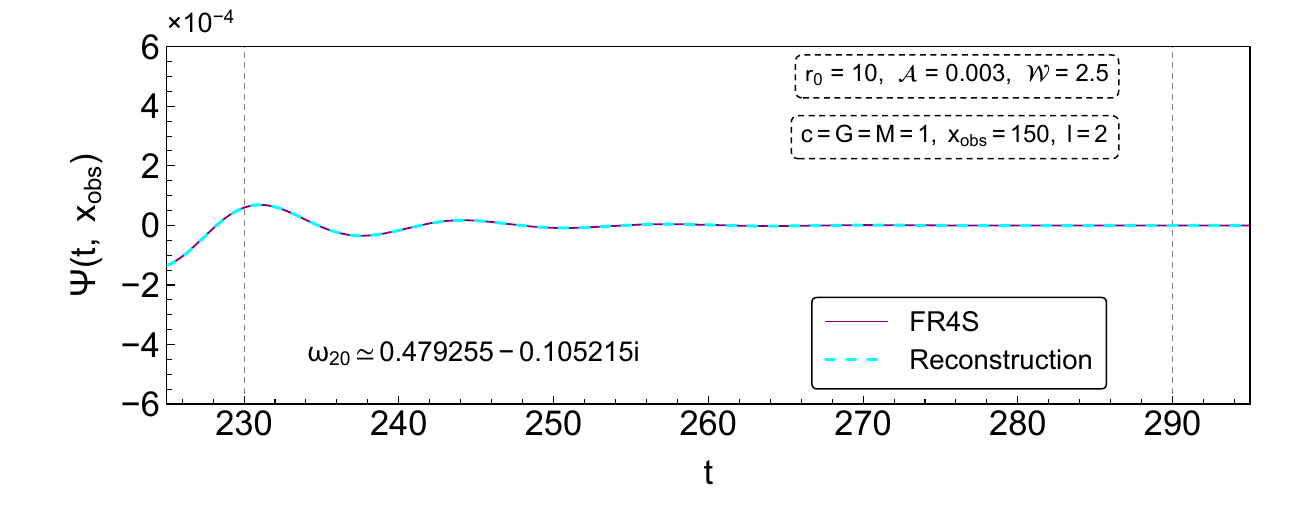}
\end{minipage}
\par\vspace{2mm}
\begin{minipage}{0.48\textwidth}
\centering
\includegraphics[width=\linewidth]{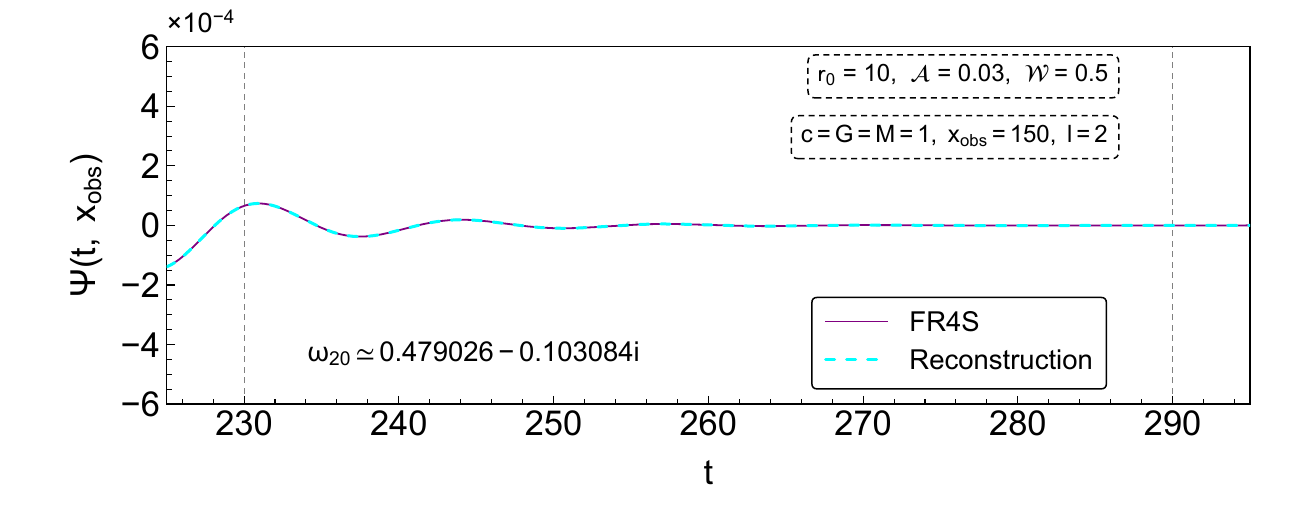}
\end{minipage}\hspace{0.02\textwidth}
\begin{minipage}{0.48\textwidth}
\centering
\includegraphics[width=\linewidth]{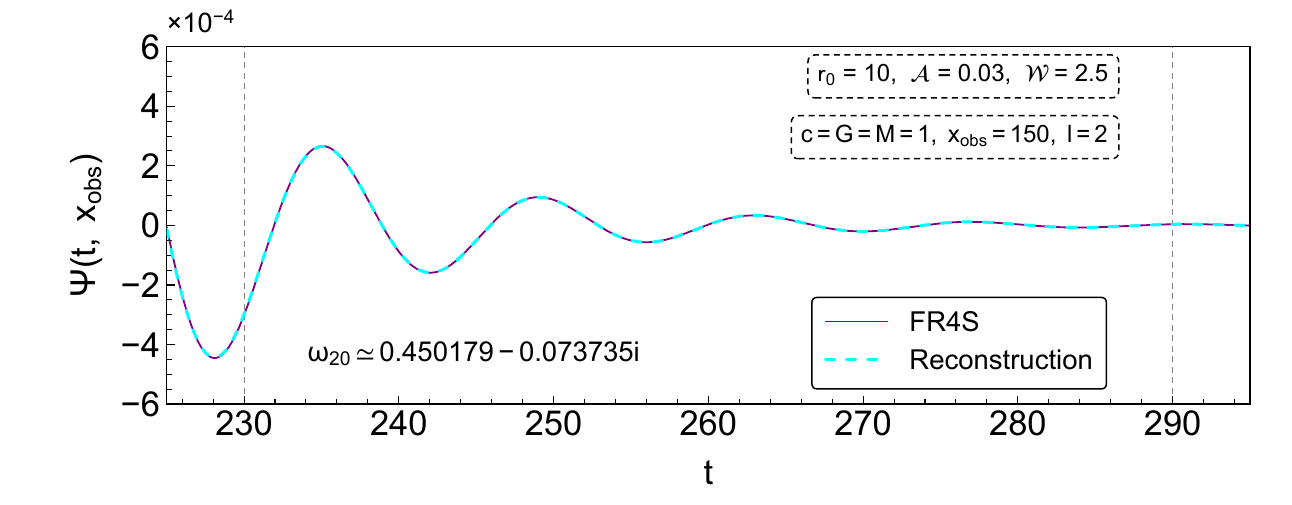}
\end{minipage}
\caption{A comparison between $\Psi(t,x_{\mathrm{obs}})$ calculated using the FR4S method and $\Psi_{\mathrm{rec}}(t,x_{\mathrm{obs}})$ reconstructed using Eq.~\eqref{Psirec}, over the time interval around the fitting window $[t_a,t_b]=[230,290]$, for the four cases shown in Fig.~\ref{plotFR4S2}. The panel arrangement and physical parameters are the same as in that figure. The purple solid and cyan dashed curves show the FR4S and reconstructed waveforms, respectively. Each reconstruction includes only the dominant mode, with the corresponding $\omega_{20}$ indicated in the panel. The vertical dashed line on the left marks $t_a$, while the one on the right marks $t_b$.}
\label{plotMPM2}
\end{figure*}

The preceding example illustrates a waveform in which two modes make comparable contributions, giving the late-time behavior of $\Psi$ a more intricate structure (cf.~Fig.~\ref{plotFR4S1}). We now examine how the bump amplitude and width affect the response at a fixed location, using the four configurations shown in Figs.~\ref{plotFR4S2} and~\ref{plotMPM2}. We choose $r_0=10$, an intermediate location between the nearer and farther bumps considered above, and compare four combinations of the amplitude $\mathcal{A}$ and the width $\mathcal{W}$. The four resulting bump profiles, shown in the insets of Fig.~\ref{plotFR4S2}, combine $\mathcal{A}=0.003$ and $0.03$ with $\mathcal{W}=0.5$ and $2.5$. The calculations retain $l=2$, $x_{\mathrm{obs}}=150$, the same initial data, and the units $c=G=M=1$. All four MPM extractions use the common fitting window $[t_a,t_b]=[230,290]$ and retain the previous settings $\delta t_s=0.2$, $\mathfrak p=10$, and $\mathfrak l=100$. Figure~\ref{plotMPM2} compares the corresponding FR4S and reconstructed waveforms over $t\in[225,295]$, with the vertical dashed lines marking the fitting-window boundaries.

As seen in Fig.~\ref{plotFR4S2}, the smallest and narrowest bump produces only a slight departure from GR, qualitatively consistent with the small frequency shifts at low amplitudes shown in Fig.~\ref{plotomegal2n0}. Increasing either the amplitude or the width reduces $\operatorname{Re}\omega_{20}$ across these four configurations, giving a longer oscillation period. 
The damping behavior is less regular. Three of the four configurations decay faster than in GR, while the fourth decays more slowly, as reflected by their oscillation envelopes lying below or above the GR envelope.
In particular, increasing the width enhances the damping at the smaller amplitude but reduces it at the larger amplitude. The largest departure occurs for the larger and broader bump, for which $\operatorname{Re}\omega_{20}$ and $|\operatorname{Im}\omega_{20}|$ are approximately $6.9\%$ and $23.8\%$ smaller than their GR counterparts, respectively. This again confirms the finding in Sec.~\ref{subsecA} that the change in $\operatorname{Im}\omega$ is more sensitive than that in $\operatorname{Re}\omega$. These results show the combined influence of the amplitude and width, without establishing that either alone controls the response. Fig.~\ref{plotMPM2} shows that all four late-time waveforms are accurately reconstructed using a single dominant mode, consistent with the usual dominance of the fundamental mode during ringdown~\cite{Berti:2009kk}, since the more strongly damped overtones can become much weaker at sufficiently late times~\cite{Giesler:2019uxc}. Frequency-domain continuation identifies this dominant mode with the continued $n=0$ branch, i.e., the $\omega_{20}$ mode, and the MPM and shooting results agree when rounded to six decimal places. Thus, a noticeable departure from GR does not necessarily involve competition between two comparable mode contributions, although more distant bumps can allow other branches to overtake the continued $n=0$ mode~\cite{Cheung:2021bol}. Moreover, the agreement in Fig.~\ref{plotMPM2} extends slightly beyond the fitting window without readjusting the coefficients. Together with the frequency-domain verification, this supports the physical interpretation of the reconstruction in terms of QNM ringing and motivates further investigation of the bump location and the timescale on which the waveform differences become appreciable.


\section{Time for deviation and the observational window}
\renewcommand{\theequation}{4.\arabic{equation}} \setcounter{equation}{0}
\label{seckey}

The waveform comparisons in Fig.~\ref{plotFR4S2} illustrate how the bump amplitude and width affect the time-domain response at a fixed $r_0$. We now turn to the bump location $r_0$, which changes the separation between the localized deformation (viz., the bump) and the main potential barrier and has received particular attention in studies of the bump effect in recent years~\cite{Cheung:2021bol,Berti:2022xfj,Yang:2024vor}. These studies show that moving a small bump outward to the positive-$x$ direction can substantially modify the QNM spectrum while leaving the early ringdown nearly unchanged (cf., Figs.\ref{plotFR4S1} and \ref{plotFR4S2}), motivating a complementary examination of the waveforms. In our cLQBH model, the spiral trajectory in Fig.~\ref{plotspiral} further illustrates the nontrivial spectral dependence on $r_0$ and raises the question of how this dependence manifests itself in the time-domain response. We therefore extend the preceding analysis by holding $\mathcal{A}$ and $\mathcal{W}$ fixed and varying $r_0$. 
By studying the bump effect in cLQBHs, we connect the general question of how localized potential deformations affect BH ringing with the dynamical consequences of quantum-corrected geometries. The dependence on $r_0$ provides a natural point of contact, allowing us to examine how the location of the correction shapes the time-domain response. Accordingly, the choices of $\{\mathcal{A},\mathcal{W}\}$ and the ranges of $r_0/M$ warrant careful consideration, both to remain within the physically admissible regime and to reveal waveform trends that can be clearly identified and interpreted. These considerations guide the following waveform comparisons using the FR4S method described in Sec.\ref{secQNM}.

\begin{figure}[t]
\centering
\includegraphics[width=0.95\linewidth]{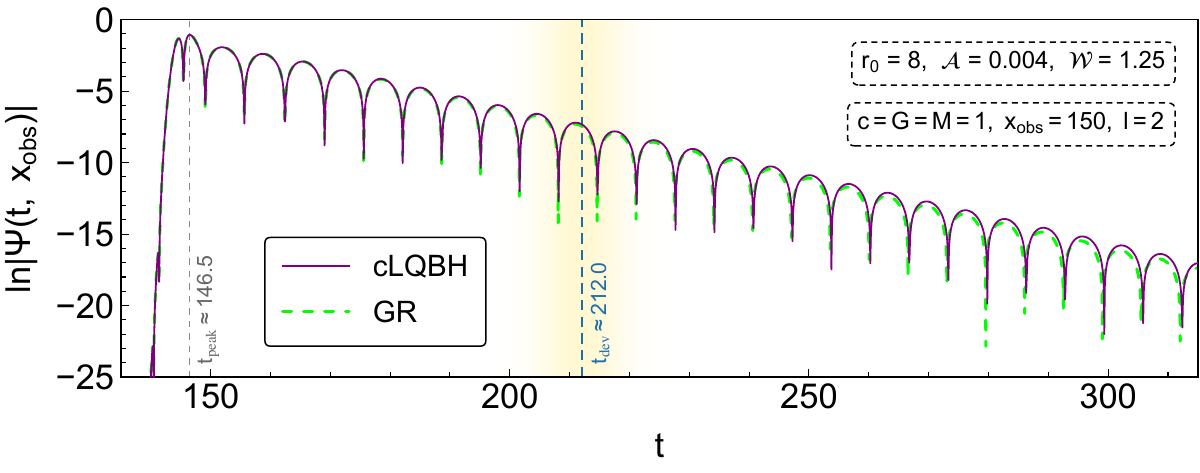}
\par\vspace{2mm}
\includegraphics[width=0.95\linewidth]{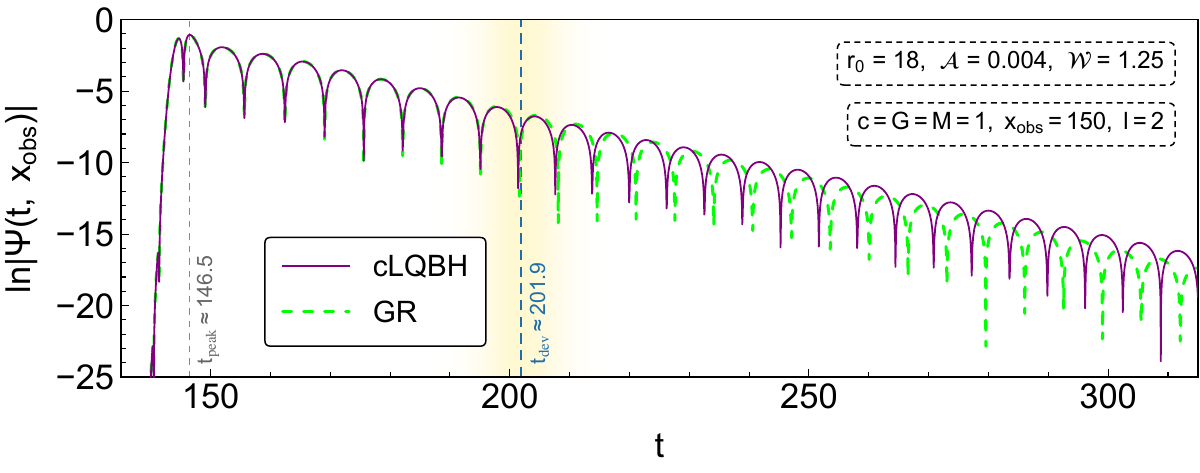}
\par\vspace{2mm}
\includegraphics[width=0.95\linewidth]{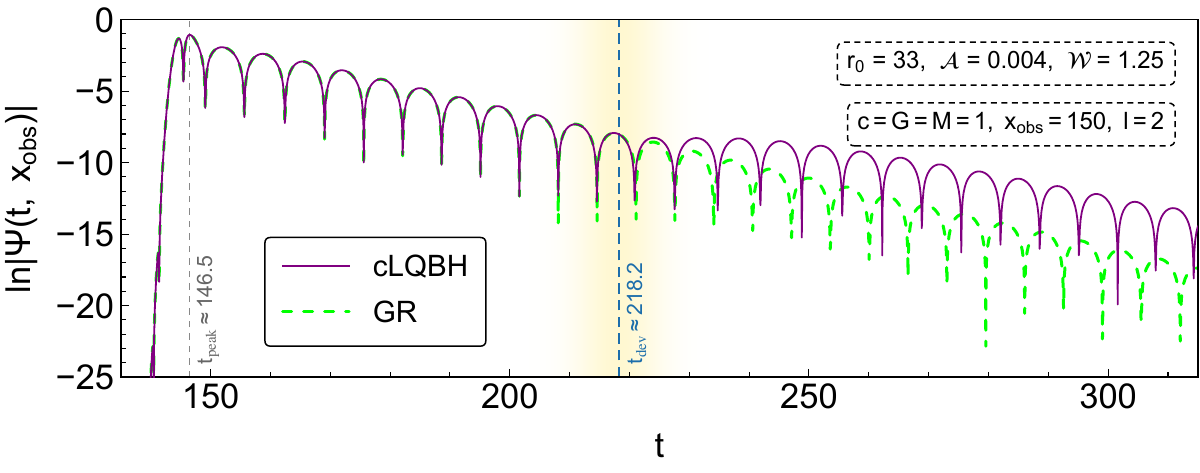}
\caption{
Time-domain waveforms $\ln|\Psi(t,x_{\mathrm{obs}})|$ for the cLQBH [case (ii)] with $r_0=8$ (top), $18$ (middle), and $33$ (bottom), obtained using the FR4S method.
We fix $l=2$, $\mathcal{A}=0.004$, $\mathcal{W}=1.25$, and $x_{\mathrm{obs}}=150$, and adopt units with $c=G=M=1$. The purple solid and green dashed curves show the cLQBH and Schwarzschild results, respectively. All panels use the same initial data and numerical settings. The vertical dashed gray line on the left marks the common $t_{\mathrm{peak}}$, at which $|\Psi_{\mathrm{GR}}|$ reaches its maximum while the vertical dashed steel-blue line in the central part marks $t_{\mathrm{dev}}$, for which we treated it as the beginning point for the two curves to start to deviate. The pale-yellow shading represents the time interval for doing the summation in \eqref{eqofDT} while determining the numerical value of $t_{\mathrm{dev}}$.
}
\label{plottdev1}
\end{figure}

To give a set of examples, we calculate $\Psi(t,x_{\mathrm{obs}})$ for three representative values of $r_0$, keeping $\{\mathcal{A},\mathcal{W}\}=\{0.004,1.25\}$ fixed. The resulting waveforms are shown in Fig.~\ref{plottdev1}, where we adopt $l=2$, $x_{\mathrm{obs}}=150$, and units with $c=G=M=1$, as before. These comparisons show that varying the bump location changes how the cLQBH waveform departs from its GR counterpart, complementing the nontrivial dependence of the QNM spectrum on $r_0$ illustrated in Fig.~\ref{plotspiral}.
Specially, at least for the late-time windows, we notice again that the departure from that of GR is more explicit with a relatively larger $r_0$.

More importantly, these waveforms draw attention to a quantity of particular interest, the approximate time at which the difference between the two responses becomes appreciable, which we denote by $t_{\mathrm{dev}}$. The onset of this appreciable departure can be estimated visually, and similar behavior is evident in previous waveform comparisons~\cite{Berti:2022xfj,Yang:2024vor}. These studies also relate the delayed response to the location of the potential perturbation through wave propagation and scattering. Building on these results, a systematic examination of $t_{\mathrm{dev}}$ in the present cLQBH model can further clarify how the bump location controls the emergence of waveform differences.

The examples in Fig.~\ref{plottdev1} suggest that $t_{\mathrm{dev}}$ need not vary monotonically with $r_0$, motivating a more detailed examination of the full dependence. We therefore undertake a semi-quantitative study of this behavior. Although the departure itself is a physical feature of the waveforms, assigning a reproducible time to its onset requires an explicit criterion. Such a criterion allows us to identify $t_{\mathrm{dev}}$ automatically and consistently across different parameter choices, while retaining its connection to visual inspection. For this purpose, we combine the normalized waveform distance used in \cite{Bassi2025} with Gaussian weighting~\cite{Smith1997book} to introduce the following local measure of waveform deviation,
\begin{widetext}
\begin{equation}
D_T(t_k)\simeq\left[\frac{\displaystyle\sum_{j=-J}^{J}e^{-\frac{(j\delta t)^2}{2T^2}}\left|\Psi(t_k+j\delta t,x_{\mathrm{obs}})-\Psi_{\mathrm{GR}}(t_k+j\delta t,x_{\mathrm{obs}})\right|^2}{\displaystyle\sum_{j=-J}^{J}e^{-\frac{(j\delta t)^2}{2T^2}}\left[\left|\Psi(t_k+j\delta t,x_{\mathrm{obs}})\right|^2+\left|\Psi_{\mathrm{GR}}(t_k+j\delta t,x_{\mathrm{obs}})\right|^2\right]}\right]^{1/2},
\label{eqofDT}
\end{equation}
\end{widetext}
where $t_k=k\delta t$, $j$ labels the time steps relative to $t_k$, and $T$ sets the width of the Gaussian weight.  This dimensionless quantity measures the local difference between the two waveforms around any given $t_k$, with the denominator providing a common normalization. For a nonzero denominator, this measure satisfies $0\leq D_T\leq\sqrt{2}$.  We evaluate the sums directly from the uniformly sampled FR4S data and examine $D_T(t_k)$ chronologically, starting at the common GR amplitude peak $t_{\mathrm{peak}}$ (indicated in Fig.\ref{plottdev1}). For the cases considered here, $D_T$ is initially below a prescribed threshold $\delta$. We locate the first sample satisfying $D_T(t_k)\geq\delta$, whose preceding sample therefore satisfies $D_T(t_{k-1})<\delta$, and determine the crossing time by a straightforward linear interpolation,
\begin{equation}
t_{\mathrm{dev}}\approx t_{k-1}+\frac{\delta-D_T(t_{k-1})}{D_T(t_k)-D_T(t_{k-1})}\,\delta t.
\label{eqoftdev}
\end{equation}
Thus, interpolation is applied only to the two adjacent diagnostic values that bracket the threshold, while the waveforms entering Eq.~\eqref{eqofDT} are used directly.

The corresponding continuous expression replaces the sums in Eq.~\eqref{eqofDT} with integrals over $[t_k-6T,t_k+6T]$. Accordingly, we set $J=6T/\delta t$, so that the discrete and continuous expressions cover the same interval. This truncation retains the Gaussian weight out to six standard deviations on either side of its center, beyond which the weight is negligible. Checks extending the interval to $\pm8T$ confirm a negligible change in the extracted deviation times for the tested cases. Within this interval, we choose $T=5M$, a width parameter comparable to half the reference GR ringing period, approximately $6.5M$. Gaussian widths on this half-period scale have also been considered in local oscillation analysis~\cite{McCormack2017}, providing a methodological precedent for this choice. With the sampling interval $\delta t=0.05M$ [cf., the context around \eqref{tdcourant}], we therefore have $J=600$.

We adopt the fixed threshold $\delta=0.075$, which is of the same order as the value $0.1$ used for a similarly normalized distance in electrical waveform comparisons~\cite{Bassi2025}. This provides a reference scale rather than a prescribed threshold for the present problem. We assess the suitability of our choices by examining the sensitivity of the extracted times to nearby values of $T$ and $\delta$, checking numerical resolution and the summation interval, and comparing the resulting markers with the waveforms themselves. These checks support the use of the adopted criterion for our semi-quantitative analysis, while the precise onset times remain dependent on the stated criterion. We keep the same $T$ and $\delta$ for all bump locations to ensure a consistent comparison.

\begin{figure}[t]
\centering
\includegraphics[width=0.95\linewidth]{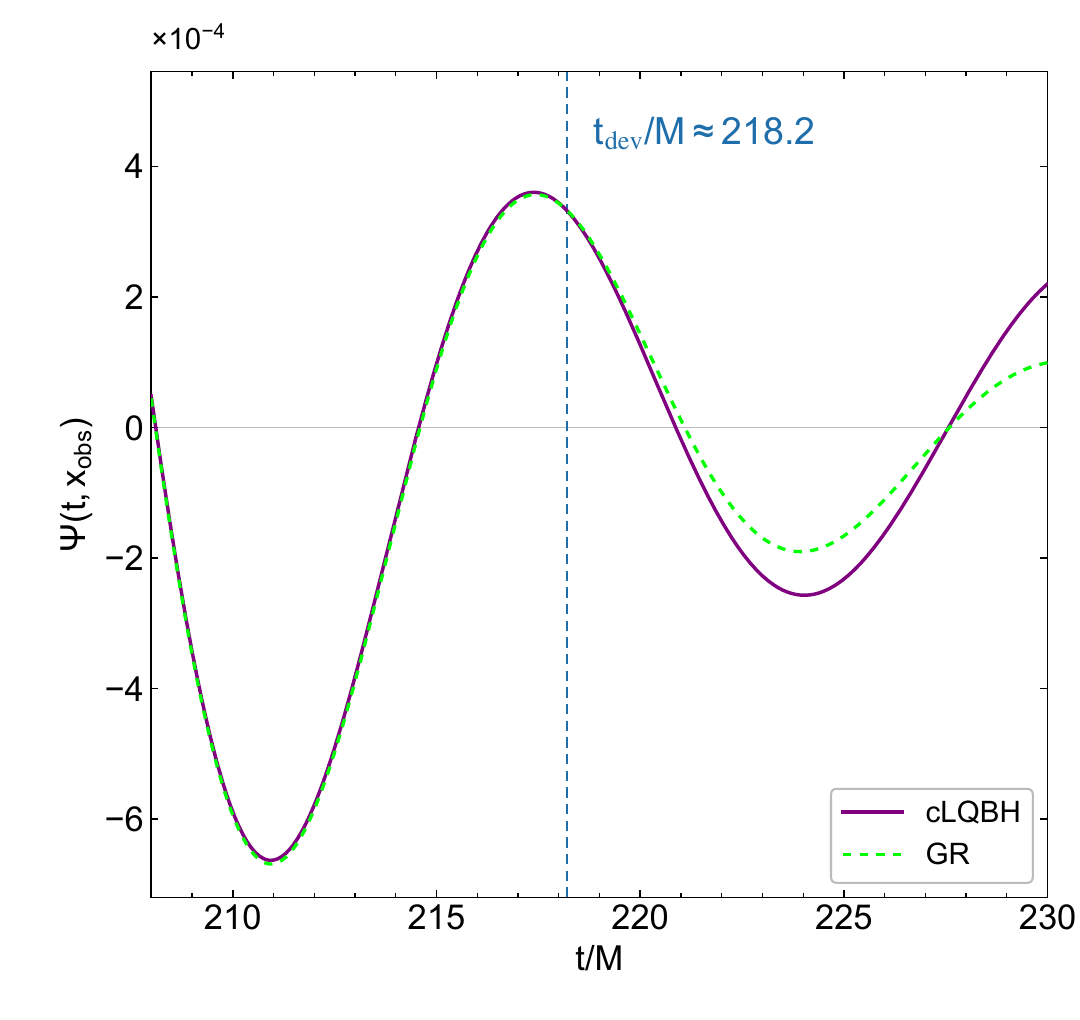}
\caption{Local comparison of the signed waveforms $\Psi(t,x_{\mathrm{obs}})$ for the $r_0/M=33$ case shown in the bottom panel of Fig.~\ref{plottdev1}, with $\mathcal{A}=0.004$, $\mathcal{W}=1.25$, $l=2$, and $x_{\mathrm{obs}}/M=150$. The purple solid and green dashed curves show the cLQBH and GR responses, respectively, using the same waveform data as in that figure. The steel-blue dashed line marks $t_{\mathrm{dev}}/M\approx218.2$, obtained from Eqs.~\eqref{eqofDT} and~\eqref{eqoftdev} with $T=5M$ and $\delta=0.075$. This enlarged view shows the development of a visible separation near the identified deviation time.}
\label{plottdev2}
\end{figure}

Applying this procedure to Fig.~\ref{plottdev1} gives the deviation times marked by the vertical steel-blue dashed lines, with their numerical values indicated alongside them. These examples again show that $t_{\mathrm{dev}}$ does not vary monotonically with $r_0$. To illustrate the time interval entering Eq.~\eqref{eqofDT}, the pale-yellow shading extends over $t_{\mathrm{dev}}\pm6T$, with its intensity proportional to the Gaussian weight. Its gradual fading shows how samples farther from the central time are assigned smaller weights. As a more direct visual cross check, Fig.~\ref{plottdev2} presents a local enlargement of the signed waveforms for the $r_0/M=33$ case in the bottom panel of Fig.~\ref{plottdev1}, with the axis ranges and aspect ratio adjusted appropriately to make their separation easier to inspect. The marker at $t_{\mathrm{dev}}/M\approx218.2$ lies near the transition from nearly overlapping waveforms to an increasingly visible separation. This agreement supports interpreting the extracted time as an approximate onset of appreciable deviation, consistent with the local averaging used in the definition of $t_{\mathrm{dev}}$.

Beyond their phenomenological significance, the results discussed above lead us to consider their observational implications. In this work, we aim not only to connect the cLQBH model with the bump effect on QNMs, but also to explore whether the resulting phenomena could have observable consequences. Existing studies have examined several complementary aspects of QNM observability, including whether the ringdown frequencies fall within the sensitive frequency bands of planned detectors \cite{Shi:2019hqa} and whether multiple modes can be detected and resolved for black-hole spectroscopy \cite{Berti:2005ys,Shi:2024ttu}, etc.

Within the present framework, there are two closely related ways to investigate this question. The first is to calculate the QNM frequencies $\omega$ and compare them with their GR counterparts through, for example, the fractional deviations defined in Eq.~\eqref{deltaomega}. Comparing these deviations with the measurement accuracies of current and planned detectors, including LVK, LISA, and TianQin, allows one to assess possible constraints on the parameters controlling the QNMs of a gravitational theory or model. Such constraints can be investigated using existing observations or forecasts in which GR is assumed to remain consistent with future data \cite{LIGOScientific:2025wao,Shi:2019hqa}. Our comparisons in Fig.~\ref{plotomegal2n0} and Table~\ref{tableomegaIV} provide an exploratory assessment in this direction. Related studies have also considered this approach using scalar QNMs \cite{Zhang:2024svj,Pani:2026yzi}. The second is to investigate whether the bump-induced changes in the QNMs and waveforms become sufficiently pronounced while the signal still retains an appreciable amplitude. In particular, even substantial changes in the QNM spectrum may leave the early ringdown nearly unchanged, with more apparent waveform differences emerging only later \cite{Berti:2022xfj,Yang:2024vor}. The observational relevance of the bump effect has already been discussed, with previous studies emphasizing that small corrections to the early ringdown and weak late-time signals can limit the observational accessibility of the modified QNM spectrum \cite{Berti:2022xfj,Laeuger:2025zgb}. These considerations motivate a more systematic examination of when waveform deviations emerge relative to signal decay in the present cLQBH model. Although closely connected to the first approach, this second approach explicitly considers whether the signal persists with sufficient amplitude for the deviations to become apparent. In this section, we pursue this direction by comparing the deviation time with the signal decay, thereby introducing an operational observation window for the scalar waveforms calculated above.

In practice, assessing the observational significance of BH ringing requires selecting a suitable time interval for the signal analysis \cite{Carullo:2018sfu,Bhagwat:2017tkm}. A natural and practical starting point is the amplitude peak, where the signal is strongest, so that the interval includes the strong part of the response while excluding the initially weak signal before its main arrival. The time of such a peak also serves as a reference time in gravitational ringdown studies, including analyses that incorporate overtones \cite{Berti:2007fi,Giesler:2019uxc,Isi:2019aib}. Accordingly, we adopt the time of the common GR reference peak, $t=t_{\mathrm{peak}}$, introduced above, as the starting point of our waveform comparison, since the cLQBH waveform reaches its peak at approximately the same time (see, e.g., Fig.~\ref{plotFR4S1}). 
For example, for $\{r_0/M,\mathcal{A},\mathcal{W}\}=\{18,0.004,1.25\}$ in the middle panel of Fig.~\ref{plottdev1}, the estimated peak-time difference is $|t_{\mathrm{peak,cLQBH}}-t_{\mathrm{peak,GR}}|/M\simeq0.0004$. After the peak, the signal generally decays, and deviations that emerge sufficiently late may therefore occur when its amplitude has already become very small. This motivates introducing an approximate faintness threshold to determine the endpoint of the interval, in the spirit of the relative-amplitude cutoffs used in numerical ringdown analyses \cite{Berti:2007fi}. We denote the first post-peak time at which a suitable amplitude measure falls below this threshold by $t_{\mathrm{faint}}$. The interval between $t_{\mathrm{peak}}$ and $t_{\mathrm{faint}}$ then provides a simple framework for examining whether the bump-induced deviations emerge while the signal retains an appreciable amplitude. Although a quantitative assessment of GW detectability requires gravitational perturbations, source properties, and the detector response and noise, the scalar waveforms obtained here allow us to explore the interplay between waveform deviation and signal decay and its dependence on the cLQBH parameters. Beyond providing physical insight into the bump effect, this analysis can serve as a prototype for more complete investigations of its observational implications.

Relative-amplitude cutoffs have been used to determine the endpoint of numerical ringdown fits. In particular, Berti et al. truncate each multipolar waveform when its amplitude falls below $10^{-3}$ or $10^{-4}$ of its own peak amplitude to limit late-time numerical contamination \cite{Berti:2007fi,Berti:2007dg}. Following this relative-amplitude prescription, we introduce a fixed fraction $\eta$ of the approximate peak amplitude of each cLQBH waveform. Writing $\Psi_{\mathrm{cLQBH}}$ for the waveform denoted by $\Psi$ above, we define $\Psi_{\mathrm{peak}}\equiv|\Psi_{\mathrm{cLQBH}}(t_{\mathrm{peak}},x_{\mathrm{obs}})|$, using the common reference time $t_{\mathrm{peak}}$, which approximately marks the peaks of both the cLQBH and GR waveforms in Fig.~\ref{plotFR4S1}. For each parameter choice, $t_{\mathrm{faint}}$ is the first post-peak time at which the local amplitude measure defined below falls below $\eta\,\Psi_{\mathrm{peak}}$. The chosen $\eta$ provides a reasonable and practical reference for comparing the waveforms, rather than an absolute boundary beyond which the signal becomes unobservable for any realistic analyses. 

To determine $t_{\mathrm{faint}}$ systematically, we estimate the local amplitude by averaging the absolute waveform values at two successive stationary points. Here we use $t_k$ to label successive stationary points (local maxima and minima identified approximately from the sampled waveform) of $\Psi_{\mathrm{cLQBH}}(t,x_{\mathrm{obs}})$, ordered chronologically. These selected times are generally not equally spaced. We define
\begin{equation}
\label{eqofPsibar}
\bar{\Psi}(t_k)\equiv\frac{\left|\Psi_{\mathrm{cLQBH}}(t_k,x_{\mathrm{obs}})\right|+\left|\Psi_{\mathrm{cLQBH}}(t_{k+1},x_{\mathrm{obs}})\right|}{2}.
\end{equation}
The average is assigned to the earlier extremum time $t_k$. Using the extrema avoids interpreting individual zero crossings of the oscillatory waveform as a loss of signal amplitude.

As illustrated in Fig.~\ref{plotFR4S1}, the cLQBH and GR waveforms nearly coincide around their maxima, so the fixed reference time $t_{\mathrm{peak}}$ provides a good approximation to their common peak time. Starting from this reference, we identify the first extremum time $t_k$ at which $\bar{\Psi}(t_k)\leq\eta\,\Psi_{\mathrm{peak}}$. For the cases considered here, the initial post-peak estimate exceeds the threshold, so the preceding estimate satisfies $\bar{\Psi}(t_{k-1})>\eta\,\Psi_{\mathrm{peak}}$. We then determine $t_{\mathrm{faint}}$ by linear interpolation between these two amplitude estimates,
\begin{equation}
\label{eqoftfaint}
t_{\mathrm{faint}}\approx t_{k-1}+\frac{\eta\,\Psi_{\mathrm{peak}}-\bar{\Psi}(t_{k-1})}{\bar{\Psi}(t_k)-\bar{\Psi}(t_{k-1})}\,(t_k-t_{k-1}),
\end{equation}
which is similar to the treatment of \eqref{eqoftdev}.
This gives the first threshold crossing of the linearly interpolated amplitude measure, allowing $t_{\mathrm{faint}}$ to lie between extrema rather than restricting it to a stationary point. We evaluate this condition separately for each cLQBH waveform, using its own approximate peak amplitude $\Psi_{\mathrm{peak}}$. If the condition is never satisfied within the available data, no $t_{\mathrm{faint}}$ is reported for that evolution.

We adopt the representative value $\eta=0.003$, corresponding to an average amplitude of $0.3\%$ of the approximate waveform peak. This is of the same order as the $10^{-3}$ relative-amplitude cutoffs discussed above and provides a practical reference for identifying a substantially decayed signal. Keeping $\eta$ fixed allows a consistent comparison across the cLQBH parameter choices. As with $t_{\mathrm{dev}}$, the extracted time depends on the stated criterion, and its sensitivity to $\eta$ can be assessed by varying the threshold while keeping the waveform data and amplitude prescription fixed.
The interval $t\in[t_{\mathrm{peak}},t_{\mathrm{faint}})$ then defines an ``observation window'' anticipated in Sec.~\ref{Introduction}. We use this window to assess whether appreciable departures from GR emerge while the cLQBH signal retains an appreciable relative amplitude. In particular, $t_{\mathrm{peak}}\leq t_{\mathrm{dev}}<t_{\mathrm{faint}}$ indicates that the deviation criterion is met before the adopted faintness threshold is reached. We now compare $t_{\mathrm{dev}}$ and $t_{\mathrm{faint}}$ as the bump parameters vary.

\begin{figure}[t]
\centering
\includegraphics[width=0.95\linewidth]{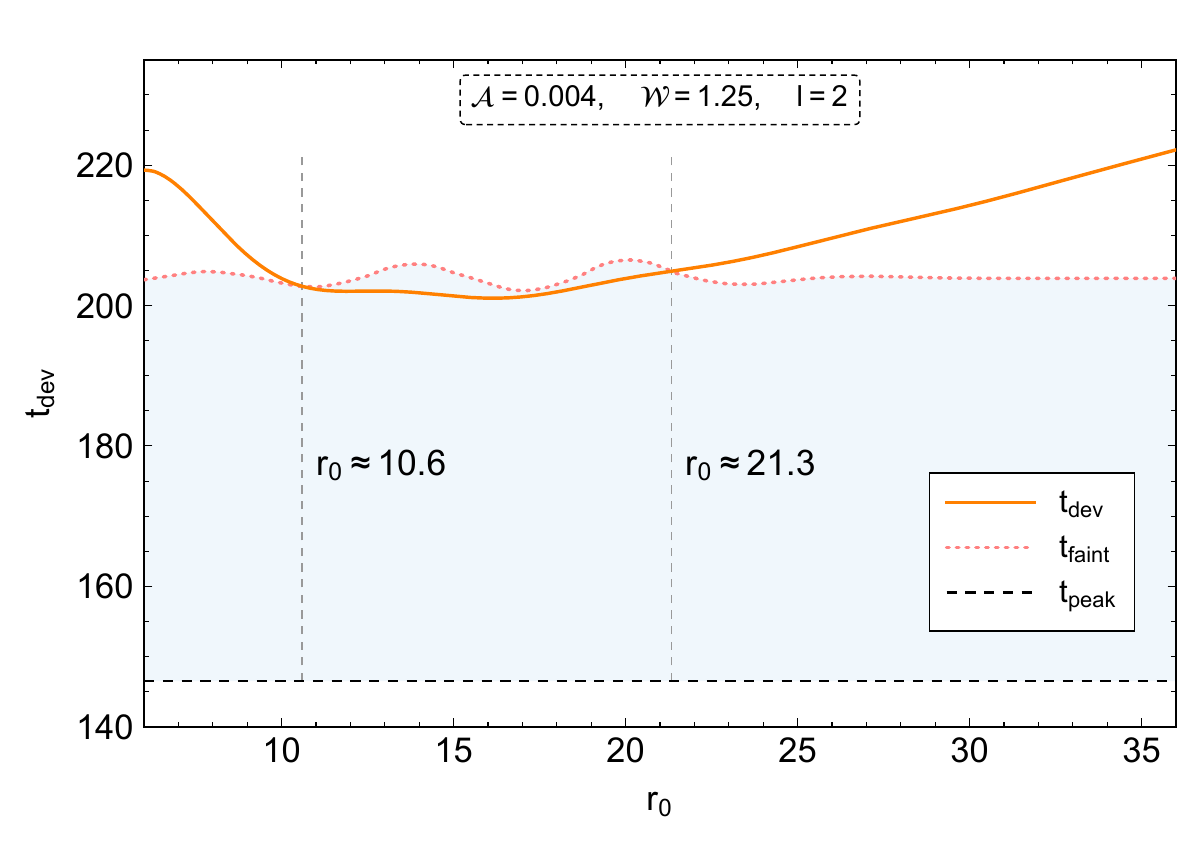}
\caption{The behavior of $t_{\mathrm{dev}}$ for $\mathcal{A}=0.004$, $\mathcal{W}=1.25$, $l=2$, and $x_{\mathrm{obs}}/M=150$, with varying $r_0/M\in[6,36]$ in units with $c=G=M=1$. The orange solid, pink dotted, and black dashed lines represent $t_{\mathrm{dev}}$ [cf. Eq.~\eqref{eqoftdev}], $t_{\mathrm{faint}}$ [cf. Eq.~\eqref{eqoftfaint}], and $t_{\mathrm{peak}}$, respectively. The two vertical dashed lines indicate the values of $r_0/M$ for which $t_{\mathrm{dev}}=t_{\mathrm{faint}}$.}
\label{plottdev3}
\end{figure}

In Fig.\ref{plottdev3}, we present our first example illustrating the behavior of $t_{\mathrm{dev}}$ for the case discussed in Fig.\ref{plottdev1}, with the curves of $t_{\mathrm{faint}}$ and $t_{\mathrm{peak}}$ also shown for reference. The light-blue shaded region marks the time interval between $t_{\mathrm{peak}}$ and $t_{\mathrm{faint}}$. According to the definitions and discussion given above, a $t_{\mathrm{dev}}$ lying within this light-blue region indicates that the associated bump effect is more likely to be observable from the QNM point of view. It is therefore clear that a ``bowl-shaped'' region, enclosed by the curves of $t_{\mathrm{dev}}$ and $t_{\mathrm{faint}}$, exists between $r_0 \approx 10.6$ and $r_0 \approx 21.3$ (in units where $c=G=M=1$). Within this region, the deviation between the cLQBH and GR signals occurs before the signal becomes too faint to be noticeable (based on our current definitions). This result provides a clue to, and an explanation of, the fact that a large deviation of the QNM from its GR counterpart alone does not guarantee its detectability. Instead, owing to the late-time features of the bump effect, we must also ask whether the corresponding $t_{\mathrm{dev}}$ can enter the favored region of possible detection, as illustrated by the light-blue region in Fig.\ref{plottdev3}. We therefore use this criterion to define the observational window mentioned above. Specifically, the bump effect can be employed to search for geometries whose parameters yield a $t_{\mathrm{dev}}$ curve that partially enters the light-blue shaded region shown in Fig.\ref{plottdev3}. Otherwise, it falls outside the observational window, even if the resultant QNM exhibits a sufficiently large deviation from its GR counterpart and is thus distinguishable by detectors at an adequate signal-to-noise ratio.

\begin{figure}[h]
\includegraphics[width=0.9\linewidth]{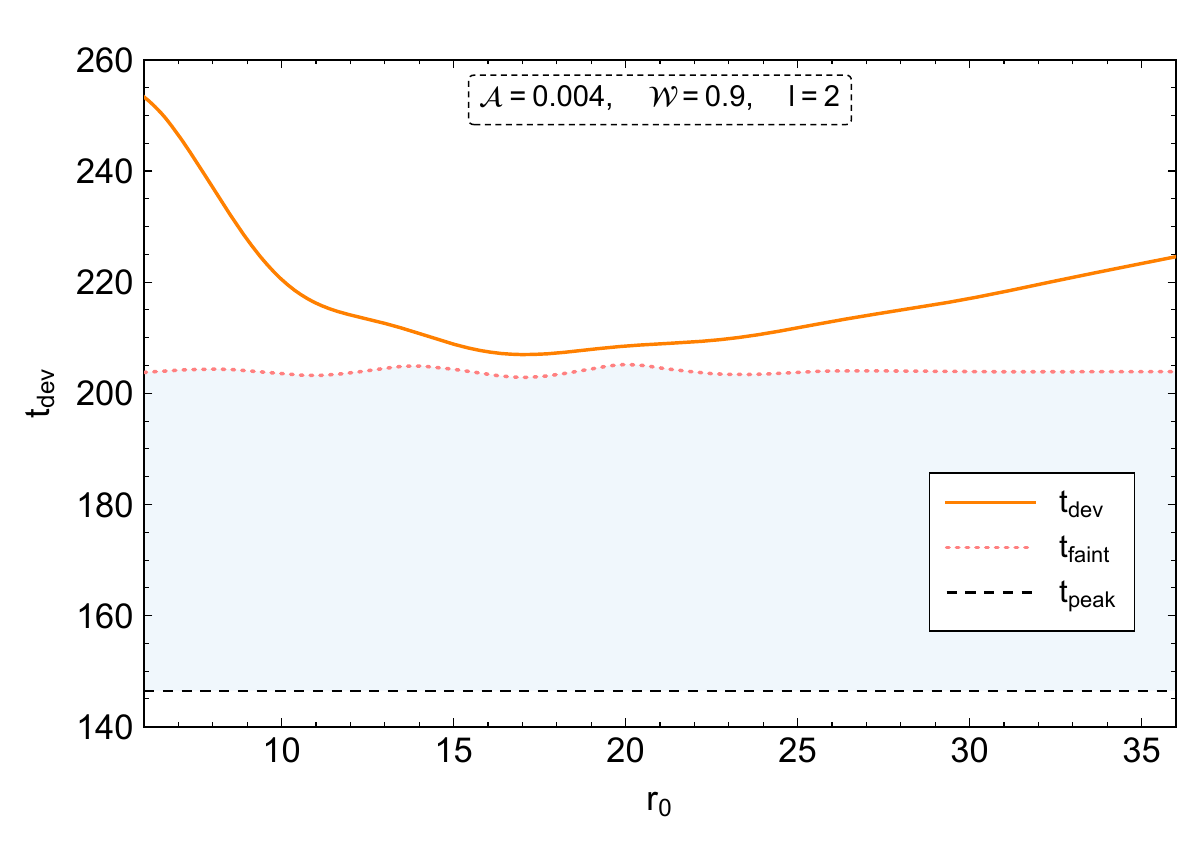}
\includegraphics[width=0.9\linewidth]{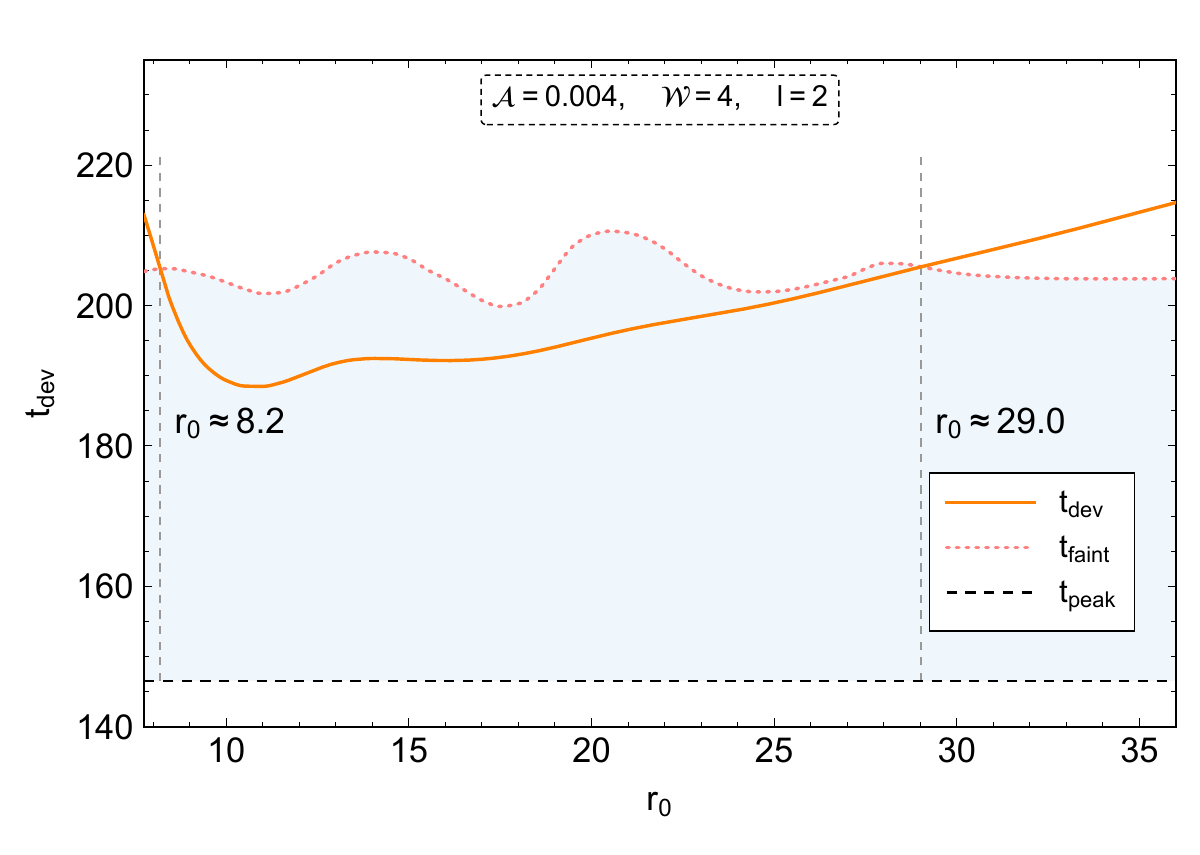}
\caption{The behavior of $t_{\mathrm{dev}}$ for the $\mathcal{A}=0.004$, $l=2$ and $x_{\mathrm{obs}}/M=150$ case with a varying $r_0$ (under the $c=G=M=1$ unit). The upper panel sues the $\mathcal{W}=0.9$ while the lower panel uses the $\mathcal{W}=4$. The orange solid line, pink dotted line and the black dashed line represent the $t_{\mathrm{dev}}$ [cf., \eqref{eqoftdev}], $t_{\mathrm{faint}}$ [cf., \eqref{eqoftfaint}] and $t_{\mathrm{peak}}$, respectively. Two vertical dashed lines are added to the lower panel so that to indicate the location of $r_0$, with which we have $t_{\mathrm{dev}}=t_{\mathrm{faint}}$.}
\label{plottdev4}
\end{figure}

As expected, varying $\mathcal{A}$ and $\mathcal{W}$ can change the extent of the bowl-shaped region discussed above. In particular, for certain combinations of $\{\mathcal{A},\mathcal{W}\}$, the $t_{\mathrm{dev}}$ curve may lie entirely above the $t_{\mathrm{faint}}$ curve, indicating that the waveform deviations become appreciable only after the signal has fallen below the adopted faintness threshold. An example is shown in the upper panel of Fig.~\ref{plottdev4}. Conversely, suitable parameter choices can enlarge the bowl-shaped region, allowing the bump effect to become potentially observable over a wider range of $r_0$, as illustrated in the lower panel of Fig.~\ref{plottdev4}.
Since localized potential bumps may also arise from matter surrounding a BH~\cite{Cheung:2021bol}, these results may have implications for environmental effects on BH ringing. In particular, the location of such matter could play an important role in whether the associated waveform deviations become appreciable before the signal becomes too faint.

\begin{figure}[t]
\centering
\includegraphics[width=0.8\linewidth]{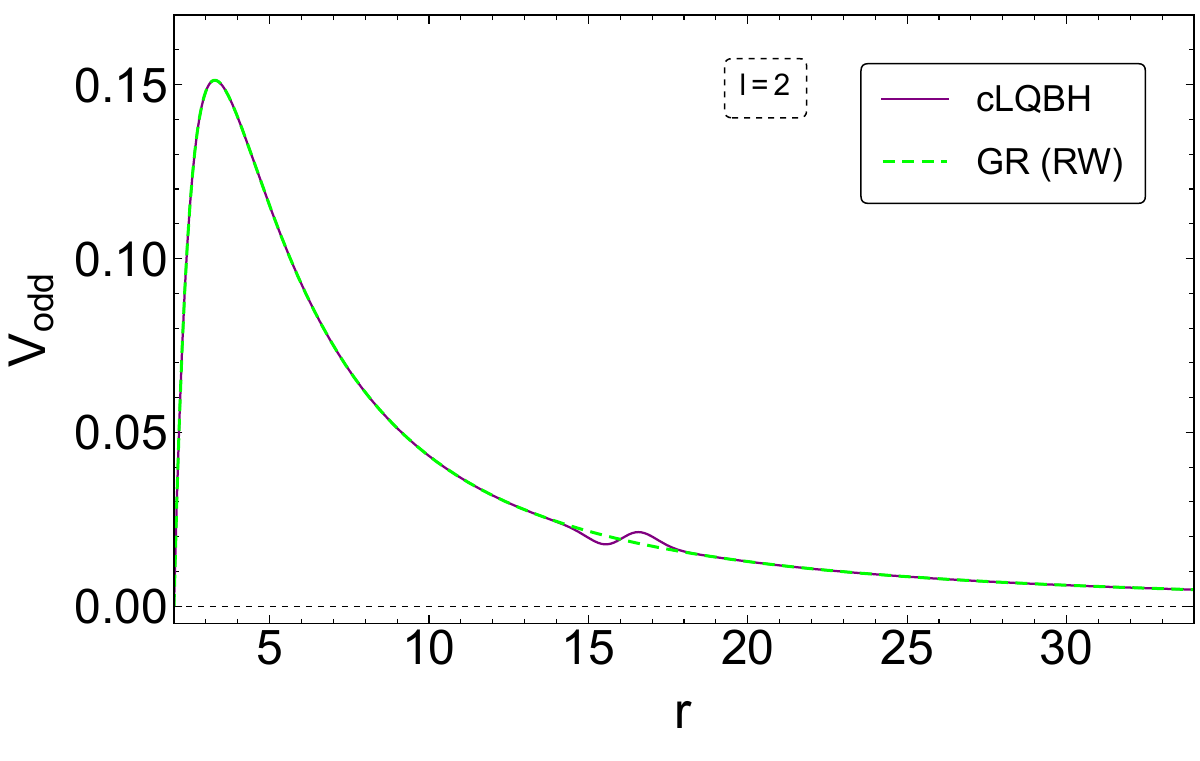}
\includegraphics[width=0.9\linewidth]{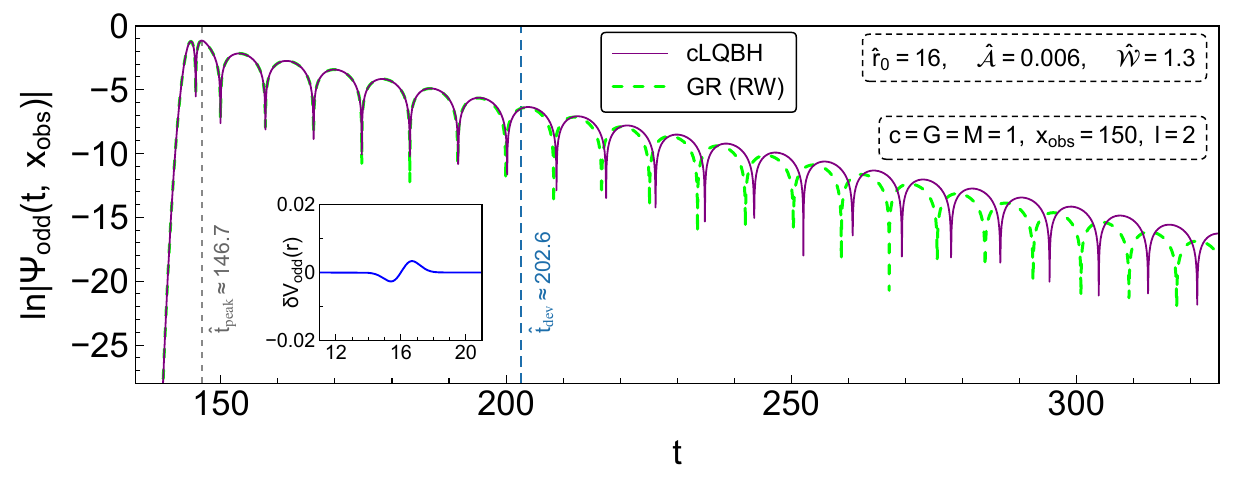}
\includegraphics[width=0.9\linewidth]{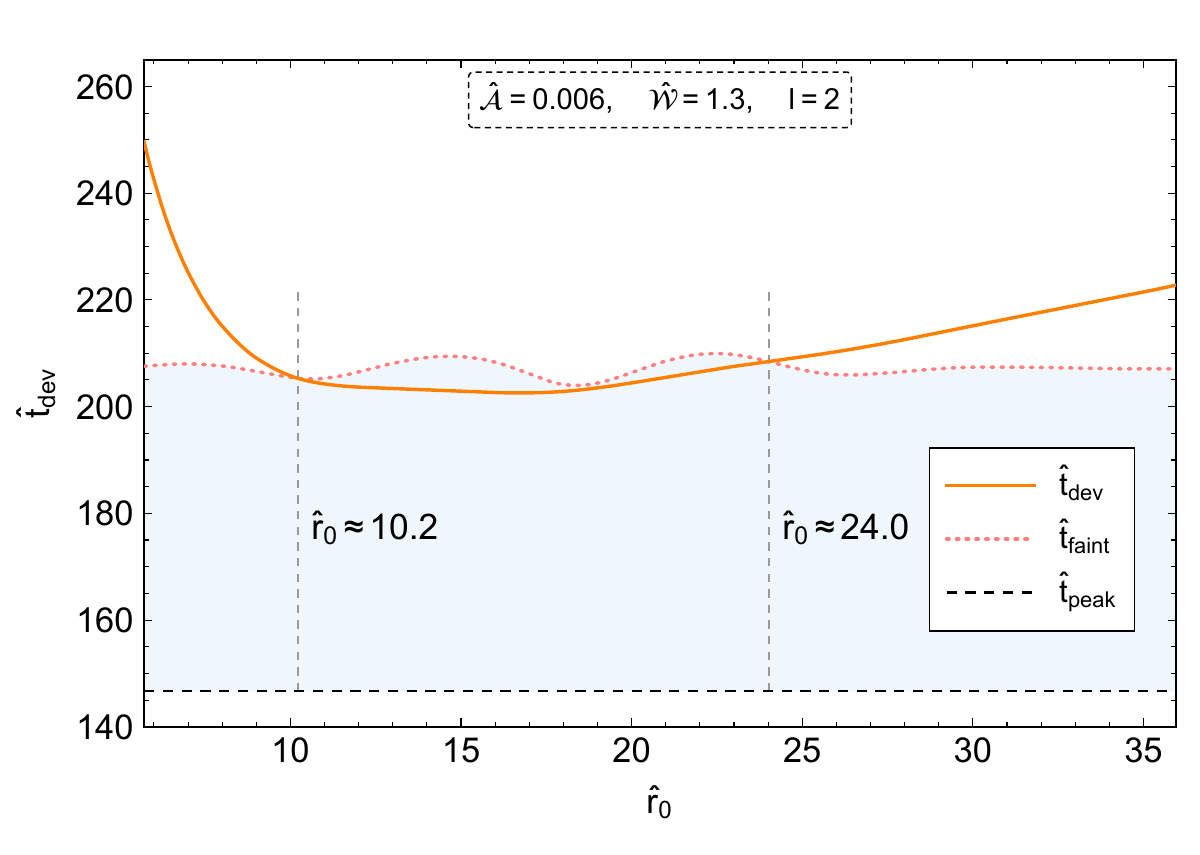}
\caption{
\textcolor{black}{
Results within the effective odd-parity prescription for $\hat{\mathcal{A}}=0.006$, $\hat{\mathcal{W}}=1.3$, and $l=2$, in units with $c=G=M=1$. The top and middle panels compare the effective potential and time-domain waveform at $\hat r_0=16$ with their GR (RW) counterparts. The inset shows the shape of $\delta V_{\mathrm{odd}}$. In the middle panel, the gray and steel-blue vertical dashed lines mark $\hat t_{\mathrm{peak}}$, defined by the maximum of $\ln|\Psi_{\mathrm{RW}}|$, and the threshold-based deviation time $\hat t_{\mathrm{dev}}$, respectively. The bottom panel shows $\hat t_{\mathrm{dev}}$ as $\hat r_0$ varies, with $\hat t_{\mathrm{faint}}$ and $\hat t_{\mathrm{peak}}$ cureves included for reference. The pale-blue region marks the adopted observation window, and the gray vertical dashed lines locate the intersections of $\hat t_{\mathrm{dev}}$ and $\hat t_{\mathrm{faint}}$. All waveforms are evaluated at $x_{\mathrm{obs}}=150$.
}}
\label{plotodd}
\end{figure}

\textcolor{black}{
Building on the above results, we now address several key points and discuss their implications in turn. First, the observation window identified above does not imply that every parameter choice within the bowl-shaped region produces QNM frequency shifts large enough to be resolved by current or future detectors. As discussed earlier in this section, the comparison of $t_{\mathrm{dev}}$ with $t_{\mathrm{faint}}$ and the fractional frequency deviations $\delta\omega_{R,I}$ in \eqref{deltaomega} provide complementary assessments of observational relevance. More generally, a detectable departure from GR requires the relevant frequencies to fall within the detector's sensitive band, the waveform differences to be distinguishable at the available signal-to-noise ratio after accounting for parameter degeneracies, and the deviations to emerge while sufficient signal remains~\cite{Berti:2005ys,Berti:2018vdi,Shi:2024ttu}. Favorable conditions in only some of these respects need not ensure detectability. Accordingly, we assess the spectral and temporal criteria jointly through representative examples, leaving systematic parameter scans, consistent gravitational-perturbation calculations, and dedicated GW data analysis to future work. For example, the parameter choice $\{r_0,\mathcal{A},\mathcal{W}\}=\{18,0.004,1.25\}$ within the bowl-shaped region of Fig.~\ref{plottdev3} yields a dominant scalar QNM frequency $M\omega_{20}\simeq0.495349-0.088255i$, corresponding to $\delta\omega_R\approx+2.42\%$ and $\delta\omega_I\approx+8.79\%$ according to \eqref{deltaomega}. Both deviation magnitudes also exceed the approximate TianQin reference accuracies adopted around Fig.~\ref{plotomegal2n0}. This example demonstrates that appreciable frequency shifts can coexist with waveform deviations emerging within the adopted observation window, providing a concrete motivation for considering the spectral and temporal criteria together when assessing the observational relevance of a QNM problem. It is therefore important to consider both aspects when evaluating the detectability of a QNM signal.
}

\textcolor{black}{
This brings us to a more careful examination of the combined assessment of the detectability of the QNM frequency shifts in our cLQBH model, although a more complete investigation of this kind should be left to future work, as mentioned above. As an illustrative example, the parameter choice $\{r_0/M,\mathcal{A},\mathcal{W}\}=\{20,0.0012,3\}$ leads to $M\omega_{20}\simeq0.477956-0.084511i$. According to \eqref{deltaomega}, the corresponding fractional deviations from GR are $\delta\omega_R\approx-1.2\%$ and $\delta\omega_I\approx+12.6\%$. Using the approximate TianQin reference accuracies introduced around Fig.~\ref{plotomegal2n0}, both deviations exceed the adopted reference intervals and are therefore potentially resolvable by TianQin, subject to the observational qualifications discussed above. In contrast, for case (i), applying the same criteria and assigning the same amplitude to $\delta V_{\mathrm{i}}$ yields only $\delta\omega_R\approx+0.02\%$ and $\delta\omega_I\approx-1.97\%$. Both deviations are smaller in magnitude than the adopted TianQin reference accuracies and would therefore remain unresolved by this criterion. This specific example illustrates a crucial point: by taking advantage of the bump effect, one can potentially amplify the spectral signatures of quantum corrections to LQBHs and their QNMs through the enhanced response of the spectrum to localized potential deformations, and importantly, despite the special form of $\lambda(r)$ adopted here, this amplification is not peculiar to that profile but reflects a generic feature of the bump effect, as argued in previous studies~\cite{Cheung:2021bol,Berti:2022xfj,Yang:2024vor}. This possibility is particularly relevant because, in many LQBH models, the quantum corrections admitted by the underlying theory or arising in physically natural parameter regimes produce only weak QNM deviations from GR, making these deviations difficult to identify observationally~\cite{Zhang:2024svj,del-Corral:2022kbk}. Although the holonomy profile in case (ii) is phenomenological and is primarily intended to produce a simple localized potential deformation whose location, amplitude, and width can be controlled separately, it is nevertheless an admissible choice within the background framework considered here. As pointed out in the preceding section, this makes it a suitable first step in this direction. More importantly, it demonstrates a possible and potentially viable route for amplifying the QNM signatures of quantum corrections to BH geometries, thereby opening an accessible observational window for distinguishing the relevant QNMs from their GR counterparts. Further applications of this mechanism merit deeper investigation.
}

\textcolor{black}{
To further investigate the generic character of the mechanism illustrated in Figs.~\ref{plottdev3} and~\ref{plottdev4}, we consider an illustrative odd-parity QNM calculation using the metric-based effective potential employed in Refs.~\cite{Bouhmadi-Lopez:2020oia,del-Corral:2022kbk} and in the dressed-metric treatment of Ref.~\cite{Garcia-Quismondo:2026tys}. Here, we incorporate quantum corrections through the background metric but do not include possible additional corrections to the perturbation dynamics. This provides an effective odd-parity model whose consistency with the complete cLQBH gravitational dynamics remains to be established.
Within this prescription, the master equation retains the form of \eqref{master1}, with $\Psi$ replaced by $\Psi_{\mathrm{odd}}$, $V_{\mathrm{eff}}$ replaced by $V_{\mathrm{odd}}=V_{\mathrm{RW}}+\delta V_{\mathrm{odd}}$, and the tortoise coordinate still defined by \eqref{rstar}. The two contributions to $V_{\mathrm{odd}}$ are given by
\begin{equation}
\label{Vodd}
\begin{aligned}
V_{\mathrm{RW}}
&=f\left(\frac{L}{r^2}-\frac{6M}{r^3}\right),\\
\delta V_{\mathrm{odd}}
&=\frac{2f^3\lambda^2}{r^2}-\frac{1}{2r}\frac{d(f^3\lambda^2)}{dr} =\frac{2f^3\lambda^2}{r^2}-V_{\mathrm{bump}}.
\end{aligned}
\end{equation}
Here, $V_{\mathrm{RW}}$ is the Schwarzschild Regge--Wheeler (RW) potential~\cite{del-Corral:2022kbk}, and $\delta V_{\mathrm{odd}}$ plays the role of the localized correction in this odd-parity problem. For the configurations considered below, it contains a negative trough followed by a positive peak, reversing the ordering found in the scalar bump. Following the conventions in Fig.~\ref{plotVbump1}, we characterize this correction by its zero-crossing location $\hat r_0$, its dimensionless peak-to-trough amplitude $\hat{\mathcal{A}}$, and its dimensionless peak-to-trough separation $\hat{\mathcal{W}}$. These quantities are determined from $\delta V_{\mathrm{odd}}$ and generally differ from the corresponding scalar quantities for the same background, although, as will be seen shortly, they can be quite close numerically.
}

\textcolor{black}{
We choose $\{\hat r_0,\hat{\mathcal{A}},\hat{\mathcal{W}}\}=\{16,0.006,1.3\}$ as a representative example, adopting $c=G=M=1$. The corresponding scalar parameterization of the same geometry is $\{r_0,\mathcal{A},\mathcal{W}\}\approx\{16.11,0.00602,1.298\}$. Thus, the two parameterizations are numerically close in this example. The top panel of Fig.~\ref{plotodd} compares $V_{\mathrm{odd}}$ with $V_{\mathrm{RW}}$ for $l=2$, while the middle panel shows the corresponding FR4S waveforms at $x_{\mathrm{obs}}=150$, using the same initial data and numerical procedures as for the scalar-perturbation case.
The middle panel shows initially close agreement with the GR waveform, denoted by $\Psi_{\mathrm{RW}}$, followed by an appreciable deviation at later times, consistent with previous time-domain studies of a RW potential supplemented by a localized bump~\cite{Berti:2022xfj} and with our scalar results (see, e.g., Fig.~\ref{plotFR4S1}). The extracted complex QNM frequency for the $\{l,n\}=\{2,0\}$ mode is $M\omega_{20}^{\mathrm{odd}}\simeq0.363764-0.081773i$. The superscript ``odd'' distinguishes this frequency from the scalar results.
Comparing this result with the Schwarzschild value $M \omega_{20}^{\mathrm{RW}} \simeq0.3736716844-0.0889623157i$~\cite{Matyjasek:2019eeu} using \eqref{deltaomega} yields $\delta\omega_R\approx-2.65\%$ and $\delta\omega_I\approx+8.08\%$. Both magnitudes comfortably exceed the approximate TianQin reference accuracies adopted around Fig.~\ref{plotomegal2n0}. This result helps motivate future investigations with consistent gravitational waveforms as a promising extension of the present analysis.
More importantly, to examine the temporal criterion, we define $\hat t_{\mathrm{dev}}$ and $\hat t_{\mathrm{faint}}$ using the same prescriptions as \eqref{eqoftdev} and \eqref{eqoftfaint}, now comparing $\Psi_{\mathrm{odd}}$ with $\Psi_{\mathrm{RW}}$ and determining the faintness time separately from each odd-parity waveform.
The reference time $\hat t_{\mathrm{peak}}$ is obtained from the RW waveform. For the example above, $\hat t_{\mathrm{peak}}\approx146.7$ and $\hat t_{\mathrm{dev}}\approx202.6$, as indicated in the middle panel. Keeping $\hat{\mathcal{A}}=0.006$ and $\hat{\mathcal{W}}=1.3$ fixed while varying the actual zero-crossing location $\hat r_0$ produces the bottom panel. A bowl-shaped region again appears, with $\hat t_{\mathrm{dev}}<\hat t_{\mathrm{faint}}$ between the intersections at $\hat r_0\approx10.2$ and $\hat r_0\approx24.0$. The representative case $\hat r_0=16$ therefore satisfies both the temporal and spectral criteria adopted here. The persistence of this behavior in the effective odd-parity calculation supports its relevance beyond the scalar examples. Although it does not establish universality for arbitrary backgrounds or potential deformations, this similarity between two distinct perturbation problems suggests that analyses based on the deviation time offer a promising direction for future detectability studies.
}

\textcolor{black}{
Finally, the results presented here highlight the bridge that our work builds between the investigation of QNMs in LQBHs and the generic bump effect. On the one hand, as discussed above, our analysis serves as a prototype for a more complete framework and provides preliminary evidence that the bump effect can aid the prospective extraction of LQBH QNMs by offering a potentially viable route to amplify the spectral signatures of quantum corrections to the geometry and the associated QNM deviations from GR, thereby increasing the prospect that these signatures fall within the reach of the detectors considered here. Moreover, Figs.~\ref{plottdev3}, \ref{plottdev4}, and~\ref{plotodd} suggest a selection mechanism. For a localized correction to the background geometry, whether induced by an effective geometric deformation or by surrounding matter~\cite{Cheung:2021bol,Laeuger:2025zgb,Lan:2025brn}, observable QNM signatures may be favored only within certain parameter ranges, even when other parameter choices produce larger departures from GR. This behavior reflects a competition between the magnitude of the departure from GR and the time at which the waveform deviation emerges. Under the criteria adopted here, parameter choices that achieve a favorable balance between these two effects are more likely to yield observationally relevant QNM signals. 
On the other hand, our case (ii) [cf.~\eqref{eq1.1}] provides a concrete BH platform for investigating the bump effect. Several characteristic features identified in previous studies of prescribed potential deformations are recovered in this setting, including spiral trajectories of the QNM frequencies in the complex plane, mode overtaking, and waveforms in the time domain that initially remain close to their GR counterparts before departing at later times~\cite{Leung:1999iq,Cheung:2021bol,Berti:2022xfj,Yang:2024vor}, as illustrated by Figs.~\ref{plotspiral} and~\ref{plotFR4S2}, together with Table~\ref{tableomegal2n012}. The cLQBH results further illustrate the value of combining spectral and temporal criteria in assessing the observational relevance of the bump effect, providing a preliminary step from theoretical modeling toward observational applications. The flexibility of case (ii) also allows other aspects of the bump response to be examined, including its amplitude, width, and location dependences, as illustrated in Figs.~\ref{plotomegal2n0}, \ref{plottrajectoryofvaryingW}, and~\ref{plottdev3}.
}

\section{Conclusions}
\renewcommand{\theequation}{5.\arabic{equation}} \setcounter{equation}{0}
\label{secconclusion}

In this paper, we have investigated the QNM spectrum and time-domain response of a massless, minimally coupled test scalar field on static, spherically symmetric cLQBH backgrounds. Our main focus has been a phenomenological holonomy function with a Gaussian-localized radial profile, with the previously studied decreasing profile serving as a reference. The localized holonomy correction modifies both the scalar effective potential and the relation between the areal radius and the tortoise coordinate. Consequently, this model provides a realization of the bump effect in which the scattering potential and wave-propagation geometry follow consistently from the same quantum-corrected background.

The resulting potential deformation contains a positive peak and a negative trough. We have characterized it through its dimensionless location \(r_0/M\), dimensionless amplitude \(\mathcal{A}\), and dimensionless width \(\mathcal{W}\), allowing the effects of these quantities to be examined separately. Our calculations concentrate on configurations for which the total effective potential remains positive throughout the black hole exterior. Within this parameter regime, localized corrections produce a variety of spectral behaviors. Varying the bump location generates spiral trajectories of the QNM frequencies in the complex plane, while changes in the amplitude and width affect both the oscillation frequencies and damping rates. The response depends on the combined parameter choices, and the damping can either increase or decrease relative to Schwarzschild.

Tracking the \(l=2\) modes continuously from their Schwarzschild limits reveals that the branches associated with the first two overtones are substantially more sensitive to the localized correction than the continued fundamental branch. In particular, the continued \(n=1\) branch can overtake the continued \(n=0\) branch and become the longest-lived among the three tracked modes. Here, the overtone indices identify continuation from Schwarzschild rather than an ordering by the damping rate at each parameter value. This change in lifetime ordering occurs without a coalescence of the tracked complex frequencies and no exceptional point was identified in the cases examined.

Our time-domain calculations complement these spectral results. For representative configurations, the early cLQBH response remains close to its Schwarzschild counterpart, whereas the subsequent ringing exhibits different oscillation and decay patterns. Frequencies extracted using the matrix pencil method agree closely with the frequency-domain shooting results, and reconstructions using the dominant extracted modes reproduce the numerical waveforms over the selected ringdown intervals. These comparisons also confirm the mode-overtaking behavior: in the representative two-mode example, the longer-lived contribution corresponds to the branch continued from the first Schwarzschild overtone. Other configurations are accurately described by a single dominant mode, demonstrating that appreciable waveform changes do not necessarily require competition between comparably excited modes.

To examine when these differences become significant relative to signal decay, we have introduced a deviation time \(t_{\mathrm{dev}}\), defined through a Gaussian-weighted, normalized waveform distance. We compare it with a faintness time \(t_{\mathrm{faint}}\), determined by a prescribed relative-amplitude threshold. Together with the reference peak time, these quantities define an operational observation window in which the deviation criterion is met before the waveform falls below the adopted amplitude threshold. The dependence of \(t_{\mathrm{dev}}\) on the bump location is nonmonotonic. For example, with \(\mathcal{A}=0.004\), \(\mathcal{W}=1.25\), and the initial data, observation prescription, and thresholds adopted here, the condition \(t_{\mathrm{peak}}\leq t_{\mathrm{dev}}<t_{\mathrm{faint}}\) holds approximately for \(10.6<r_0/M<21.3\). Changing the width can enlarge this interval or remove it over the sampled range. Thus, substantial QNM frequency shifts alone do not establish that the associated waveform differences emerge while the signal retains an appreciable relative amplitude.
\textcolor{black}{(Conversely, waveform deviations appearing before substantial signal decay do not by themselves imply sufficiently large frequency shifts to be resolved.)}

\textcolor{black}{
The examples in Sec.\ref{seckey} demonstrate the value of assessing spectral shifts and waveform timing together. Potentially resolvable frequency shifts can coexist with waveform deviations emerging before the signal becomes too faint, and our matched-amplitude comparison shows that case (ii) can produce substantially larger shifts than case (i). This suggests that the bump effect may provide a route to amplify the weak quantum signatures expected in physically natural regimes of many LQBH models. Although case (ii) is phenomenological, it is admissible within the cLQBH framework and illustrates a generic response to localized potential deformations rather than an artifact of the particular profile. The bowl-shaped windows further suggest a selection mechanism that favors parameter choices balancing the magnitude of the departure from GR with its emergence time, rather than simply maximizing the frequency shift. The appearance of a similar window in the effective odd-parity calculation supports its relevance beyond the scalar examples without establishing universality. In turn, the cLQBH model provides a concrete BH setting for advancing bump physics from spectral characterization toward observationally oriented waveform analysis.
}

The observation window is a diagnostic of the calculated scalar response, and its boundaries depend on the stated deviation and faintness criteria. Likewise, our comparisons with reference accuracies motivated by LIGO observations and TianQin forecasts are exploratory and do not establish observational exclusions of the model. Quantitative gravitational-wave predictions require gravitational perturbation equations consistent with the cLQBH dynamics, extensions to rotating backgrounds, and analyses incorporating mode excitation, source properties, and detector noise. Alongside these developments, further investigations could extend the scalar analysis to nonminimal coupling, systematically search for exceptional points, and explore the possible connection between localized potential deformations and effective reflectivity \cite{Oshita:2025ibu}. The present results provide a framework for connecting localized holonomy corrections, QNM spectral changes, and the temporal development of waveform deviations, helping identify configurations that warrant more complete investigations.



\section*{Acknowledgments}

C.Z. would like to thank Profs. WANG Peng, LIN Kai and YANG Runqiu, and Drs. GAN Wencong,  MAI Zhanfeng and WU Liangbi for valuable discussions.  A.W. and C.Z. also want to thank Prof. Shinji Mukohyama for suggestions and valuable discussions. 
C.Z. is supported in part by the National Natural Science Foundation of China (NSFC) under Grant No. 12205254, and the Science Foundation of China University of Petroleum, Beijing, under Grant No.~2462024BJRC005. 
A.W. is partially  supported by the US National Science Foundation (NSF) through the grant: PHY-2308845.




\end{document}